%% file: bordiga_lattice_15-06-2018.tex
\documentclass[11pt,a4paper]{article}

\usepackage[T1]{fontenc}
\usepackage{amssymb,amsbsy}
\usepackage{amsmath}
\usepackage{amsthm}
\usepackage{amsfonts}
\usepackage{graphicx}
\usepackage[textfont=footnotesize,labelfont={footnotesize,bf}]{caption}
\usepackage{subcaption}
\usepackage{stackengine}
\usepackage{tikz}
\usepackage[colorinlistoftodos,textwidth=25mm,textsize=footnotesize]{todonotes}
\usepackage{booktabs}
\usepackage{color}
\usepackage{enumitem}
\usepackage[affil-it,auth-sc]{authblk}
\usepackage{stmaryrd}           
\usepackage[top=25mm,right=25mm,left=25mm,bottom=25mm]{geometry}
\usepackage{hyperref}
\hypersetup{colorlinks=true,linkcolor=blue,urlcolor=blue}
\usepackage[normalem]{ulem}     

\usepackage[autostyle]{csquotes}
\usepackage[
    backend=bibtex,
    style=numeric,
    citestyle=numeric-comp,
    maxbibnames=99,
    minbibnames=99,
    maxcitenames=2,
    mincitenames=1,
    date=year,
    giveninits=true,
    sorting=none,
    sortlocale=auto,
    natbib=true,
    url=false, 
    isbn=false,
    doi=true,
    eprint=true
]{biblatex}
\input{definitions}		        
\graphicspath{{./}{./figures/}} 

\title{Free and forced wave propagation in a Rayleigh-beam grid: flat bands, Dirac cones, and vibration localization vs isotropization}

\author[1]{G. Bordiga}
\author[1]{L. Cabras}
\author[1]{D. Bigoni\footnote{Corresponding author: e-mail: \href{mailto:bigoni@ing.unitn.it}{bigoni@ing.unitn.it}; phone: +39\,0461\,282507.}}
\author[1]{A. Piccolroaz}

\affil[1]{Department of Civil, Environmental and Mechanical Engineering, University of Trento, Italy}

\date{}

\begin{document}

\maketitle

\begin{abstract}
\noindent
In-plane wave propagation in a periodic rectangular grid beam structure, which includes rotational  inertia (so-called \lq Rayleigh beams'), is analyzed both with a Floquet-Bloch exact formulation for free oscillations and with a numerical treatment (developed with PML absorbing boundary conditions) for forced vibrations (including Fourier representation and energy flux evaluations), induced by a concentrated force or moment. 
A complex interplay is observed between axial and flexural vibrations (not found in the common idealization of out-of-plane motion), giving rise to several forms of vibration localization: \lq X-', \lq cross-' and \lq star-' shaped, and channel propagation.  
These localizations are triggered by several factors, including rotational inertia and slenderness of the beams and the type of forcing source (concentrated force or moment). 
Although the considered grid of beams introduces an orthotropy in the mechanical response, a surprising \lq isotropization' of the vibration is observed at special frequencies. Moreover, rotational inertia is shown to \lq sharpen' degeneracies related to Dirac cones (which become more pronounced when the aspect ratio of the grid is increased), while 
the slenderness can be tuned to achieve a perfectly flat band in the dispersion diagram. The obtained results can be exploited in the realization of metamaterials designed to control wave propagation. 
\end{abstract}


\section{Introduction}
\label{sec:introduction}

Research on metamaterials (employed to guide and control elastic waves for applications in microstructured devices~\citep{sigmund_2003, wang_2015, wang_2015a, lim_2015, bacigalupo_2017, lepidi_2018, Antonakakis_2013} and earthquake resistant structures~\citep{Brun_2012, Brun_2013, Carta_2016, Colombi_2016, Achaoui_2017}) has focused a strong research effort to time-harmonic vibrations of periodic beam networks. 
These networks can be analyzed via Floquet-Bloch analysis for free vibrations of an infinite domain (which can be either \lq exact', when performed with a symbolic computation program~\citep{leamy_2012} or approximated, when solved numerically~\citep{phani_2006}), or using the f.e. methodology for forced vibrations of finite-size structures \citep{Piccolroaz_2017}.

Several topologies, vibration conditions and beam models have been considered for wave propagation in two-dimensional lattices, namely, hexagonal, triangular, and square honeycombs, re-entrant and Kagom\'e lattices~\citep{gonella_2008, spadoni_2009}, subject to out-of-plane motion with~\citep{Piccolroaz_2017, Piccolroaz_2017_2, Cabras_2017} or without~\citep{ruzzene_2003} rotational inertia (the so-called \lq Rayleigh correction', introducing a bound to  the phase and group velocity of a beam~\citep{kolsky_1963,Piccolroaz_2014}).

Forced vibrations of grid of beams has been considered for a two-dimensional mass/spring periodic structure~\citep{langley_1996}, while asymptotic approximations of lattice Green's functions have been given~\citep{movchan_2014, vanel_2016}, close to standing wave frequencies, with the purpose of revealing the directional anisotropy in two and three-dimensional periodic lattices.

Although in-plane vibrations of a rectangular grid of Rayleigh (axially and flexurally deformable) beams may be considered a mature research field, for which the governing equations and the solution techniques are well-known, many interesting features still remain to be explored. 
This exploration is provided in the present article, where an exact Floquet-Bloch analysis is performed and complemented with a numerical treatment of the forced vibrations induced by the application of a concentrated force or moment, including presentation of the Fourier transform and energy flow (treated in~\citep{bacigalupo_2018} for free vibrations).
It is shown that (i.) aspect ratio of the grid, (ii.) slenderness and (iii.) rotational inertia of the beams decide the emergence of several forms of highly-localized waveforms, namely, \lq channel propagation', \lq X-', \lq cross-', \lq star-' shaped vibration modes. 
Moreover, these mechanical properties of the grid can be designed to obtain flat bands and degeneracies related to Dirac cones in the dispersion diagram and directional anisotropy or, surprisingly, dynamic \lq isotropization', for which waves propagate in a square lattice with the polar symmetry characterizing propagation in an isotropic medium.

The presented results open the way to the design of vibrating devices with engineered properties, to achieve control of elastic wave propagation.

\section{In-plane Floquet-Bloch waves in a rectangular grid of beams}
\label{sec:formulation_problem}

\begin{figure}[htb!]
\centering
\begin{subfigure}{0.5\textwidth}
\centering
\caption{\label{fig:lattice}}
\includegraphics[width=0.95\linewidth]{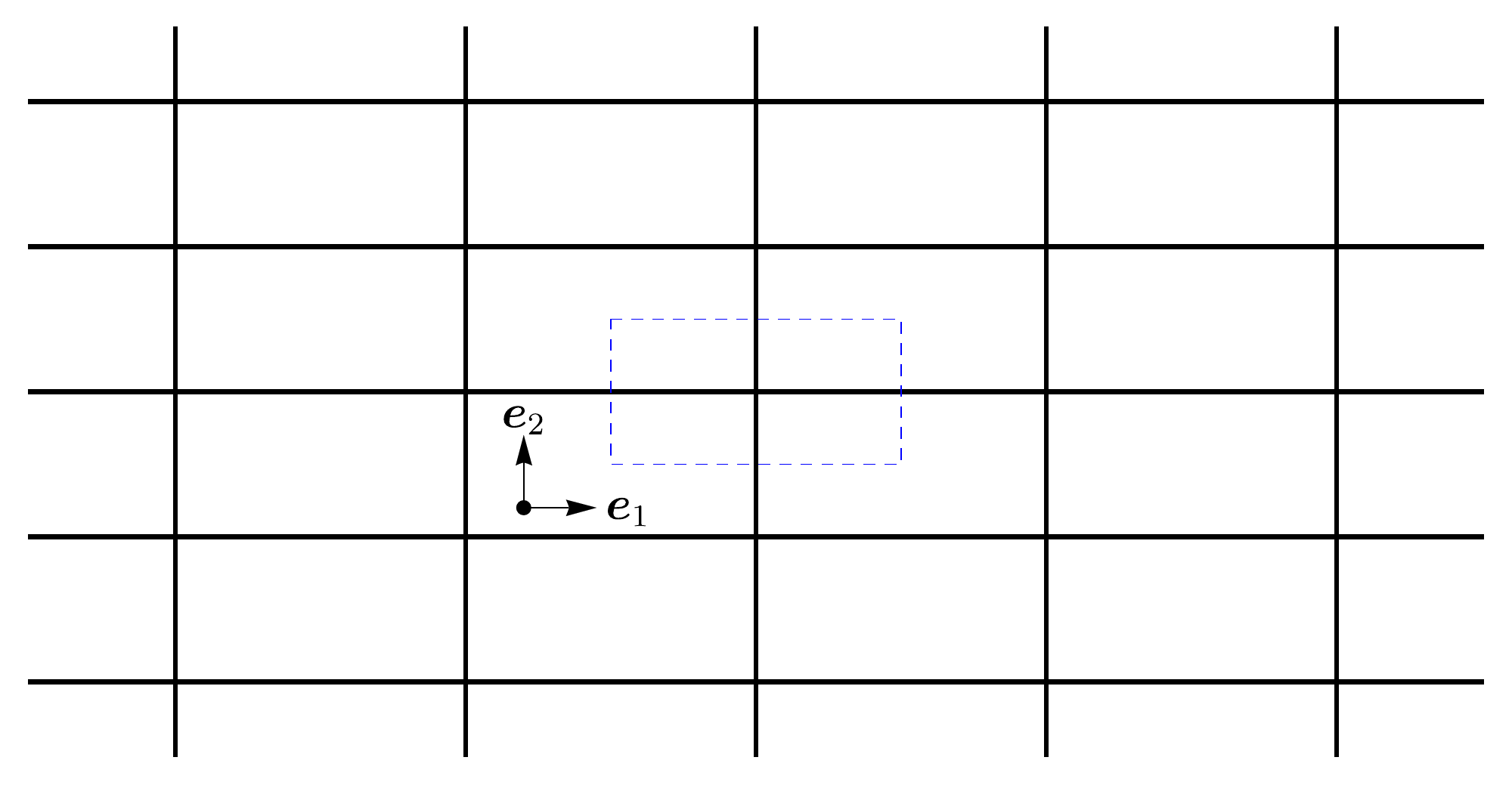}
\end{subfigure}%
\begin{subfigure}{0.5\textwidth}
\centering
\caption{\label{fig:unit_cell}}
\includegraphics[width=0.95\linewidth]{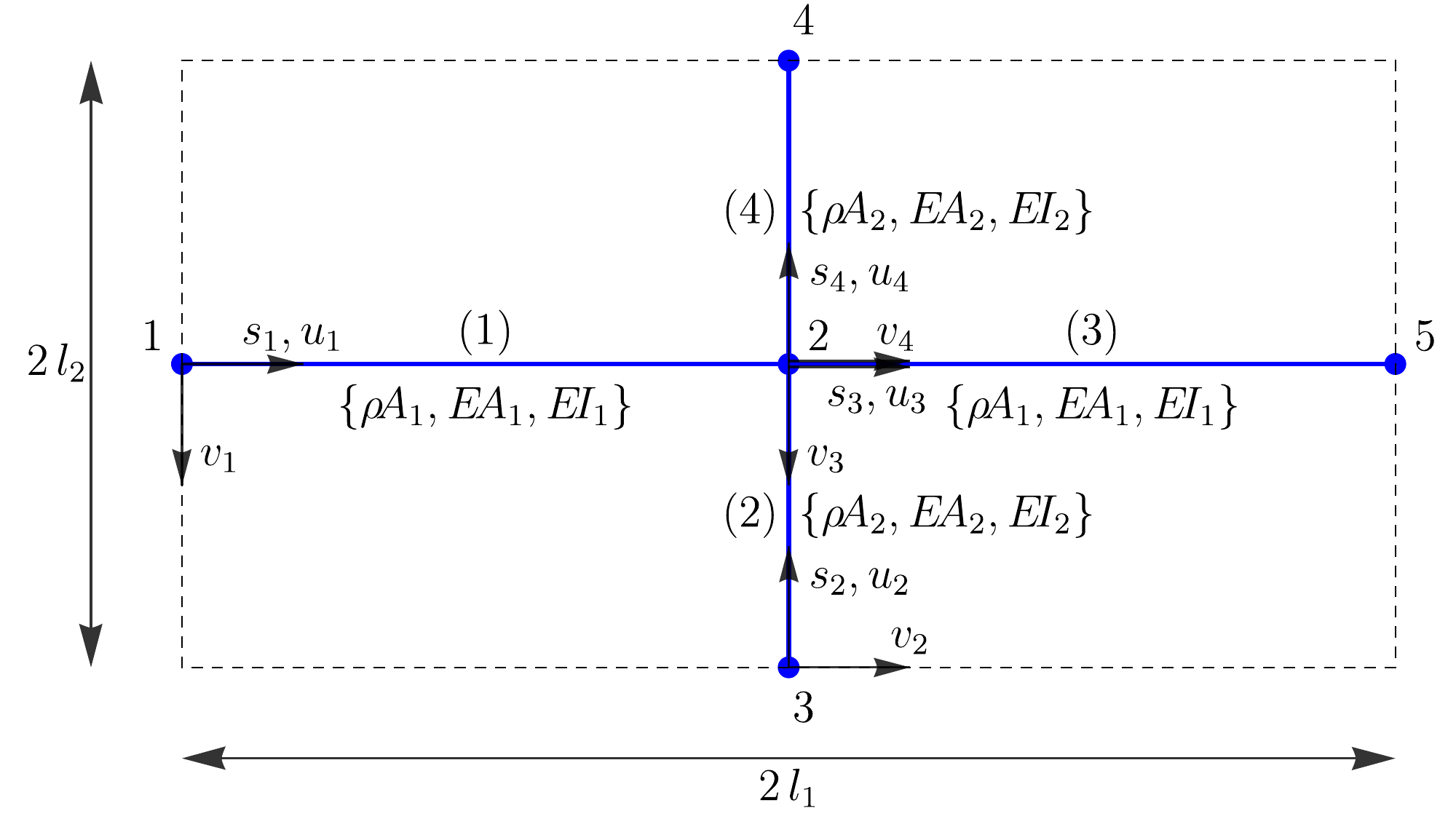}
\end{subfigure}
\caption{\label{fig:geometry}
Geometry of the grid-beam structure (\subref{fig:lattice}) and the relevant unit cell (\subref{fig:unit_cell}).}
\end{figure}

An infinite lattice of Rayleigh beams is considered, periodically arranged in a rectangular geometry as shown in Fig.~\ref{fig:lattice}, together with the unit cell, Fig.~\ref{fig:unit_cell}.

Each beam is assumed to be axially extensible and flexible, so that the equations governing the time-harmonic (in-plane) response are 
\begin{equation}
\label{eq:governing_equations}
{E A}\, \nderiv{u(s)}{s}{2} + \rho A\,\omega^2 u(s) = 0, \qquad
{E I}\,\nderiv{v(s)}{s}{4} + {\rho I}\,\omega^2 \nderiv{v(s)}{s}{2} - {\rho A}\,\omega^2 v(s) = 0,
\end{equation}
where $\rho$ is the mass density, $E$ the Young modulus, $A$ and $I$ are respectively the area and the second moment of inertia of the beam's cross-section, and $\omega$ is the angular frequency of the time-harmonic vibration.
The axial and transverse displacements are denote with $u(s)$ and $v(s)$, respectively, Fig.~\ref{fig:unit_cell}.
By setting $\xi=s/l$, with $l$ being the length of the beam, Eqs.~\eqref{eq:governing_equations} can be cast in the following dimensionless form
\begin{equation}
\label{eq:governing_equations_nondimensional}
u''(\xi) + \tilde{\omega}^2 u(\xi) = 0, \qquad 
v''''(\xi) + \tilde{\omega}^2 v''(\xi) - \frac{\lambda^2}{4}\,\tilde{\omega}^2 v(\xi) = 0,
\end{equation}
where $\tilde{\omega}=\omega\,l\sqrt{\rho/E}$ is a non-dimensional angular frequency, $\lambda=2\,l\sqrt{A/I}$ represents the slenderness of the beam, and the prime denotes differentiation with respect to $\xi$.

The general solution of Eqs.~\eqref{eq:governing_equations_nondimensional} is sought in the form
\begin{equation}
u(\xi) = \sum_{j=1}^2 C_j e^{i\,\eta_j\,\xi}, \qquad v(\xi) = \sum_{j=1}^4 D_j e^{i\,\gamma_j\,\xi},
\end{equation}
where the $C_j$ and $D_j$ denote 6 complex constants, while the $\eta_j$ and $\gamma_j$ are characteristic roots 
\begin{equation}
\label{eq:characteristic_roots}
\eta_{1,2} = \pm \tilde{\omega}, \qquad \gamma_{1,2,3,4} = \pm\sqrt{\frac{\tilde{\omega}}{2}\left(\tilde{\omega} \pm \sqrt{\lambda^2 + \tilde{\omega}^2}\right)},
\end{equation}

With the local coordinates shown in Fig.~\ref{fig:unit_cell}, the displacement field on each beam of the unit cell is 
\begin{equation}
\label{eq:solution_u_v_beams}
u_p(\xi_p) = \sum_{q=1}^2 C_{pq} e^{i\,\eta_q\,\xi_p}, \quad v_p(\xi_p) = \sum_{q=1}^4 D_{pq} e^{i\,\gamma_q\,\xi_p}, \quad \xi_p = s_p/l_p, \quad \forall \, p\in\{1,...,4\},
\end{equation}
where the 24 undetermined constants, $C_{pq}$ and $D_{pq}$, can be found by imposing kinematic compatibility and equilibrium conditions at the central junction, plus the Floquet-Bloch boundary conditions between corresponding sides of the unit cell.

Assuming for simplicity the elastic modulus $E$ and the mass density $\rho$ to be equal in all the beams, by choosing the following dimensionless variables
\begin{equation}
\label{eq:nondimensional_variables}
\tilde{\omega}_1 = \omega\,l_1\sqrt{\rho/E}, \quad \lambda_1 = 2\,l_1\sqrt{A_1/I_1}, \quad \lambda_2 = 2\,l_2\sqrt{A_2/I_2}, \quad \alpha = l_1/l_2, \quad \chi = A_1/A_2,
\end{equation}
the linear system governing the time-harmonic oscillation of the lattice is defined as follows.

\begin{itemize}
\item Compatibility of displacements and rotations at the central node of the unit cell 
\begin{equation}
\begin{aligned}
\label{eq:bvp_unit_cell_compatibility}
v_1(1) &= v_3(0),		& v_2(1) &= v_4(0),		& v_1(1) &= -u_2(1), \\
u_1(1) &= u_3(0),		& u_2(1) &= u_4(0),		& v_2(1) &= u_1(1), \\
v_1'(1) &= \alpha\,v_2'(1),		& \alpha\,v_2'(1) &= v_3'(0),		& v_3'(0) &= \alpha\,v_4'(0),
\end{aligned}
\end{equation}
\item equilibrium of the central node 
\begin{equation}
\begin{aligned}
\label{eq:bvp_unit_cell_equilibrium}
u_3'(0) - u_1'(1) - \frac{4\,\alpha}{\chi \lambda_2^2} \,v_4'''(0) - \frac{4}{\chi \alpha \lambda_2^2}\,\tilde{\omega}_1^2 \,v_4'(0) + \frac{4\,\alpha}{\chi \lambda_2^2} \,v_2'''(1) + \frac{4}{\chi \alpha \lambda_2^2}\tilde{\omega}_1^2 \,v_2'(1) &= 0, \\
u_4'(0) - u_2'(1) - \frac{4\,\chi}{\alpha \lambda_1^2} \,v_1'''(1) - \frac{4\,\chi}{\alpha \lambda_1^2}\tilde{\omega}_1^2 \,v_1'(1) + \frac{4\,\chi}{\alpha \lambda_1^2} \,v_3'''(0) + \frac{4\,\chi}{\alpha \lambda_1^2}\tilde{\omega}_1^2 \,v_3'(0) &= 0, \\
v_3''(0) + \frac{\lambda_1^2}{\chi \lambda_2^2} \,v_4''(0) - v_1''(1) - \frac{\lambda_1^2}{\chi \lambda_2^2} \,v_2''(1) &= 0,
\end{aligned}
\end{equation}
\item Floquet-Bloch boundary conditions
\begin{equation}
\begin{aligned}
\label{eq:bvp_unit_cell_bloch}
u_3(1) &= u_1(0)\, e^{i\,K_1}, \\
v_3(1) &= v_1(0)\, e^{i\,K_1}, \\
v_3'(1) &= v_1'(0)\, e^{i\,K_1}, \\
u_3'(1) &= u_1'(0)\, e^{i\,K_1}, \\
v_3'''(1) + \tilde{\omega}_1^2 \,v_3'(1) &= (v_1'''(0) + \tilde{\omega}_1^2 \,v_1'(0))\, e^{i\,K_1}, \\
v_3''(1) &= v_1''(0)\, e^{i\,K_1}, \\
u_4(1) &= u_2(0)\, e^{i\,K_2/\alpha}, \\
v_4(1) &= v_2(0)\, e^{i\,K_2/\alpha}, \\
v_4'(1) &= v_2'(0)\, e^{i\,K_2/\alpha} \\
u_4'(1) &= u_2'(0)\, e^{i\,K_2/\alpha}, \\
v_4'''(1) + \frac{\tilde{\omega}_1^2}{\alpha^2} \,v_4'(1) &= \left(v_2'''(0) + \frac{\tilde{\omega}_1^2}{\alpha^2} \,v_2'(0)\right)\, e^{i\,K_2/\alpha}, \\
v_4''(1) &= v_2''(0)\, e^{i\,K_2/\alpha},
\end{aligned}
\end{equation}
where $K_1$ and $K_2$ are dimensionless components of the Bloch wave vector $\bk = k_1\be_1 + k_2\be_2$, namely, $K_1 = k_1 2l_1$, $K_2 = k_2 2l_1$.
\end{itemize}

Equations \eqref{eq:bvp_unit_cell_compatibility}--\eqref{eq:bvp_unit_cell_bloch} provide the complete set of equations governing the propagation of in-plane Floquet-Bloch waves for an infinite and periodic Rayleigh beam lattice.
The governing equations for the corresponding Euler-Bernoulli approximation can be easily obtained by neglecting the rotational inertia terms, $\rho I_1 = \rho I_2 = 0$, and by retaining only the low-frequency term of the flexural characteristic roots $\gamma_j$, i.e.\ $\gamma_{1,2,3,4} = \pm\sqrt{\pm\tilde{\omega}\lambda/2}$.

\section{Dispersion properties and Bloch waveforms}
\label{sec:dispersion_properties}

\subsection{Dispersion equation}
\label{sec:dispersion_equation}
A substitution of representation~\eqref{eq:solution_u_v_beams} into the boundary conditions~\eqref{eq:bvp_unit_cell_compatibility}--\eqref{eq:bvp_unit_cell_bloch} leads to an algebraic homogeneous linear system of the type
\begin{equation}
\label{eq:system}
\bA(\tilde{\omega}_1,\bK,\lambda_1,\lambda_2,\alpha, \chi) \, \bc = \b0,
\end{equation}
where $\bA$ is a $24\times24$ complex matrix, function of the dimensionless angular frequency $\tilde{\omega}_1$ and wave vector $\bK$, slenderness $\lambda_1$ and $\lambda_2$, aspect ratio $\alpha$ and geometric ratio $\chi$. 
Vector $\bc$ collects the 24 complex constants, $C_{pq}$ and $D_{pq}$, appearing in the displacement field, Eqs.~\eqref{eq:solution_u_v_beams}.

Introducing the following normalization 
\begin{equation}
\label{eq:normalization}
\Omega = \frac{4l_1^2\omega}{\pi^2\sqrt{E I_1/(\rho A_1)}}= \frac{2\lambda_1 \tilde{\omega}_1}{\pi^2}, 
\end{equation}
where the angular frequency $\omega$ has been made dimensionless through division by the first flexural natural frequency of a simply supported Euler-Bernoulli beam, 
the non-trivial solutions of the system~\eqref{eq:system} are found when the matrix $\bA$ is rank-deficient
\begin{equation}
\label{eq:dispersion}
\det\bA(\Omega,\bK, \lambda_1, \lambda_2, \alpha, \chi) = 0,
\end{equation}
which is the \textit{dispersion equation}, implicitly defining the relation between the angular frequency $\Omega$ and the wave vector $\bK$, namely,~the so-called \textit{dispersion relation}.
Furthermore, for each point of the $\{\Omega,\bK\}$-space satisfying Eq.~\eqref{eq:dispersion}, the corresponding eigenvector $\bc(\Omega,\bK)$ can be computed from~\eqref{eq:system}.

Hence, the propagation of Floquet-Bloch waves is governed by the generalized eigenvalue problem~\eqref{eq:system}, where the \textit{eigenfrequencies} are determined by the dispersion relation $\Omega(\bK)$, periodic with period $\left[0,2\pi\right]{\times}\left[0,2\pi\alpha \right]$, and the \textit{eigenmodes} (or waveforms) are defined by the eigenvectors $\bc(\Omega,\bK)$, to be inserted into Eqs.~\eqref{eq:solution_u_v_beams}.

\subsection{Dispersion surfaces: Euler-Bernoulli vs Rayleigh}
\label{sec:dispersion_surfaces}

\begin{figure}[htb!]
\centering
\begin{subfigure}{0.33\textwidth}
\centering
\caption{\label{fig:disp_surf_5_EB}}
\includegraphics[width=0.95\linewidth]{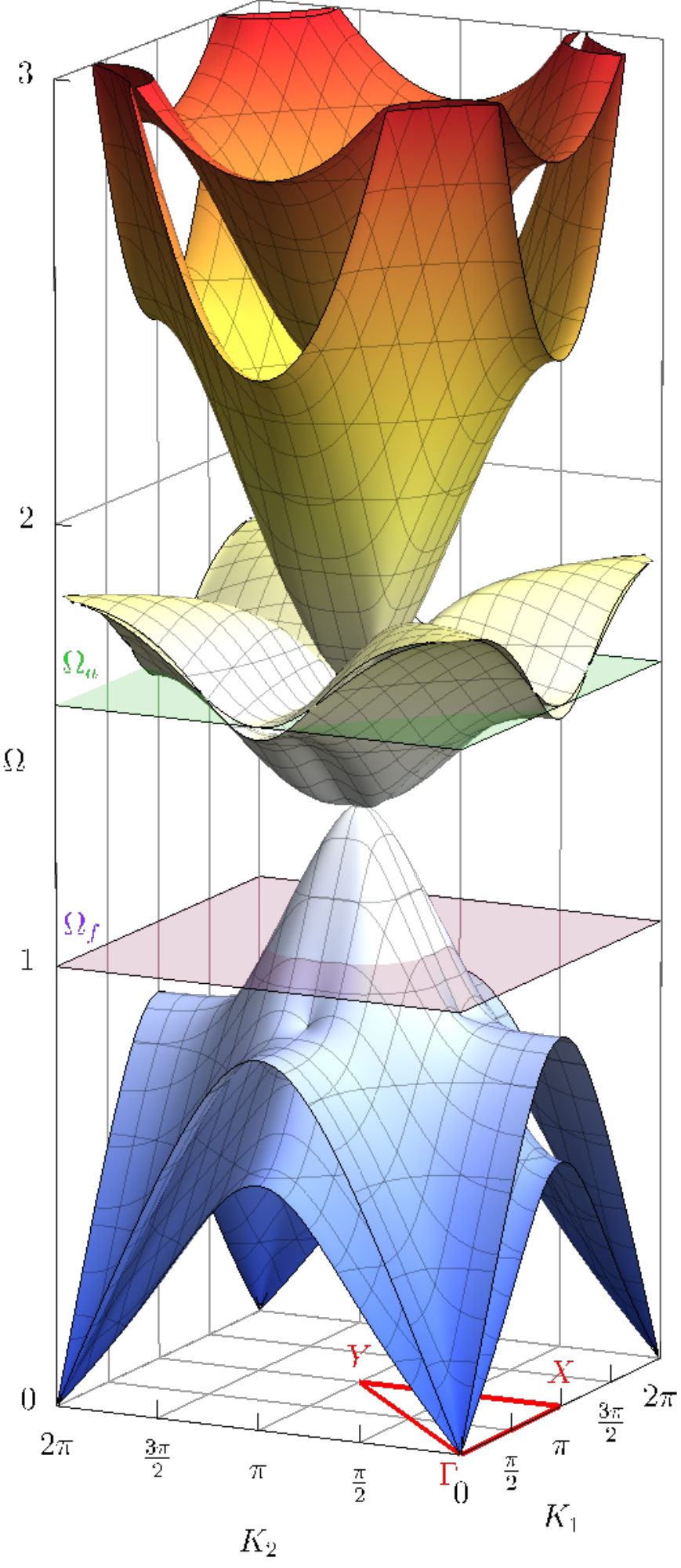}
\end{subfigure}%
\begin{subfigure}{0.33\textwidth}
\centering
\caption{\label{fig:disp_surf_5_Ray}}
\includegraphics[width=0.95\linewidth]{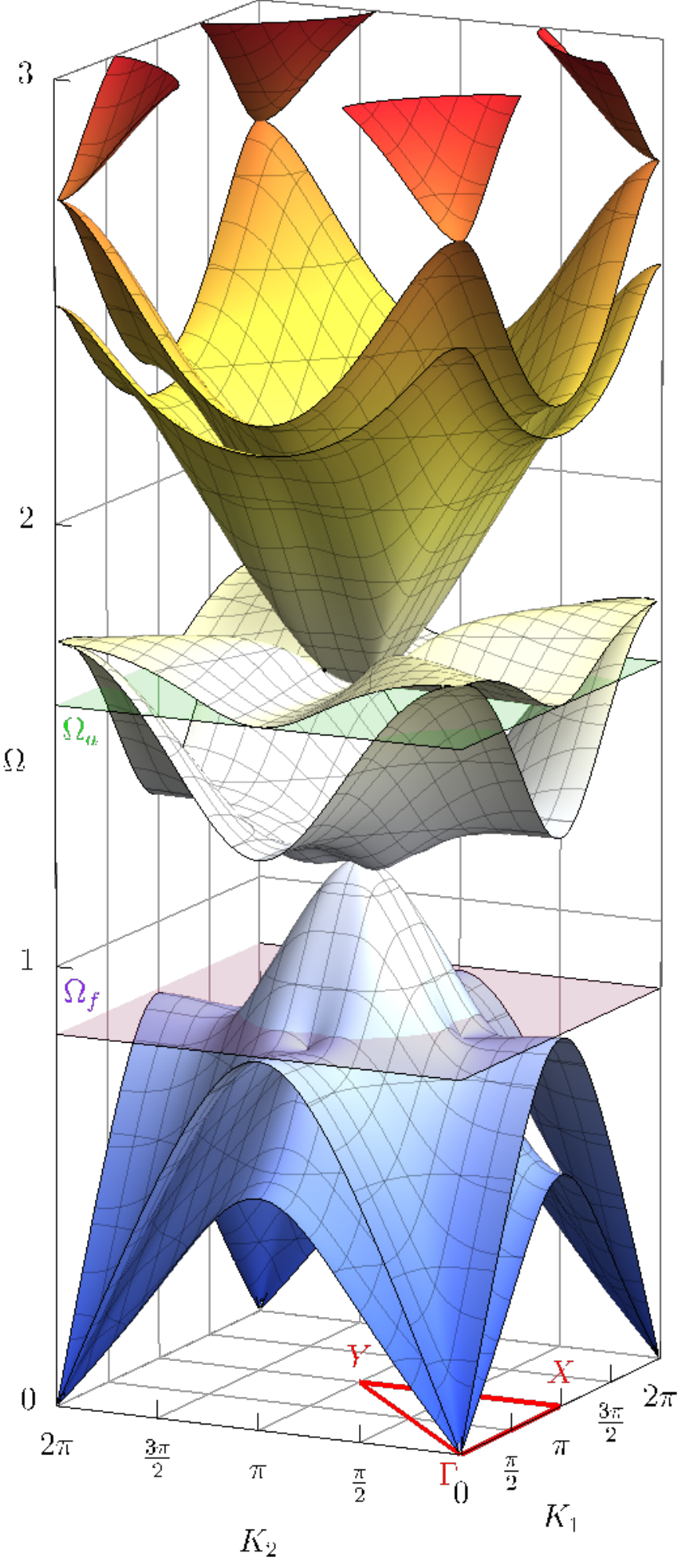}
\end{subfigure}%
\begin{subfigure}{0.33\textwidth}
\centering
\caption{\label{fig:disp_surf_10_Ray}}
\includegraphics[width=0.95\linewidth]{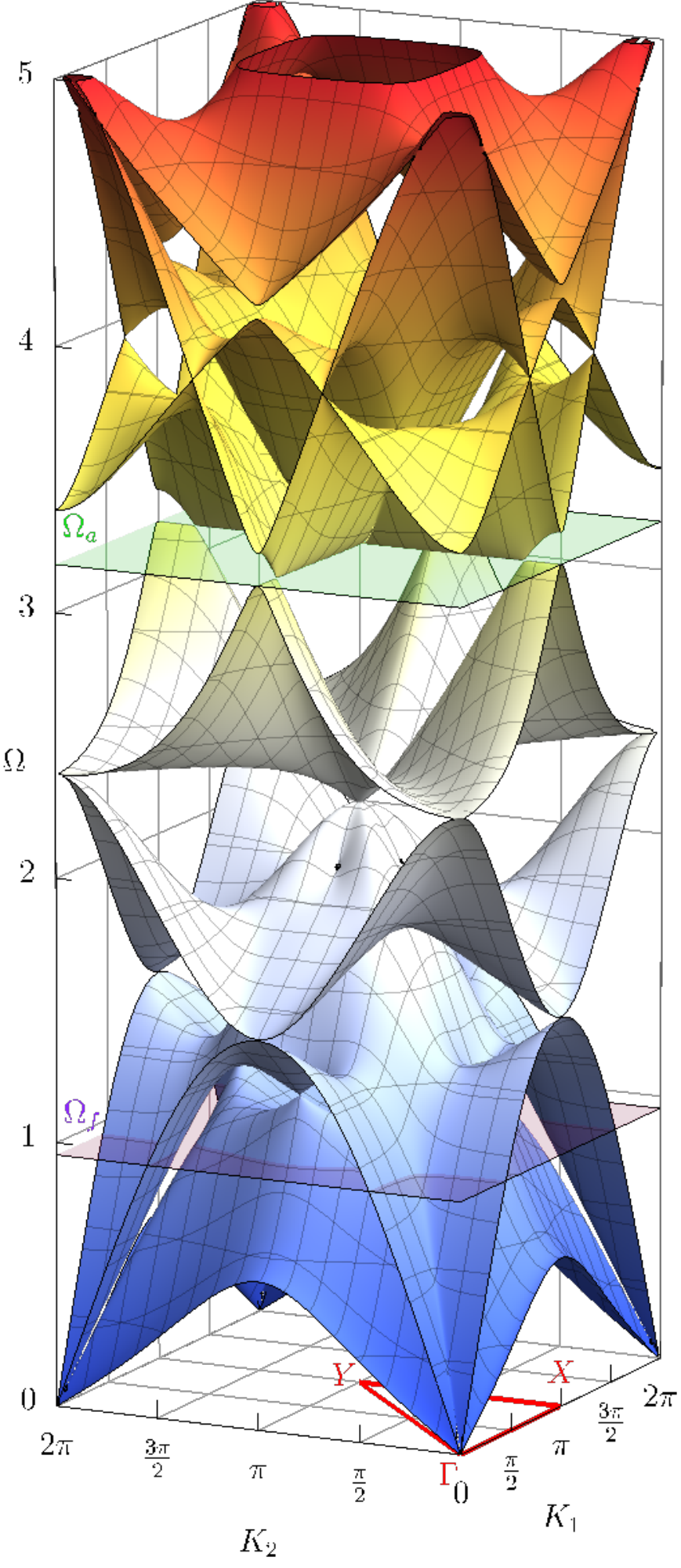}
\end{subfigure}
\caption{\label{fig:disp_surf_square}
The effect of rotational inertia (becoming important at high frequency) is evidenced by the differences between the 
dispersion surfaces for a square lattice made up of Euler-Bernoulli (\subref{fig:disp_surf_5_EB}) and Rayleigh (\subref{fig:disp_surf_5_Ray}) beams, both relative to a slenderness $\lambda=5$. 
The influence of the slenderness (already evident at low frequency) may be appreciated by comparing the case  $\lambda=5$ (\subref{fig:disp_surf_5_Ray}) with the case $\lambda = 10$ (\subref{fig:disp_surf_10_Ray}).
The green and pink horizontal planes denote two first natural frequencies of a double-pinned beam, namely, $\Omega_f$ corresponds to the flexural vibration and identifies the stationary point of the lowest dispersion surface (occurring at $\{K_1,K_2\}=\{\pi,\pi\}$).
$\Omega_a$ corresponds to the axial vibration and identifies the set of stationary points of the fourth and fifth surface located respectively at $K_1=\pi,\ \forall K_2$ and $K_2=\pi,\ \forall K_1$. 
The band diagrams corresponding to the path $\Gamma$--$X$--$Y$--$\Gamma$ sketched in the figure are reported in Fig.~\ref{fig:disp_diagram}.
}
\end{figure}

\begin{figure}[htb!]
\centering
\begin{subfigure}{0.5\textwidth}
\centering
\caption{\label{fig:disp_surf_6_Ray}}
\includegraphics[height=0.4\textheight]{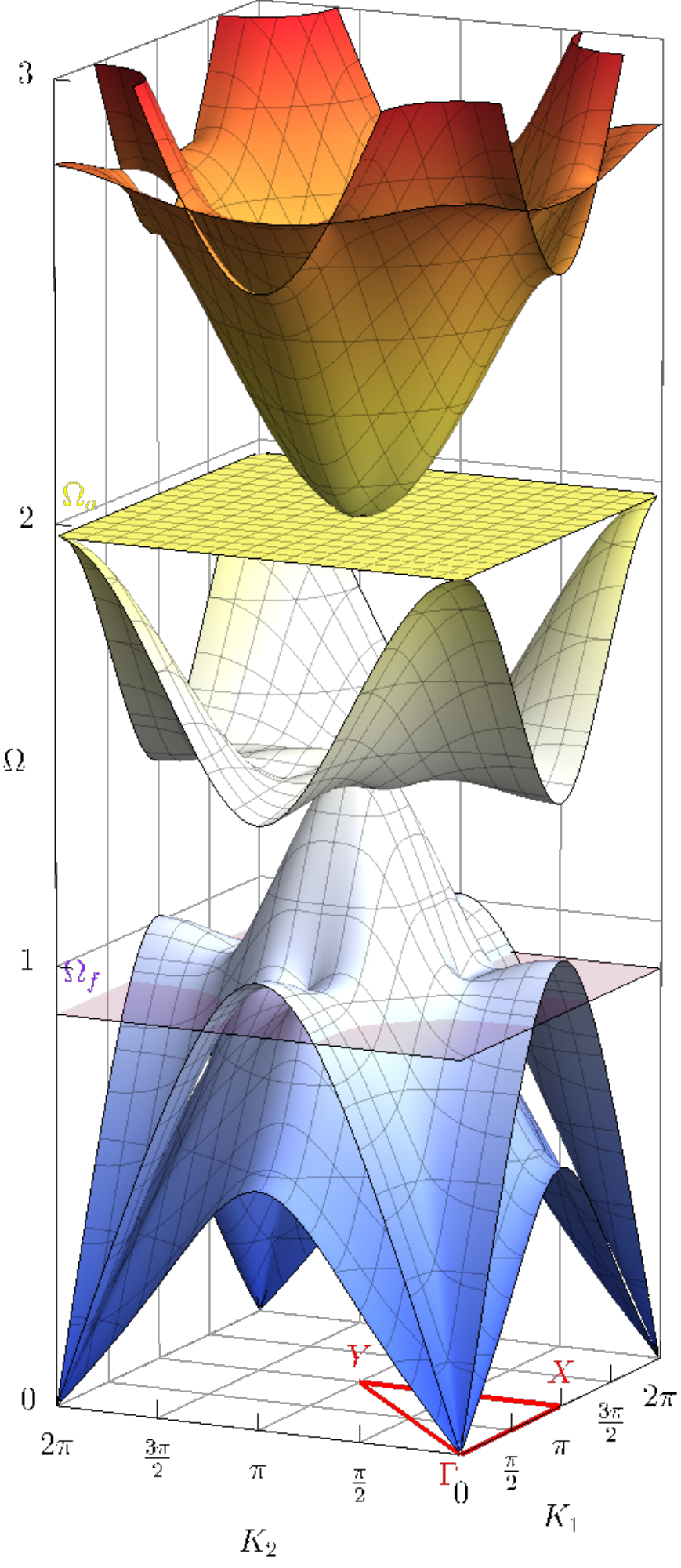}
\end{subfigure}%
\centering
\begin{subfigure}{0.5\textwidth}
\centering
\caption{\label{fig:disp_surf_10_5_Ray}}
\includegraphics[height=0.4\textheight]{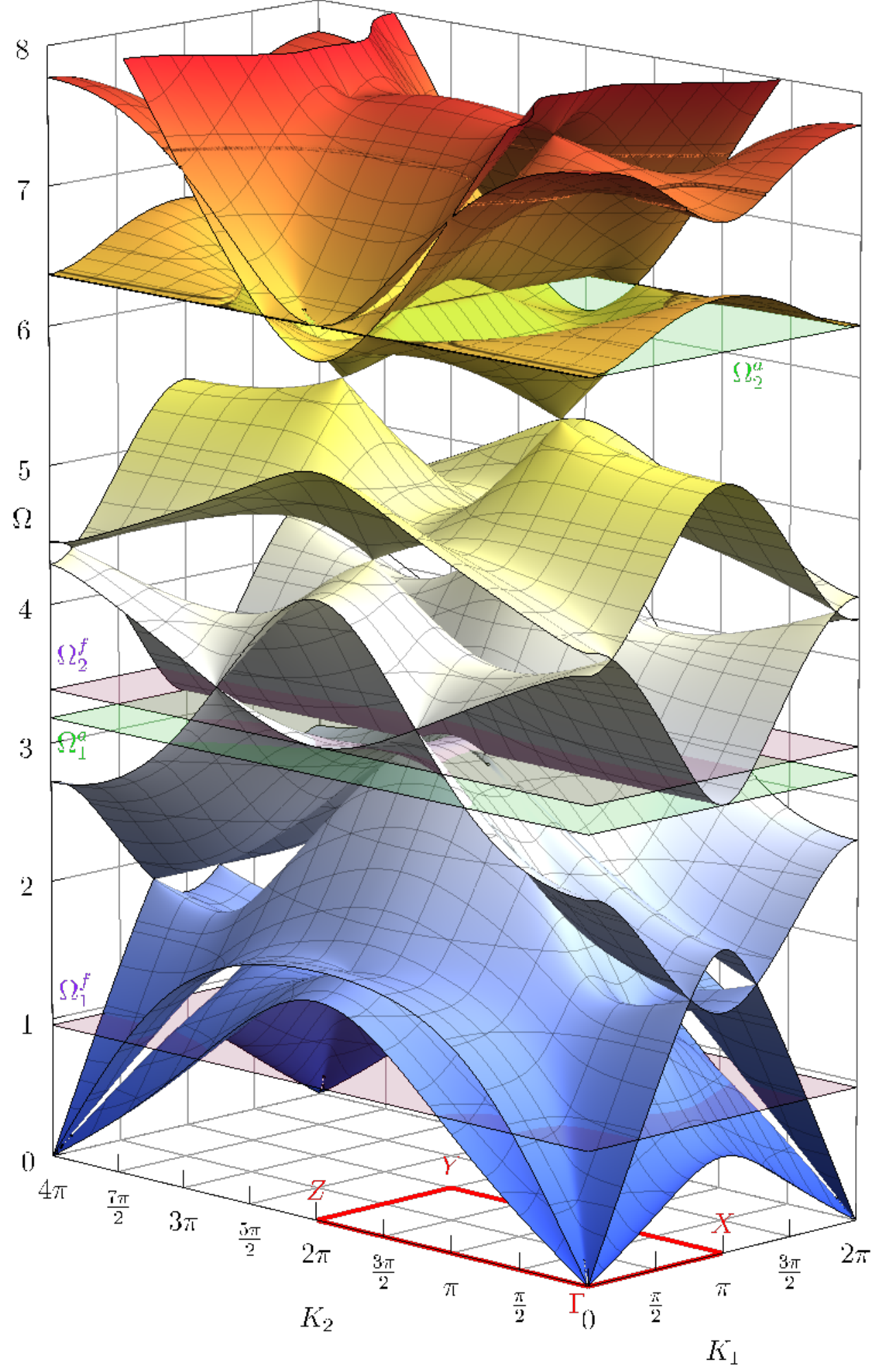}
\end{subfigure}
\caption{\label{fig:disp_surf_rect}
A completely flat band of the fourth dispersion surface is produced in a Rayleigh beam square lattice at the slenderness $\lambda=6.192$, 
so that an infinite set of standing waves propagate at the same frequency $\Omega=1.971$, part (\subref{fig:disp_surf_6_Ray}). 
Dirac cones are clearly visible in the dispersion surfaces of a rectangular lattice with contrasting slenderness with $\lambda_1=10$ and $\lambda_2=5$, part (\subref{fig:disp_surf_10_5_Ray}). Note that the frequencies $\Omega_f$ and $\Omega_a$ found for the square lattice (Figs.~\ref{fig:disp_surf_5_Ray} and~\ref{fig:disp_surf_10_Ray}) are now split in the four frequencies
$\Omega_{1,2}^f$ and $\Omega_{1,2}^a$.
The band diagrams corresponding to the paths $\Gamma$--$X$--$Y$--$\Gamma$ and $\Gamma$--$X$--$Y$--$Z$--$\Gamma$  sketched in the figure are reported in 
Fig.~\ref{fig:disp_diagram2}.}
\end{figure}

Dispersion surfaces are provided for the Euler-Bernoulli as well as the Rayleigh lattices, with an emphasis on the effects of both the rotational inertia and the slenderness of the beams.
To this end, a lattice made up of beams of equal characteristics, except the length, is addressed, $\chi = 1$, $I_1=I_2$. 
A square grid, $\alpha=1, \lambda_1=\lambda_2=\lambda=5$, and a rectangular, $\alpha=2, \lambda_1=2\lambda_2=10$, are considered. 
Results are reported in Figs.~\ref{fig:disp_surf_square} and~\ref{fig:disp_surf_6_Ray} for a square and in Fig.~\ref{fig:disp_surf_10_5_Ray} for a rectangular grid.

\begin{figure}[htb!]
\centering
\begin{subfigure}{0.45\textwidth}
\caption{\label{fig:Brillouin_5}}
\includegraphics[width=0.98\linewidth]{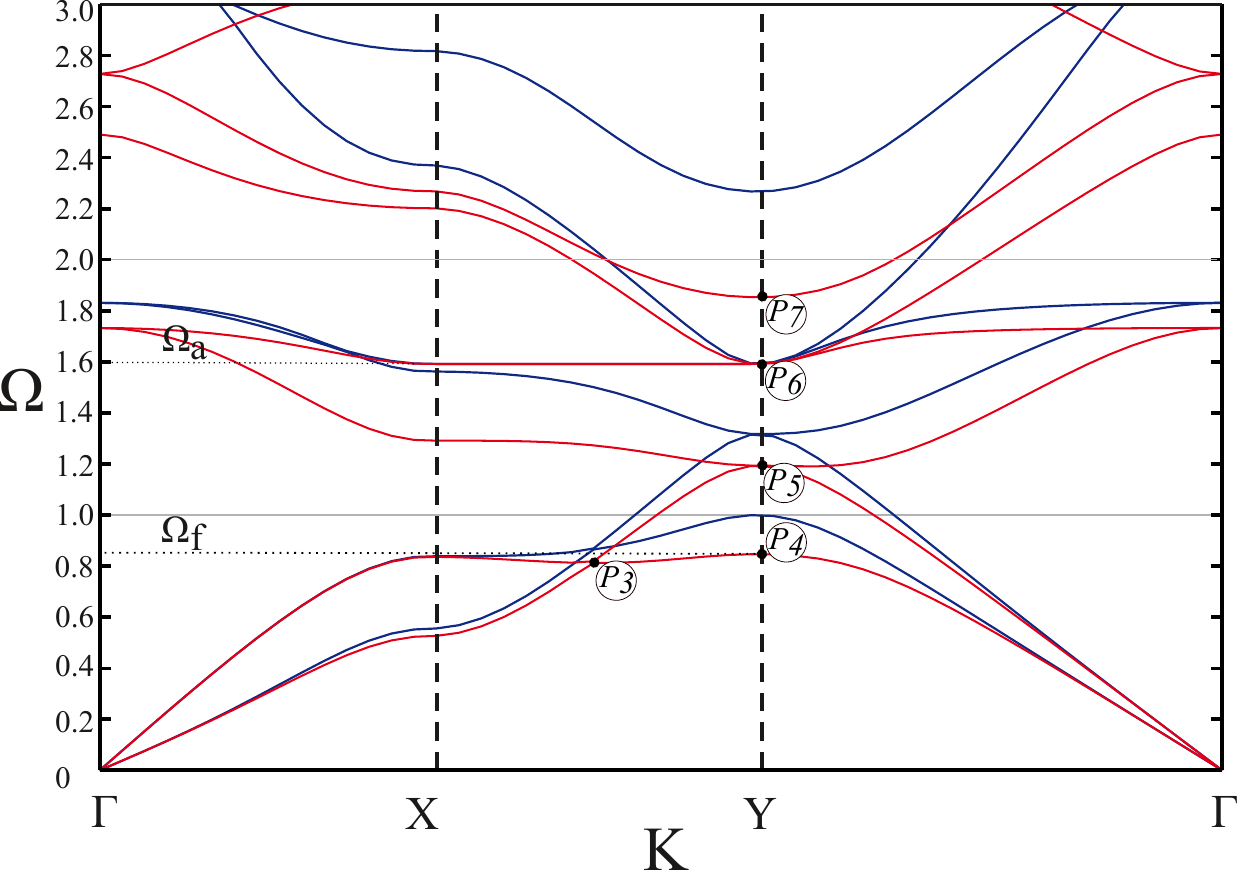}
\end{subfigure}%
\begin{subfigure}{0.45\textwidth}
\centering
\caption{\label{fig:Brillouin_10}}
\includegraphics[width=0.98\linewidth]{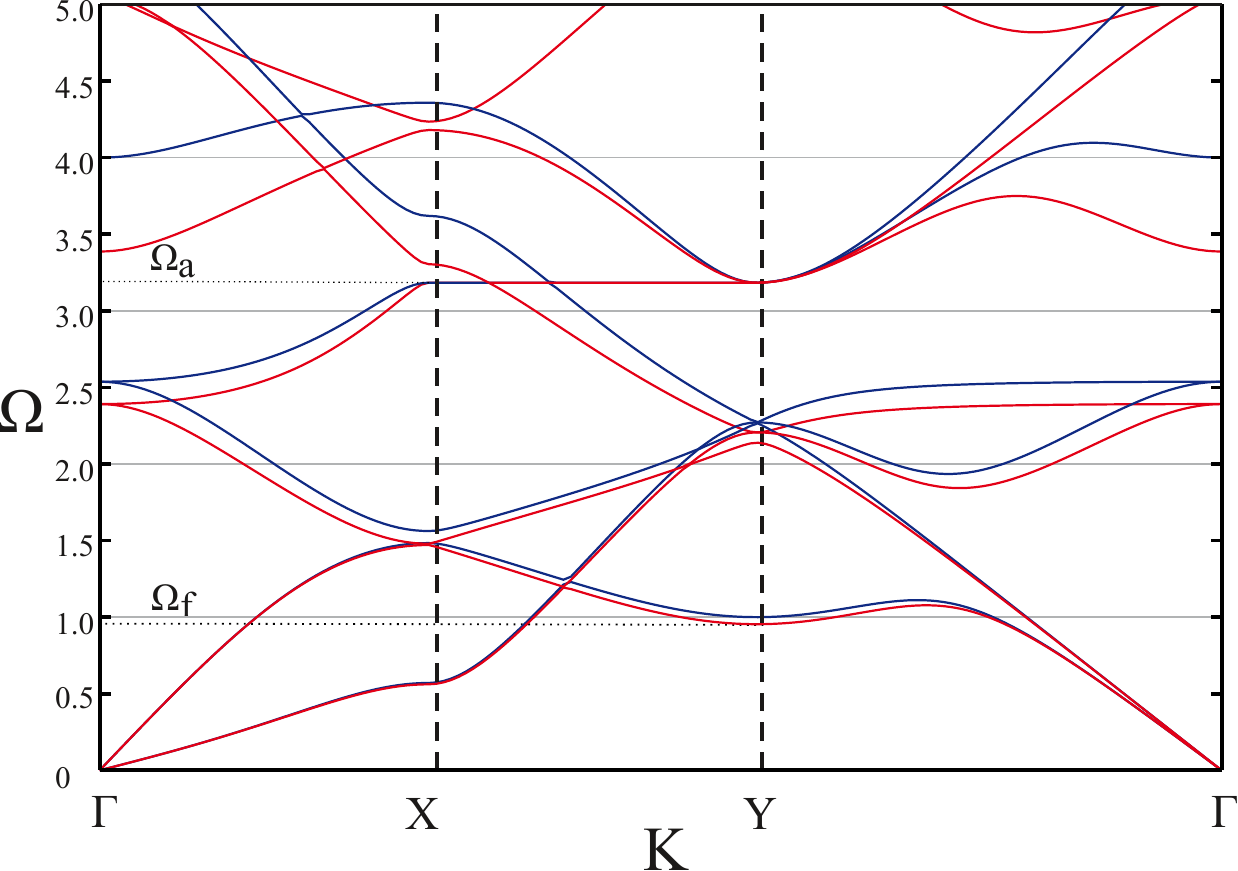}
\end{subfigure}%
\caption{\label{fig:disp_diagram}
A lowering of the dispersion frequency induced by the rotational inertia is visible in the band diagrams for a square lattice made up of Euler-Bernoulli (blue curves) and Rayleigh beams (red curves), with slenderness $\lambda=5$~(\subref{fig:Brillouin_5}) and $\lambda=10$~(\subref{fig:Brillouin_10}). 
The labels $P_i$ (marked also in Fig.~\ref{fig:surf_contour_5_Ray}, see also Tab.~\ref{tab:points_modes}) denote the points where the corresponding waveforms have been computed and shown in Figs.~\ref{fig:mode_1_2_5}--\ref{fig:mode_6_8_5}.
Note that $\Omega_f$ always corresponds to the vertex of the first dispersion surface, denoted by $P_4$, whereas $\Omega_a$ corresponds to the flat band, denoted by $P_6$. 
Furthermore, the rotational inertia leaves the flat band at $\Omega=\Omega_a$ unaltered along the $X\!-\!Y$ path, due to the fact that purely axial vibrations occur. 
The diagrams have been evaluated along the boundary of the first irreducible Brillouin zone (path $\Gamma$--$X$--$Y$--$\Gamma$ sketched in Fig.~\ref{fig:disp_surf_square}).}
\end{figure}

\begin{figure}[htb!]
\centering
\begin{subfigure}{0.45\textwidth}
\centering
\caption{\label{fig:disp_diagram_6_Ray}}
\includegraphics[width=0.98\linewidth]{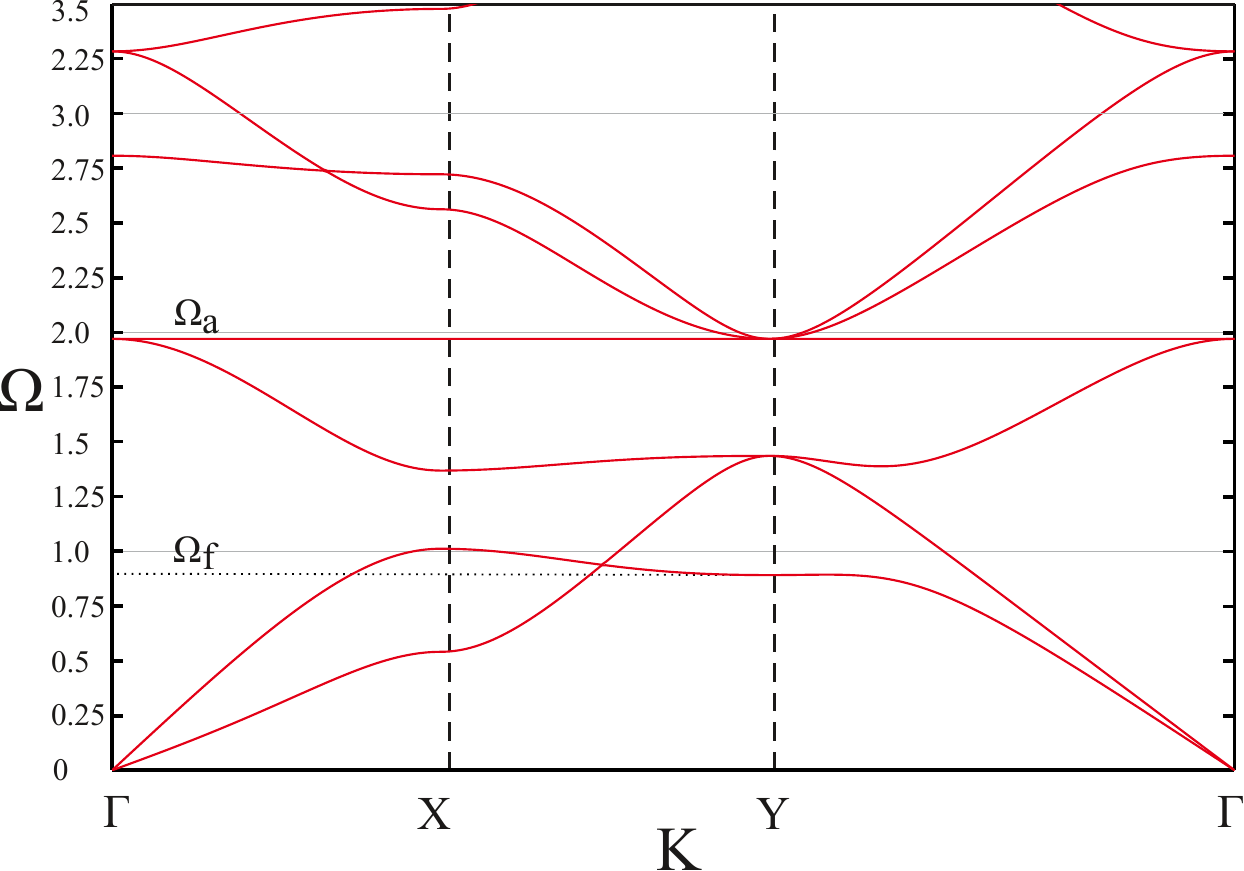}
\end{subfigure}%
\centering
\begin{subfigure}{0.45\textwidth}
\centering
\caption{\label{fig:Brillon_rettangolare}}
\includegraphics[width=0.98\linewidth]{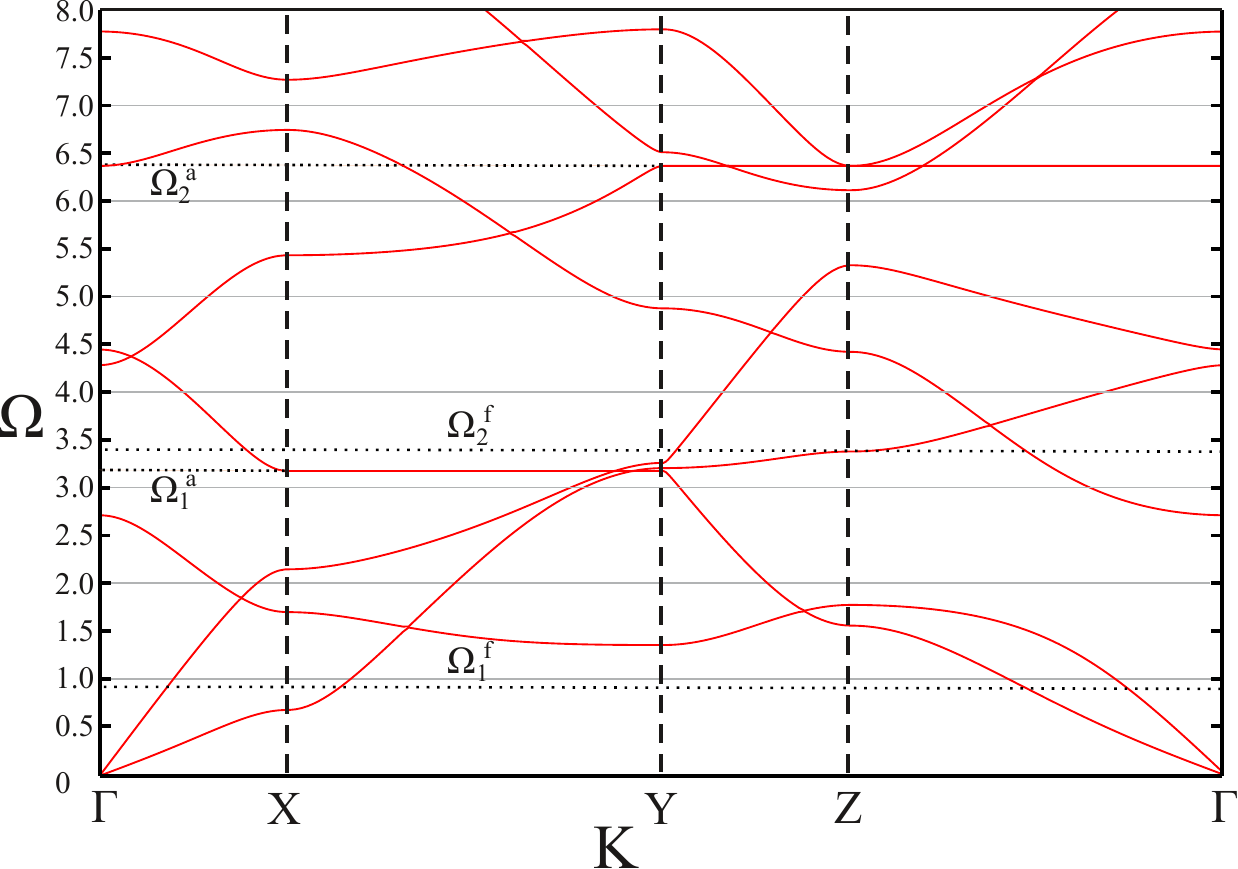}
\end{subfigure}
\caption{\label{fig:disp_diagram2}
The band diagrams relative to the paths $\Gamma$--$X$--$Y$--$\Gamma$ and $\Gamma$--$X$--$Y$--$Z$--$\Gamma$ sketched in Fig.~\ref{fig:disp_surf_rect} show: (\subref{fig:disp_diagram_6_Ray}) the perfectly flat band occurring at the slenderness $\lambda=6.192$ in a Rayleigh square lattice; (\subref{fig:Brillon_rettangolare}) the Dirac cones (particularly evident in the fourth and fifth band) present in a rectangular lattice with slenderness contrast $\lambda_1=10$, $\lambda_2=5$.}
\end{figure}

The dispersion surfaces shown in the figures are complemented by the band diagrams reported in Figs.~\ref{fig:disp_diagram} and~\ref{fig:disp_diagram2}, relative to the paths $\Gamma$--$X$--$Y$--$\Gamma$ and $\Gamma$--$X$--$Y$--$Z$--$\Gamma$ reported in the Figs.~\ref{fig:disp_surf_square} and~\ref{fig:disp_surf_10_5_Ray}, permitting the appreciation of details which remain undetected from the dispersion surfaces. 
%

%

The dispersion surfaces reported in Figs.~\ref{fig:disp_surf_square} and~\ref{fig:disp_surf_rect} have been marked with the following two particular frequencies (respectively with a pink and green plane)
\begin{equation}
    \Omega_j^f = \frac{\lambda_1}{\sqrt{\pi^2 + \lambda_j^2}}\frac{l_1}{l_j}, 
    \qquad
    \Omega_j^a = \frac{\lambda_1}{\pi}\frac{l_1}{l_j},
    \qquad \forall \, j\in\{1,2\}
\end{equation}
which are the lowest natural frequencies of, respectively, the flexural and axial mode of a double-pinned Rayleigh beam.
It is worth noting that $\Omega_j^f$ is always lower than $\Omega_j^a$, and
that, in the particular case of the Euler-Bernoulli model, the dimensionless natural frequencies become $\Omega_1^f = 1$ and $\Omega_2^f = \alpha \lambda_1/\lambda_2$. 

\paragraph{The beam slenderness} (which measures the relative importance between flexural and axial deformations along the beams in the grid) is expected to play an important role in the in-plane wave propagation and thus in the dispersion relation $\Omega(K_1,K_2)$. This is in fact a consequence of 
the compatibility and equilibrium equations to be satisfied at the central node,  Eqs.~\eqref{eq:bvp_unit_cell_compatibility}--\eqref{eq:bvp_unit_cell_equilibrium}, which produce a coupling between axial and transverse displacements along the beams, simply absent in the case of out-of-plane motion \citep{Piccolroaz_2017}.
The influence of the slenderness can be easily appreciated by comparing results reported in Fig.~\ref{fig:disp_surf_5_Ray} with those reported in Fig.~\ref{fig:disp_surf_10_Ray}, relative to a slenderness $\lambda=5$ in the former figure and $\lambda=10$ in the latter. 
It can be for instance noticed that the second and third dispersion surfaces are strongly separated by an increase of stiffness, while a seventh surface enters the frequency response in Fig.~\ref{fig:disp_surf_10_Ray}.

\paragraph{Rotational inertia} produces a lowering of the propagation frequency, so that for any fixed value of slenderness, each dispersion surface of the Rayleigh beam lattice is \textit{lower} than the corresponding surface for Euler-Bernoulli (compare Fig.~\ref{fig:disp_surf_5_EB} to Fig.~\ref{fig:disp_surf_5_Ray} and see Fig. 
\ref{fig:disp_diagram}). Moreover, a separation is observed between the dispersion surfaces, except at low frequency, the so-called \lq acoustic branches', where it is known that the two Rayleigh and Euler-Bernoulli models predict the same response. 

An interesting feature emerging from the dispersion surfaces is the presence of sets of points independent of the rotational inertia, so that their position remains the same for both beam models.
These points can be  seen by comparing the fourth surface in Fig.~\ref{fig:disp_surf_5_EB}~and~\ref{fig:disp_surf_5_Ray}, where it can be noticed that the points corresponding to $K_1=\pi,\ \forall K_2$ or $K_2=\pi,\ \forall K_1$ are located at the same frequency $\Omega_a=\lambda/\pi$ in both figures (highlighted with an horizontal green plane), which is the frequency corresponding to the first \textit{axial} mode of vibration of a double-pinned beam.
Here the dispersion relation is stationary, so that the corresponding waveforms has a null group velocity, and, in fact (see Section~\ref{sec:contours_waveforms}), these waves do not involve flexion, so that the joints of the entire lattice remain fixed.

\paragraph{A flattening of the fourth dispersion surface,} giving rise to an {\it infinite set of standing waves propagating at the same frequency with an arbitrary wave vector}, can be produced through a tuning of slenderness for both the Euler-Bernoulli and Rayleigh beam models.
This can be deduced by noting the reversal in the curvature of the fourth dispersion surface relative to $\lambda =5$ (Fig.~\ref{fig:disp_surf_5_Ray}) compared to that relative to $\lambda =10$ (Fig.~\ref{fig:disp_surf_10_Ray}), suggesting the existence of a flat surface for an intermediate value of slenderness.
Indeed the flat surface is present when the first flexural and axial mode of a double-clamped beam have the same natural frequency, which, for the Rayleigh model, occurs for a value of $\lambda$ satisfying the following equation
\begin{equation}
\label{eq:flat_band}
    \left[ \cos(2\,\kappa(\tilde{\omega}, \lambda)) \cosh\left(\frac{\tilde{\omega}\lambda}{\kappa(\tilde{\omega},\lambda)}\right) + \frac{\tilde{\omega}}{\lambda} \sin(2\,\kappa(\tilde{\omega}, \lambda)) \sinh\left(\frac{\tilde{\omega}\lambda}{\kappa(\tilde{\omega}, \lambda)}\right) -1 \right]_{\tilde{\omega}=\pi/2} = 0,
\end{equation}
where 
\begin{equation*}
\kappa(\tilde{\omega},\lambda)=\sqrt{\tilde{\omega} \left(\tilde{\omega} + \sqrt{\lambda^2 + \tilde{\omega}^2}\right)/2}.
\end{equation*}
A numerical solution of Eq.~\eqref{eq:flat_band} (in the interval $5\leq\lambda\leq10$) yields $\lambda\approx
6.192$. For this value of slenderness, the dispersion surfaces and the band diagram reported respectively in Figs.~\ref{fig:disp_surf_6_Ray} and~\ref{fig:disp_diagram_6_Ray} 
show the presence of a flat dispersion surface. Note in particular that at the point $Y$ (i.e. $K_1=K_2=\pi$) a triple root of the dispersion equation exists, corresponding to the intersection between the fourth, fifth and sixth dispersion surfaces.
%

%

\paragraph{The aspect ratio $(\alpha=l_1/l_2)$ of the rectangular lattice} has an important effect on the vibrational characteristics of the grid. In particular, Figs.~\ref{fig:disp_surf_10_5_Ray} and~\ref{fig:Brillon_rettangolare} pertain to a Rayleigh beam with $\alpha=2, \lambda_1=2\lambda_2=10$ and these results can be compared to those reported in Figs.~\ref{fig:disp_surf_5_Ray},~\ref{fig:disp_surf_10_Ray} and~\ref{fig:disp_diagram}.
Besides the fact that the surfaces are different, two aspects can be noticed:
(i.) that the two frequencies $\Omega_a$ and $\Omega_f$ split into the four $\Omega^a_{1,2}$ and $\Omega^f_{1,2}$ and (ii.) that Dirac cones become clearly visible \citep{Piccolroaz_2017, McPhedran_2015}. 

\paragraph{Several singularities and Dirac cones} connect the complex multiple dispersion surfaces 
in Figs.~\ref{fig:disp_surf_square} and~\ref{fig:disp_surf_rect}, but Dirac cones become particularly evident in  the rectangular grid, Fig.~\ref{fig:disp_surf_10_5_Ray}. At these singular points the dispersion relation may become non-smooth.

\subsection{Isofrequency contours, Dirac cones and standing waves}
\label{sec:contours_waveforms}

Detailed features of the individual dispersion surfaces are analyzed by computing their level sets, also referred to as \textit{slowness contours}.
As these contours provide valuable information on the kind of anisotropy to be expected in the time-harmonic response of the lattice, their analysis allows us to identify the frequency regimes corresponding to different dynamic behaviours.
In fact, this tool has already been proved to be successful at predicting the preferential directions of the forced vibrations for the out-of-plane problem~\citep{Piccolroaz_2017}.

As discussed in the previous section, the in-plane wave propagation problem involves more complex dispersion characteristics than the out-of-plane, due to the coupling between the axial and flexural beam vibrations.
Furthermore, the vectorial nature of the problem allows the application of different types of in-plane concentrated loads, namely two orthogonal point forces and a concentrated bending moment, and hence the shape of the slowness contours alone cannot provide a comprehensive description of the forced lattice vibrations.
Therefore, a complete investigation of the lattice vibration properties, involves determination of the following aspects:
\begin{enumerate}[label=(\roman*)]
    \item identification of frequency ranges displaying the non-convexity of the slowness contours, for possible detection of negative refraction;
    \item computation of waveforms corresponding to double roots and standing waves, as connected to resonance under forced vibrations;
    \item identification of waveforms evidencing a purely extensional or flexural response, corresponding to vibration modes of a finite-length beam.
\end{enumerate}

\begin{figure}[htb!]
\centering
    \begin{subfigure}{0.2\textwidth}
        \centering
        \caption{\label{fig:surf_1_5_Ray}}
        \includegraphics[width=\linewidth]{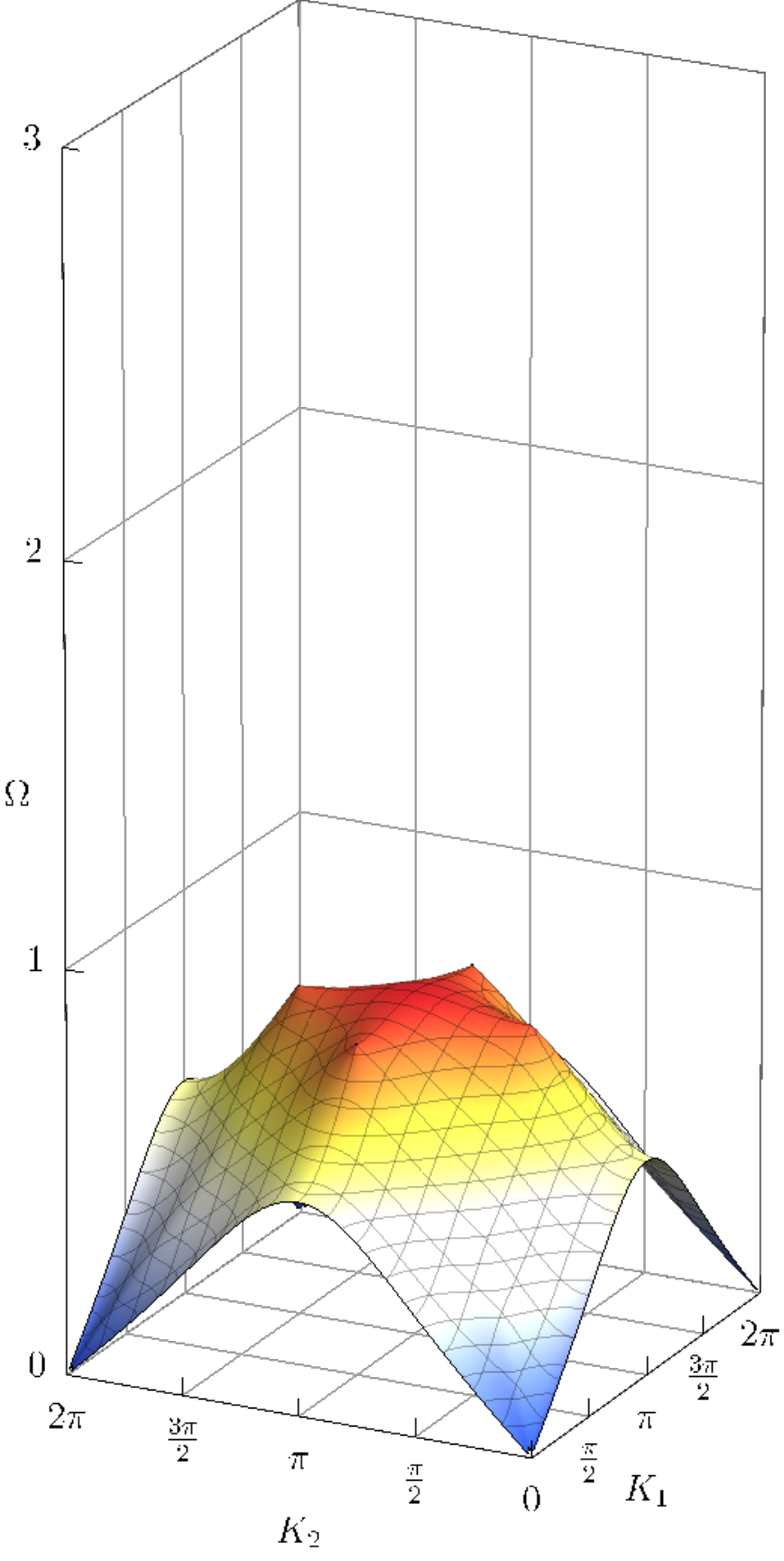}
    \end{subfigure}%
    \begin{subfigure}{0.2\textwidth}
        \centering
        \caption{\label{fig:surf_2_5_Ray}}
        \includegraphics[width=\linewidth]{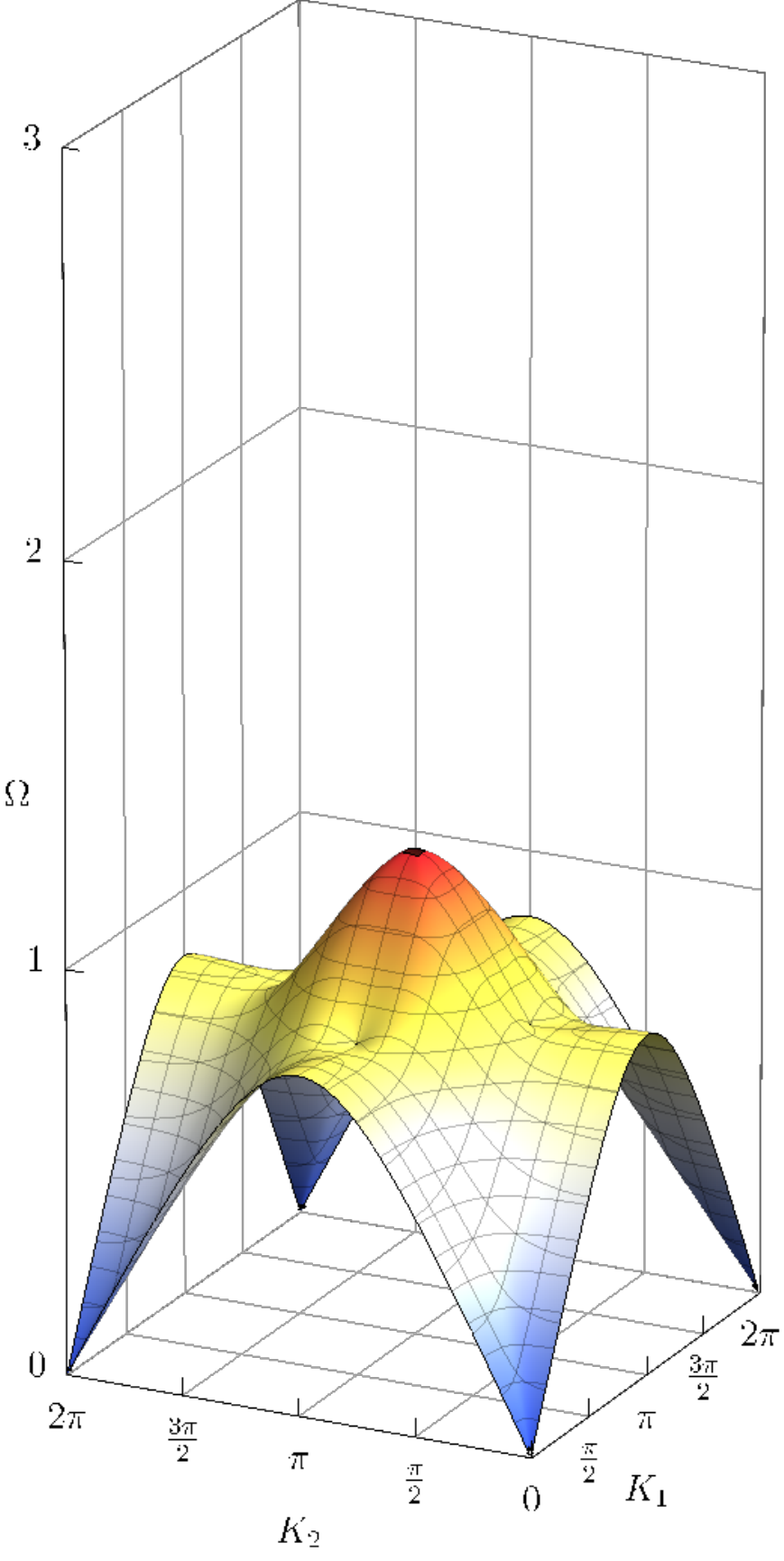}
    \end{subfigure}%
    \begin{subfigure}{0.2\textwidth}
        \centering
        \caption{\label{fig:surf_3_5_Ray}}
        \includegraphics[width=\linewidth]{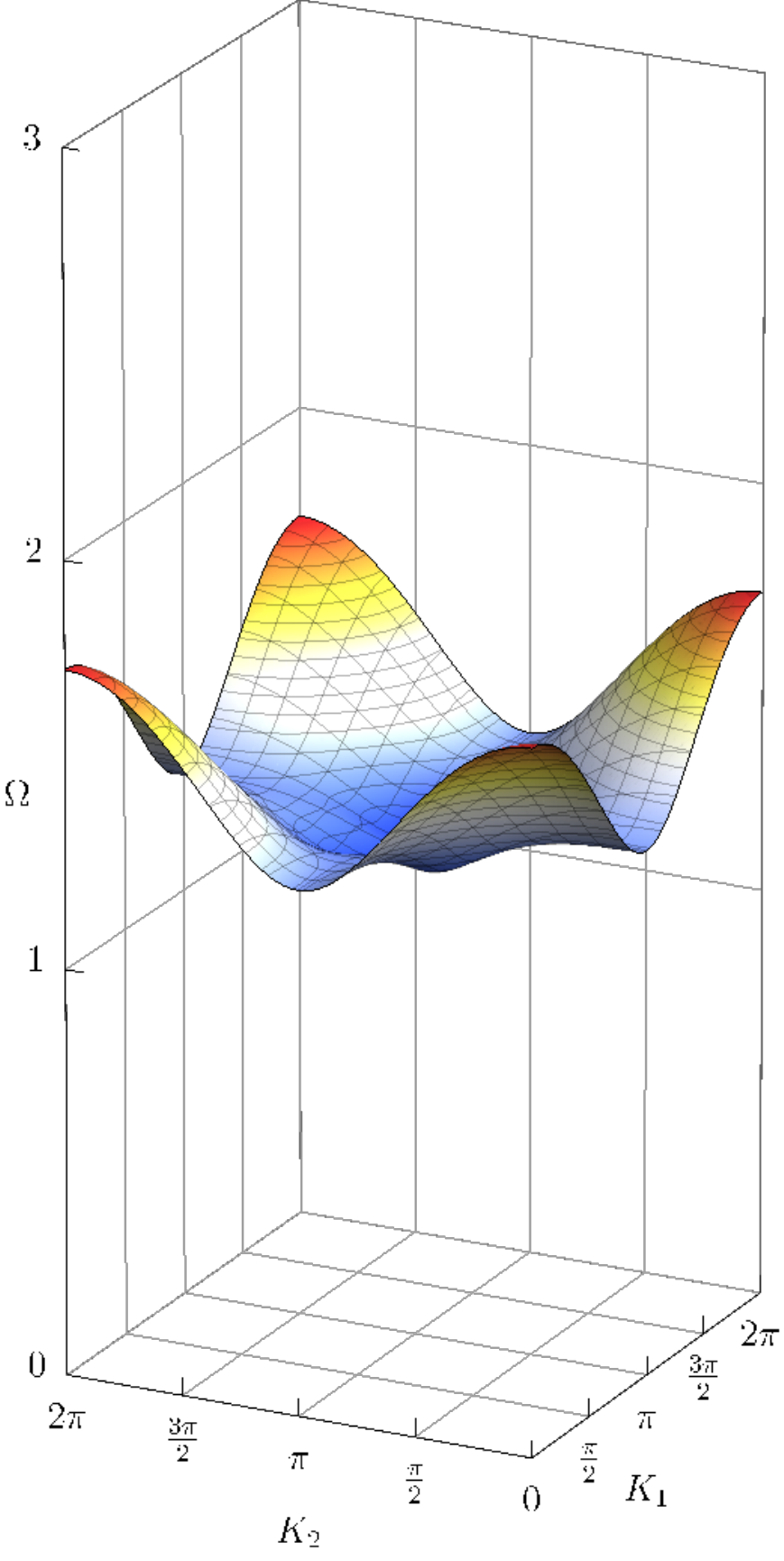}
    \end{subfigure}%
    \begin{subfigure}{0.2\textwidth}
        \centering
        \caption{\label{fig:surf_4_5_Ray}}
        \includegraphics[width=\linewidth]{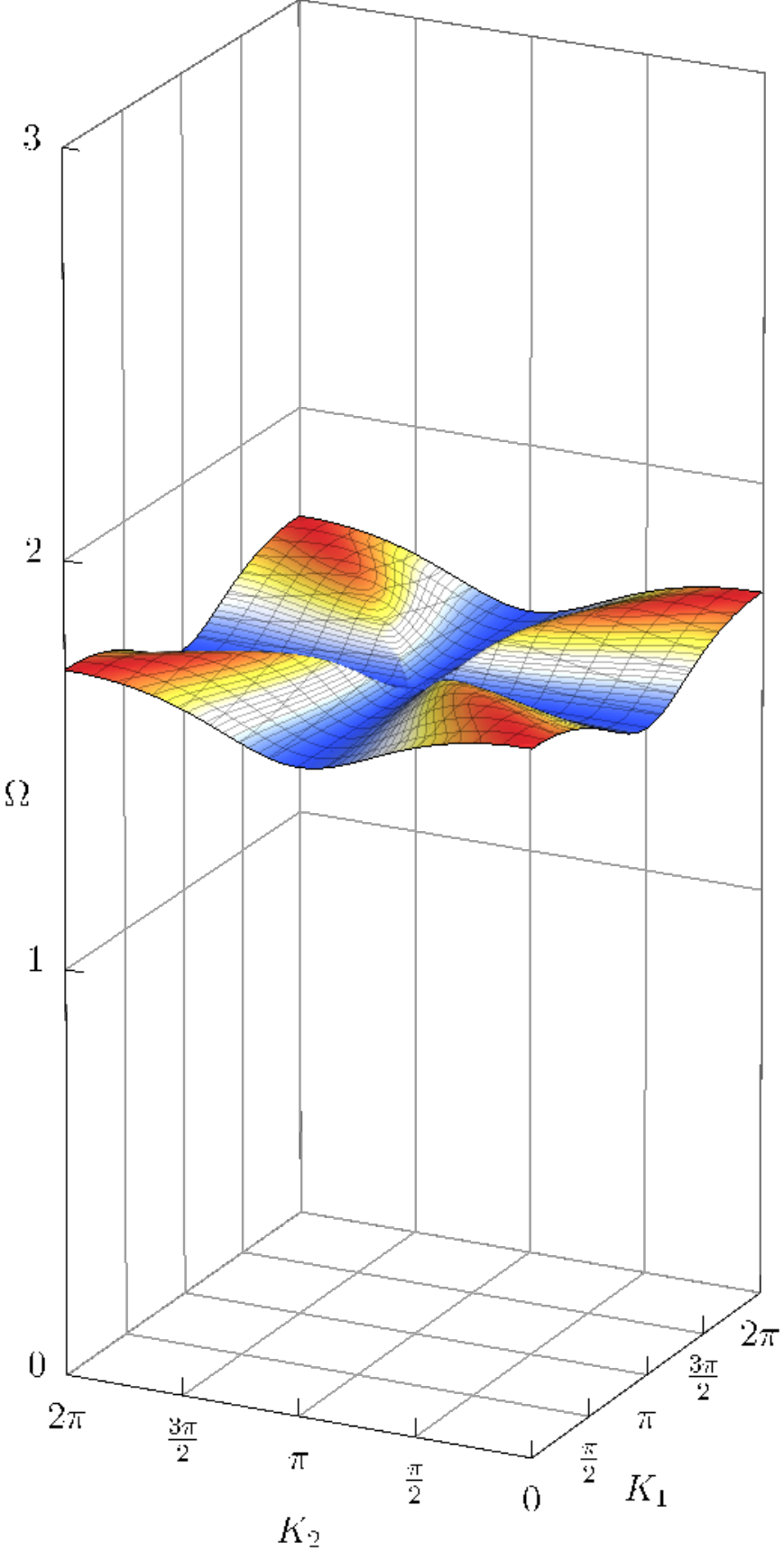}
    \end{subfigure}%
    \begin{subfigure}{0.2\textwidth}
        \centering
        \caption{\label{fig:surf_5_5_Ray}}
        \includegraphics[width=\linewidth]{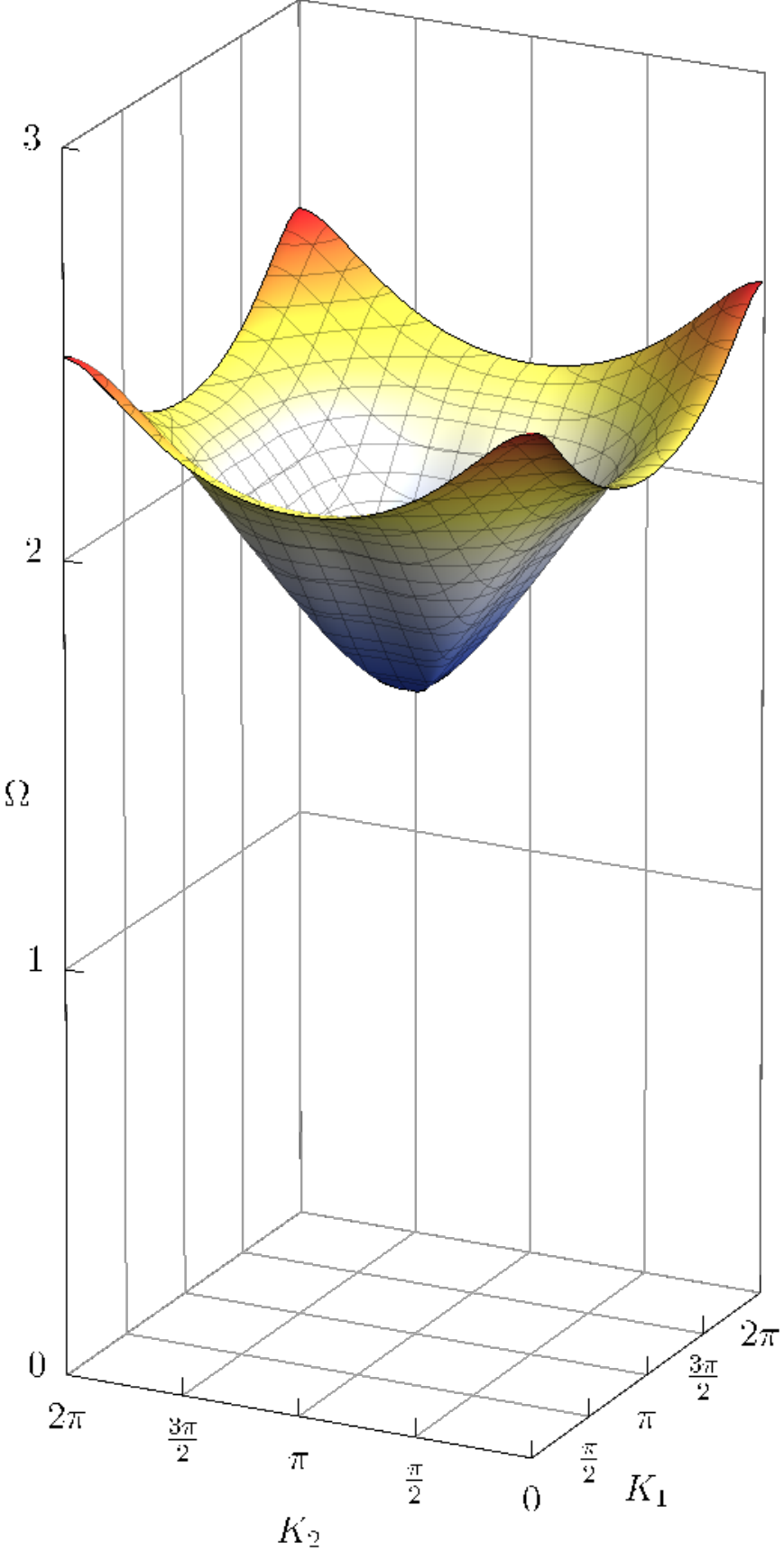}
    \end{subfigure}
    \begin{subfigure}{0.2\textwidth}
        \centering
        \caption{\label{fig:surf_contour_1_5_Ray}}
        \includegraphics[width=\linewidth]{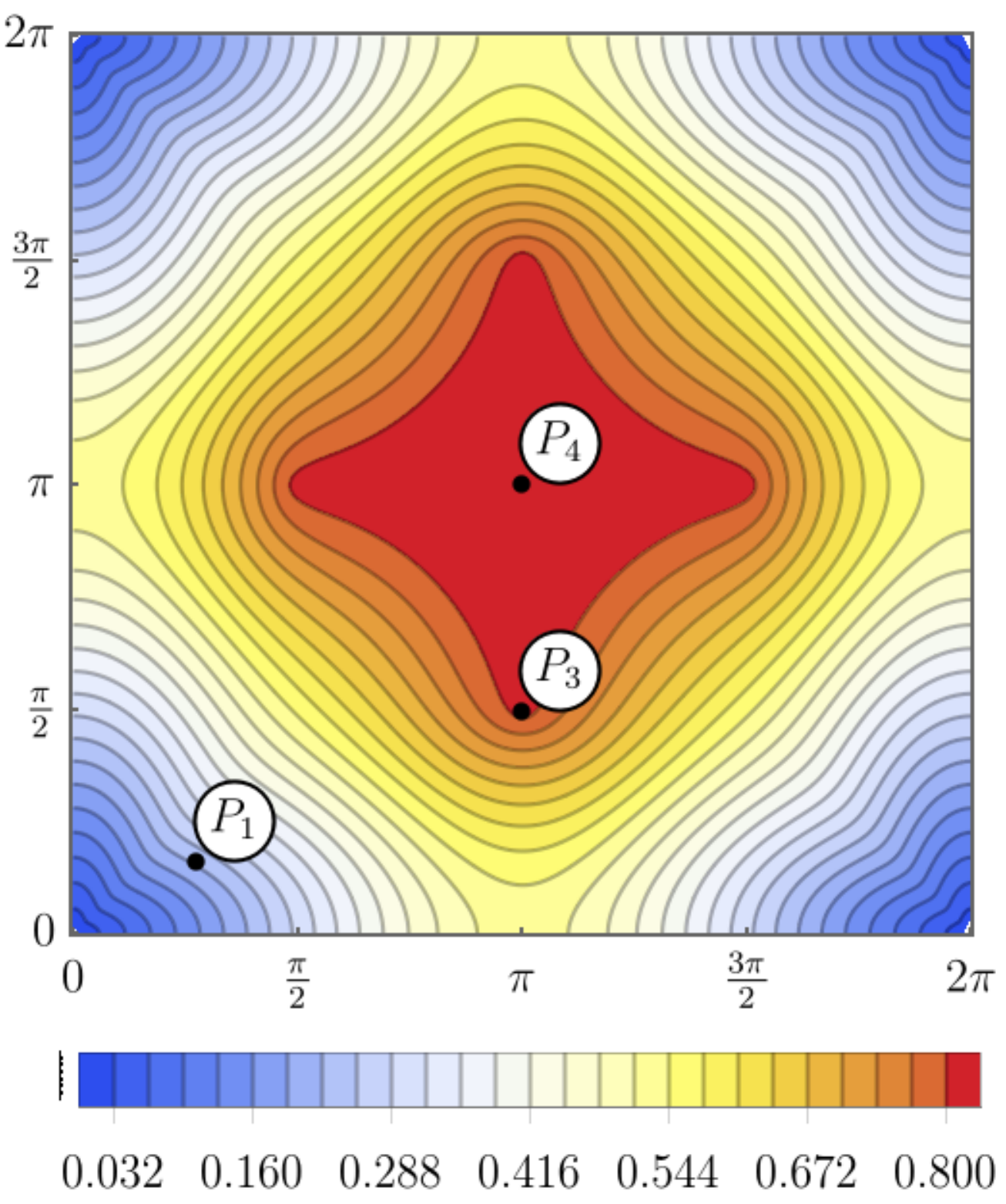}
    \end{subfigure}%
    \begin{subfigure}{0.2\textwidth}
        \centering
        \caption{\label{fig:surf_contour_2_5_Ray}}
        \includegraphics[width=\linewidth]{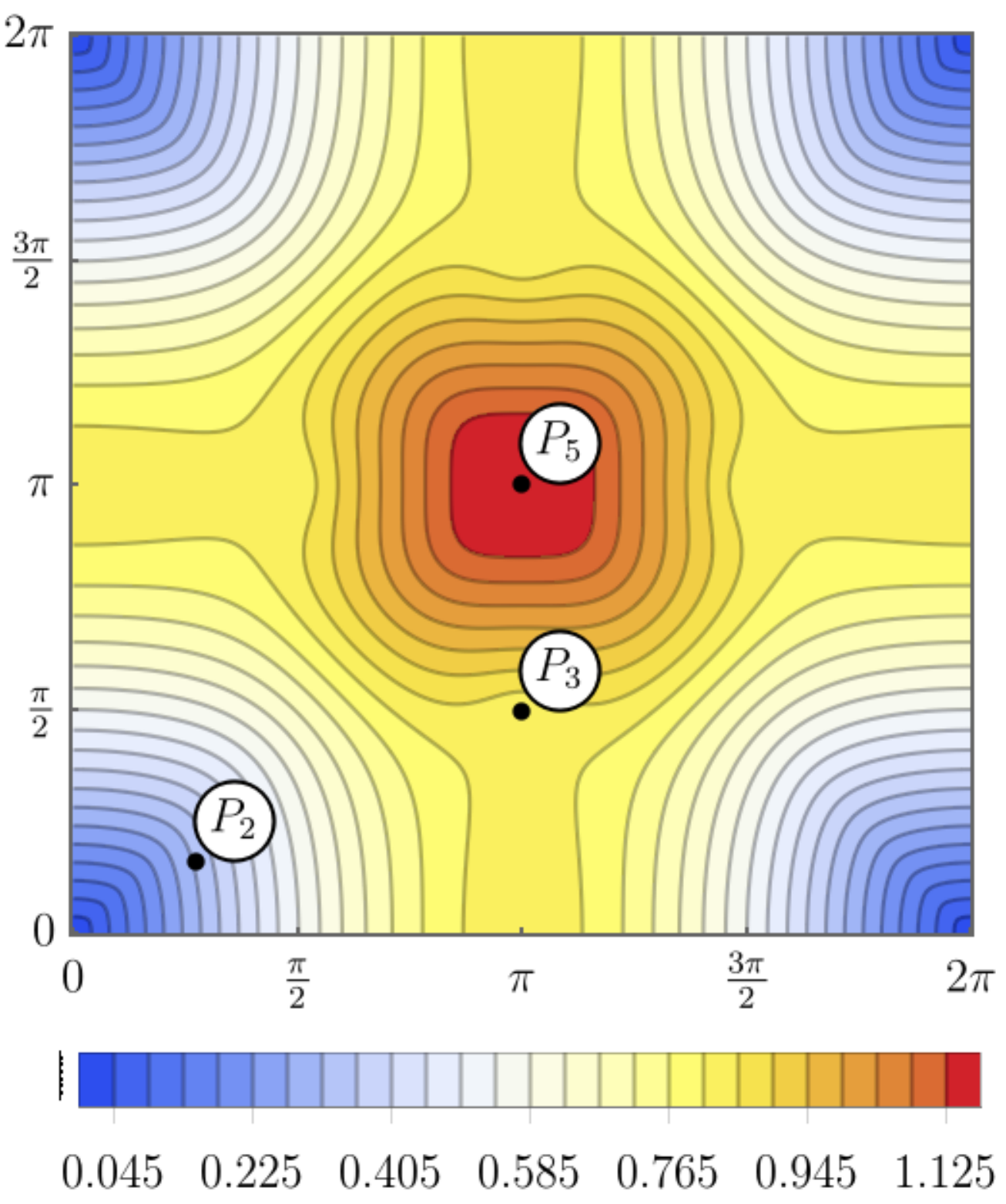}
    \end{subfigure}%
    \begin{subfigure}{0.2\textwidth}
        \centering
        \caption{\label{fig:surf_contour_3_5_Ray}}
        \includegraphics[width=\linewidth]{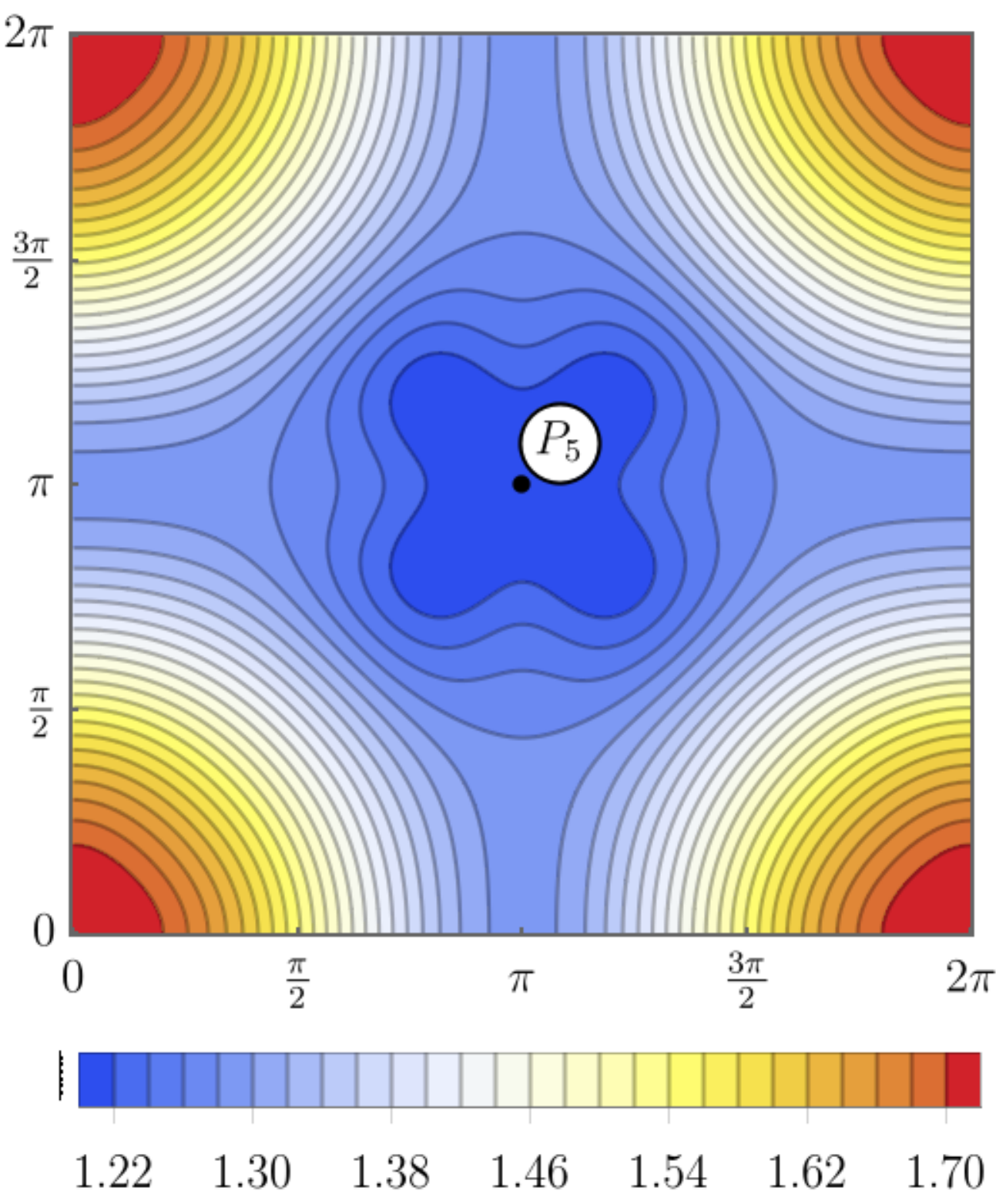}
    \end{subfigure}%
    \begin{subfigure}{0.2\textwidth}
        \centering
        \caption{\label{fig:surf_contour_4_5_Ray}}
        \includegraphics[width=\linewidth]{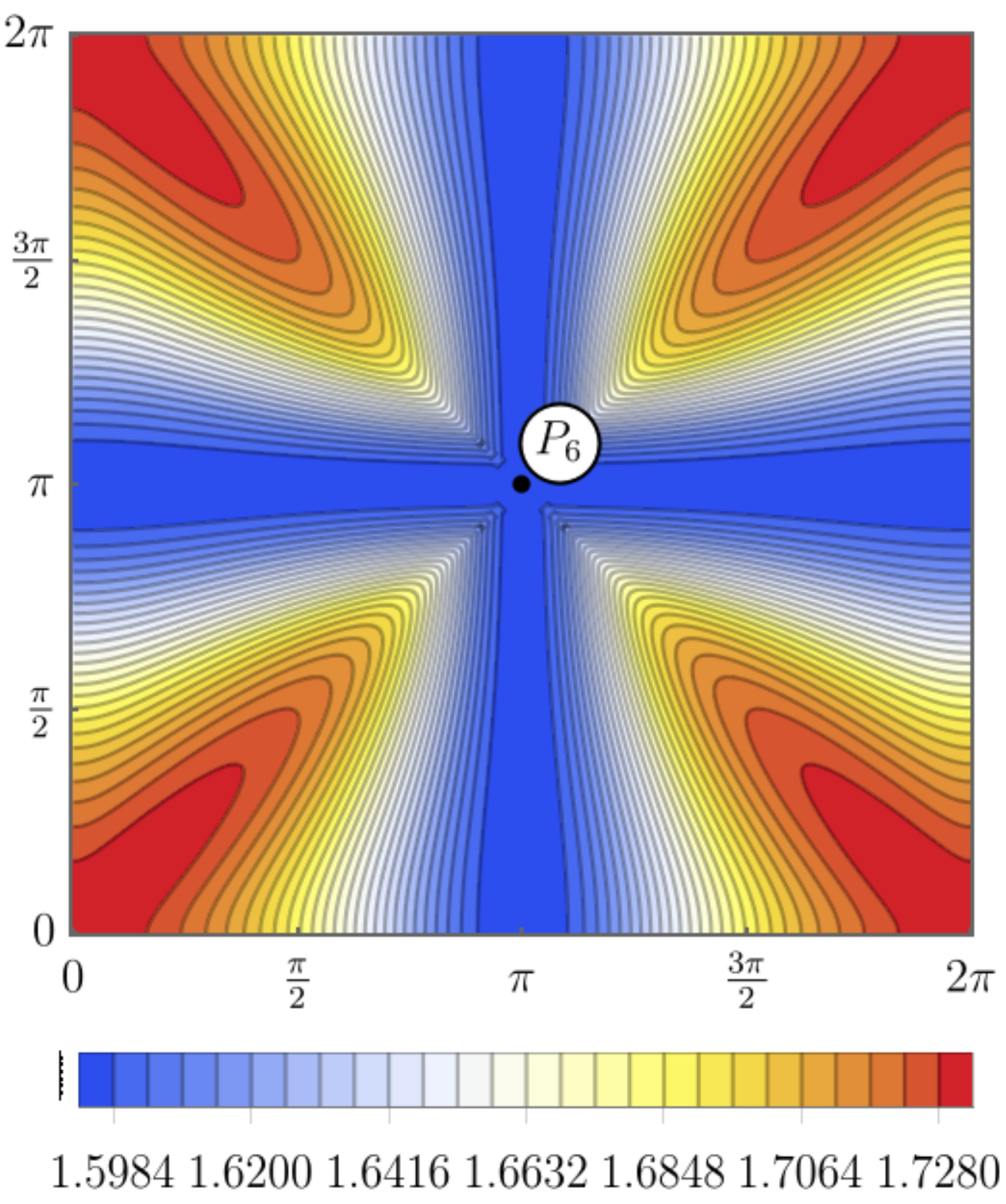}
    \end{subfigure}%
    \begin{subfigure}{0.2\textwidth}
        \centering
        \caption{\label{fig:surf_contour_5_5_Ray}}
        \includegraphics[width=\linewidth]{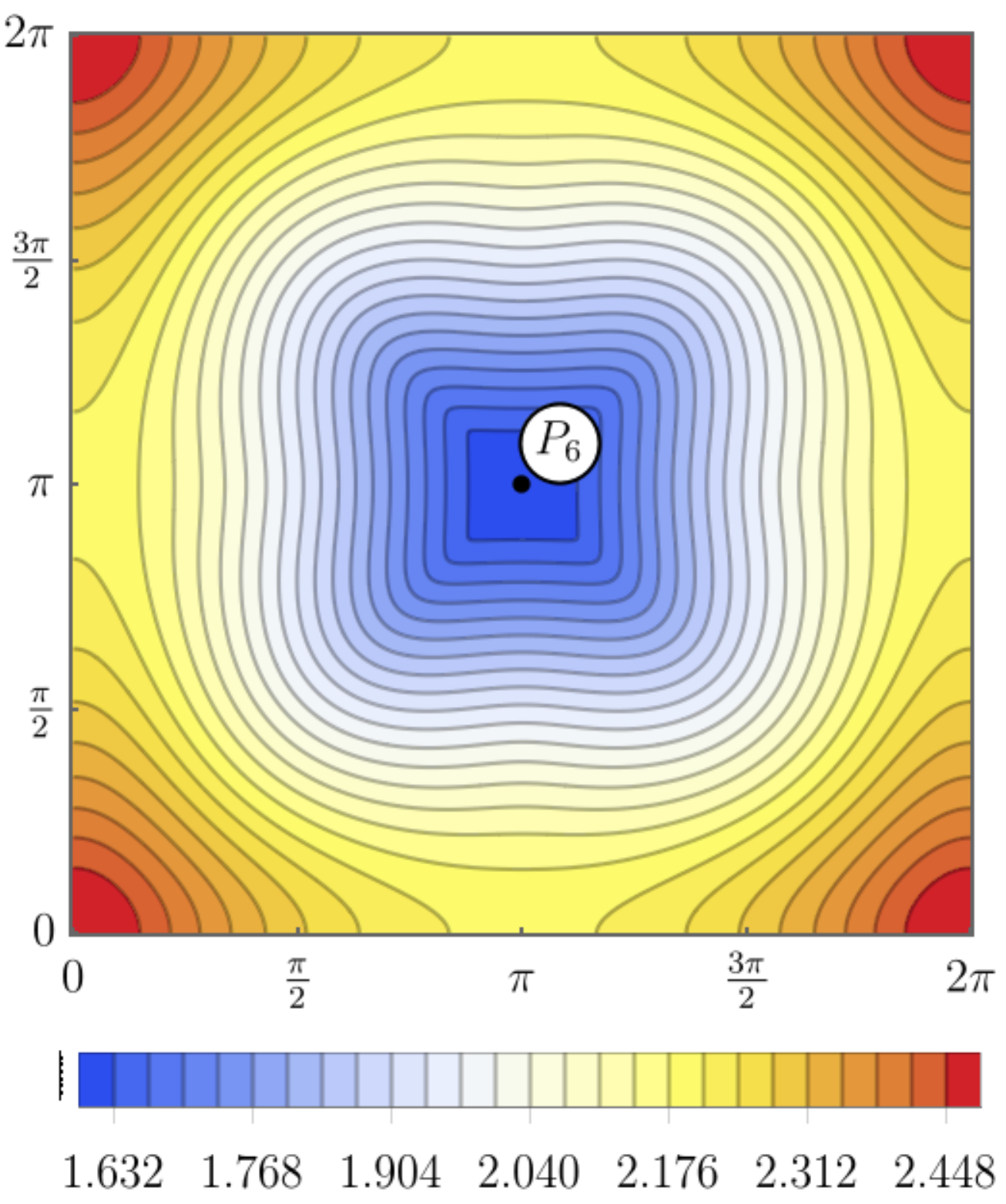}
    \end{subfigure}
\caption{\label{fig:surf_contour_5_Ray}
The cubic anisotropy is evident from the slowness contours (\subref{fig:surf_contour_1_5_Ray})--(\subref{fig:surf_contour_5_5_Ray}) associated to the dispersion surfaces (\subref{fig:surf_1_5_Ray})--(\subref{fig:surf_5_5_Ray}), for a square grid of Rayleigh beams with slenderness $\lambda=5$. The labels $P_i$ (marked also in Fig.~\ref{fig:Brillouin_5}, see also Tab.~\ref{tab:points_modes}) denote the points where the corresponding waveforms have been computed and shown in Figs.~\ref{fig:mode_1_2_5}--\ref{fig:mode_6_8_5}. 
Linear dispersion at low frequency is visible in the acoustic branches (the first two dispersion surfaces), while the dispersion relation becomes nonlinear at high frequency and the isofrequency contours  dramatically change and display several double-root points such as the four Dirac cones (one of them is labelled $P_3$) connecting the first two surfaces as well as the stationary points connecting the second and the third ($P_5$) or the fourth and fifth surface ($P_6$).
}
\end{figure}
%

\begin{figure}[htb!]
\centering
    \begin{subfigure}{0.2\textwidth}
        \centering
        \caption{\label{fig:surf_1_10_Ray}}
        \includegraphics[width=1.00\linewidth]{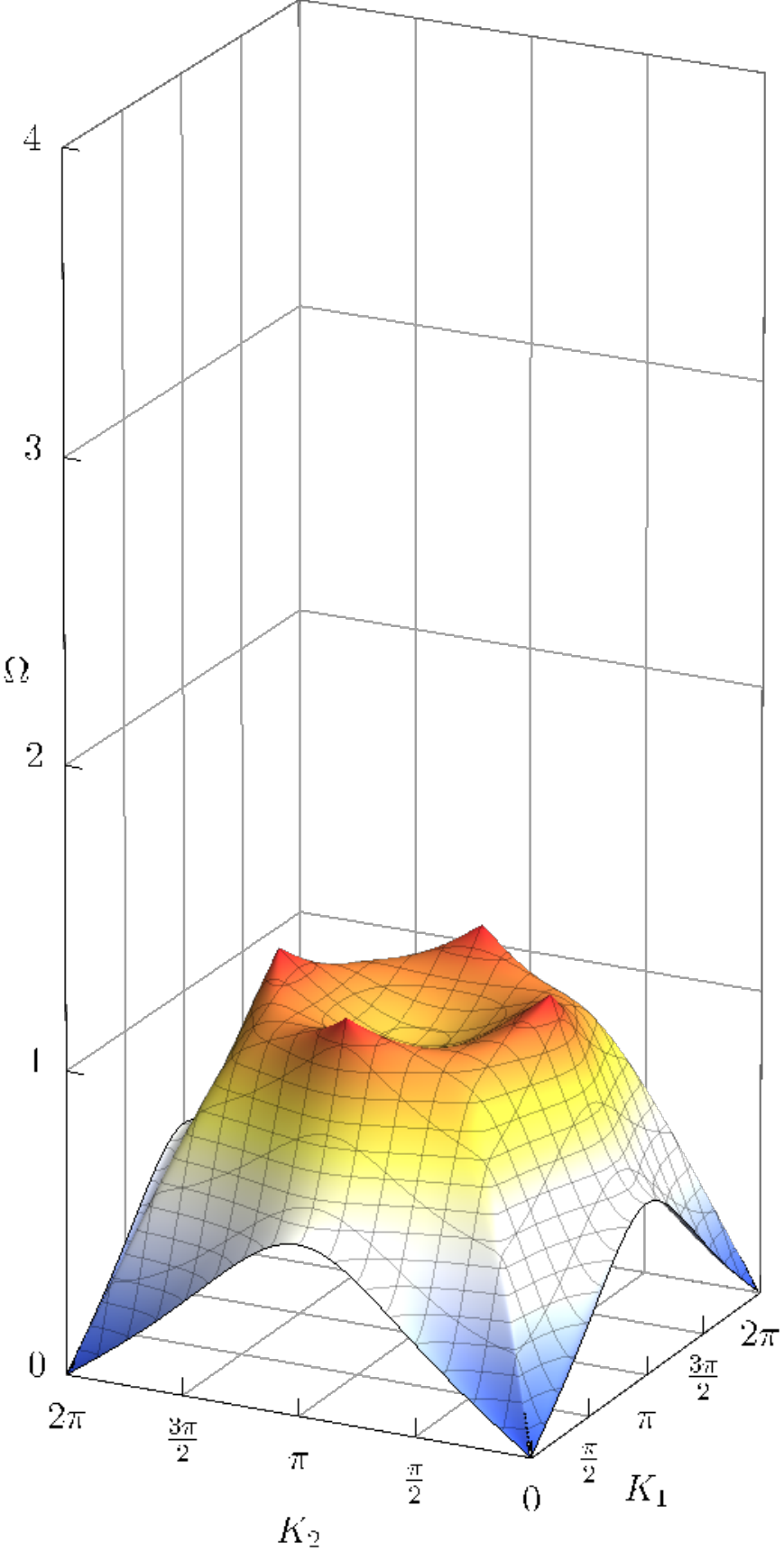}
    \end{subfigure}%
    \begin{subfigure}{0.2\textwidth}
        \centering
        \caption{\label{fig:surf_2_10_Ray}}
        \includegraphics[width=1.00\linewidth]{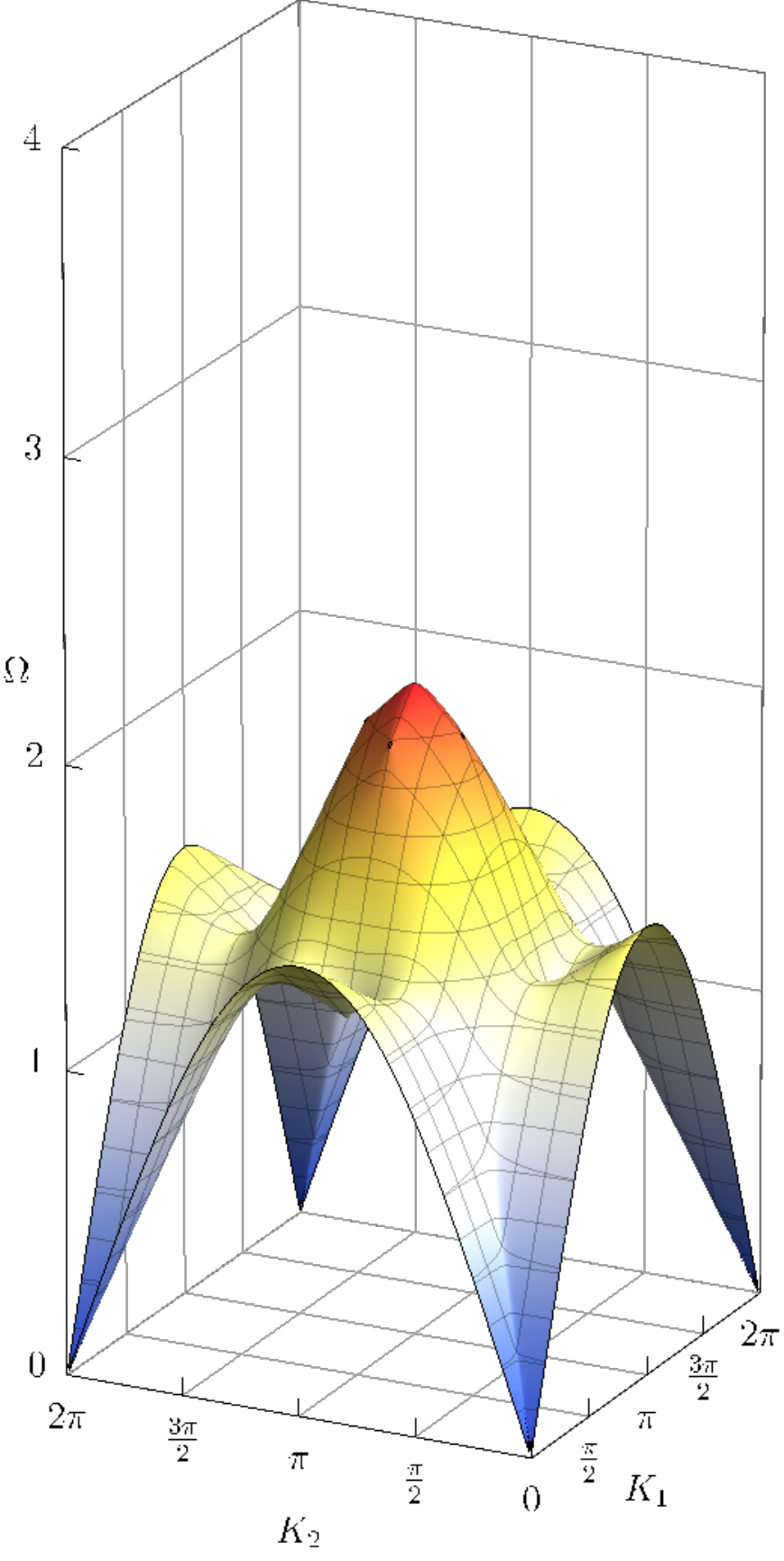}
    \end{subfigure}%
    \begin{subfigure}{0.2\textwidth}
        \centering
        \caption{\label{fig:surf_3_10_Ray}}
        \includegraphics[width=1.00\linewidth]{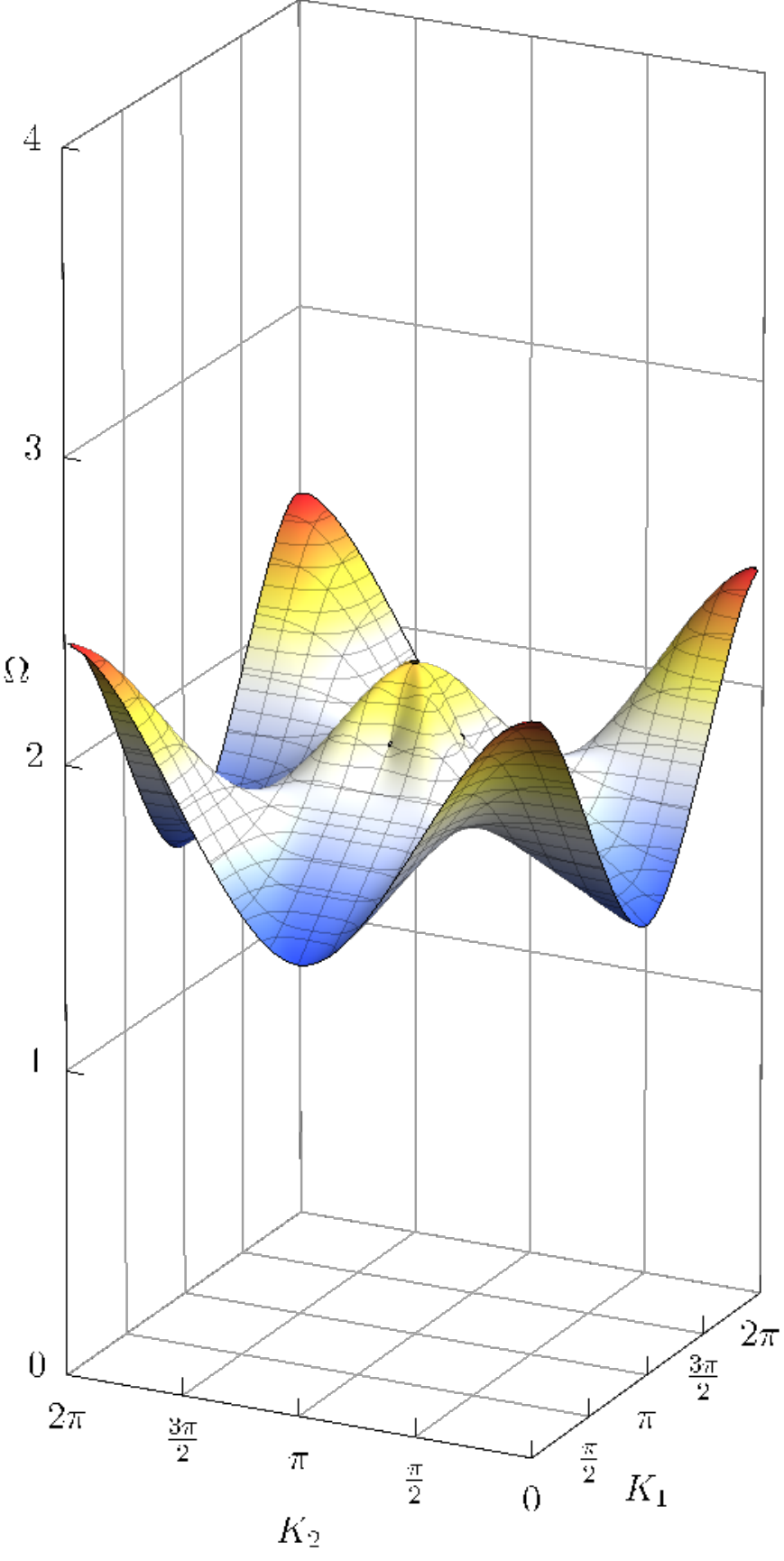}
    \end{subfigure}%
    \begin{subfigure}{0.2\textwidth}
        \centering
        \caption{\label{fig:surf_4_10_Ray}}
        \includegraphics[width=1.00\linewidth]{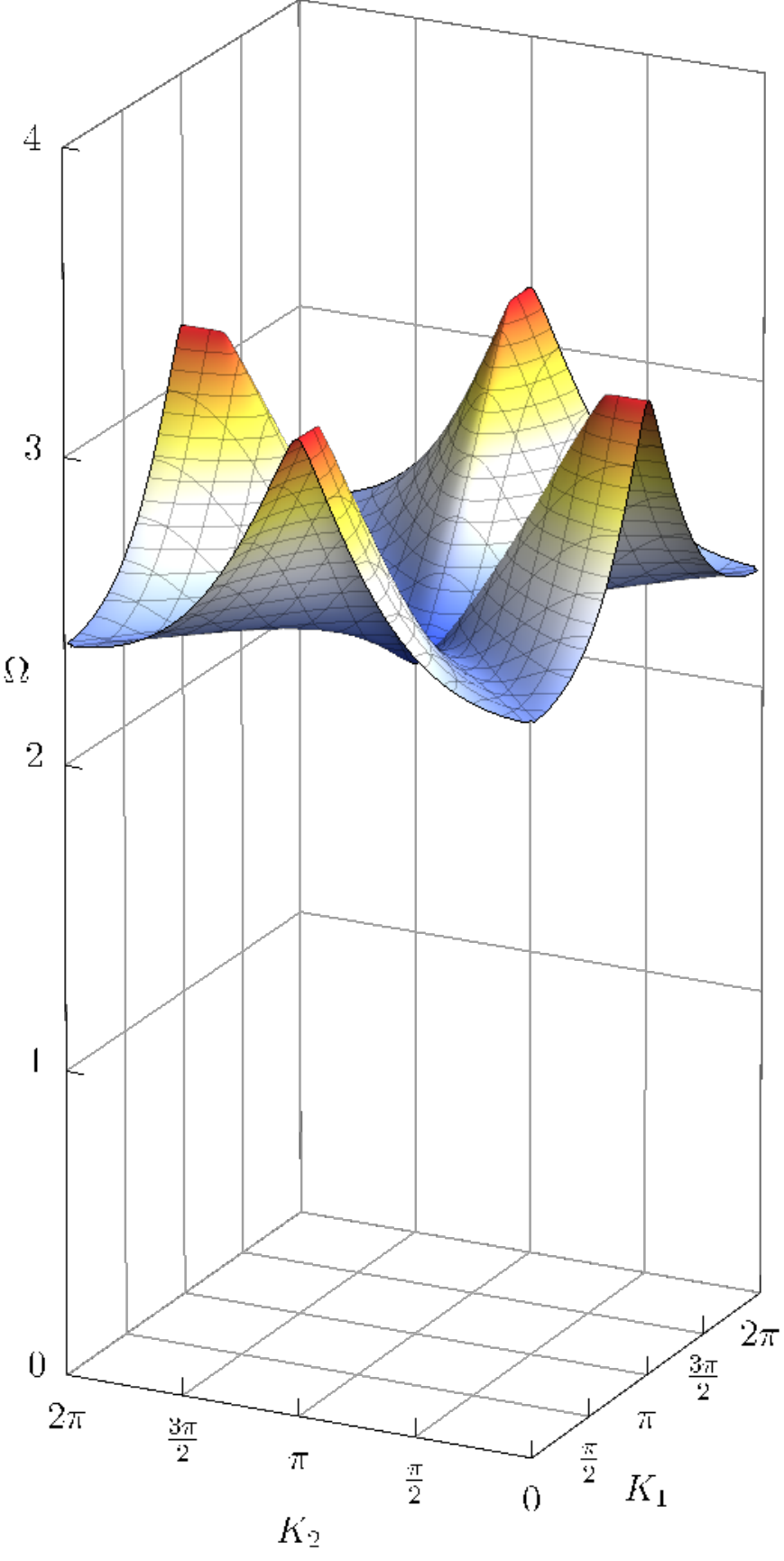}
    \end{subfigure}%
    \begin{subfigure}{0.2\textwidth}
        \centering
        \caption{\label{fig:surf_5_10_Ray}}
        \includegraphics[width=1.00\linewidth]{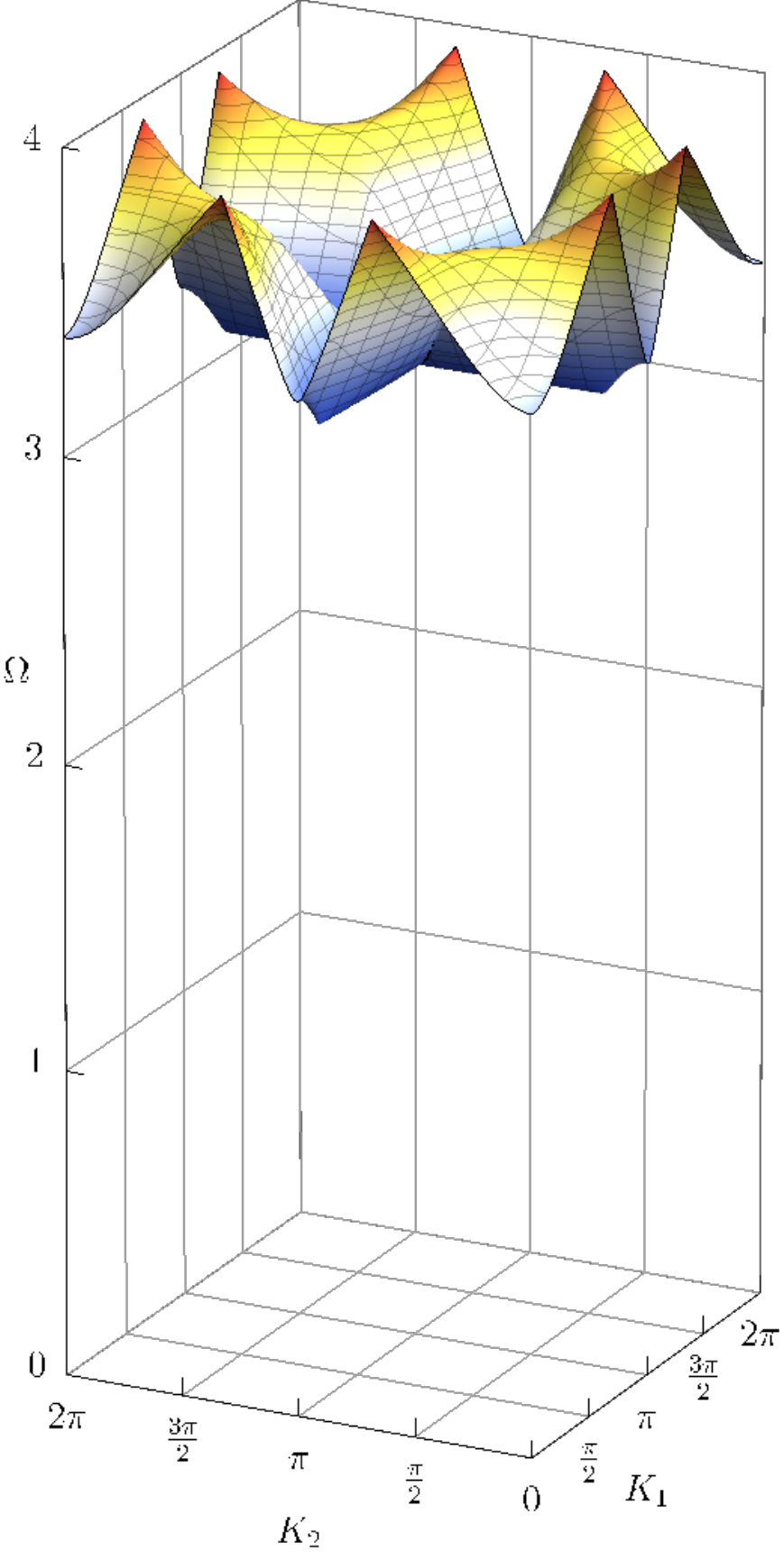}
    \end{subfigure}
    \begin{subfigure}{0.2\textwidth}
        \centering
        \caption{\label{fig:surf_contour_1_10_Ray}}
        \includegraphics[width=1.00\linewidth]{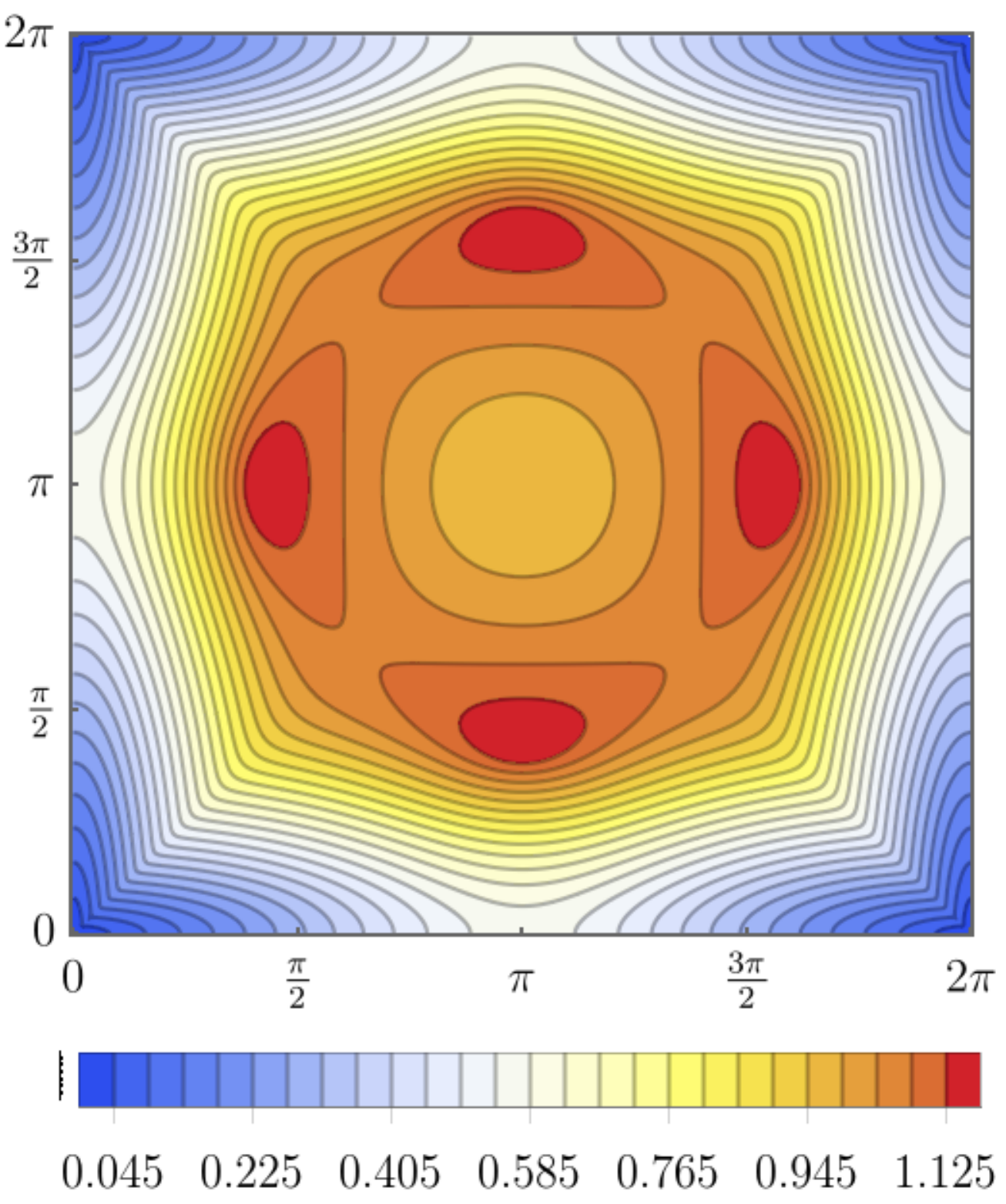}
    \end{subfigure}%
    \begin{subfigure}{0.2\textwidth}
        \centering
        \caption{\label{fig:surf_contour_2_10_Ray}}
        \includegraphics[width=1.00\linewidth]{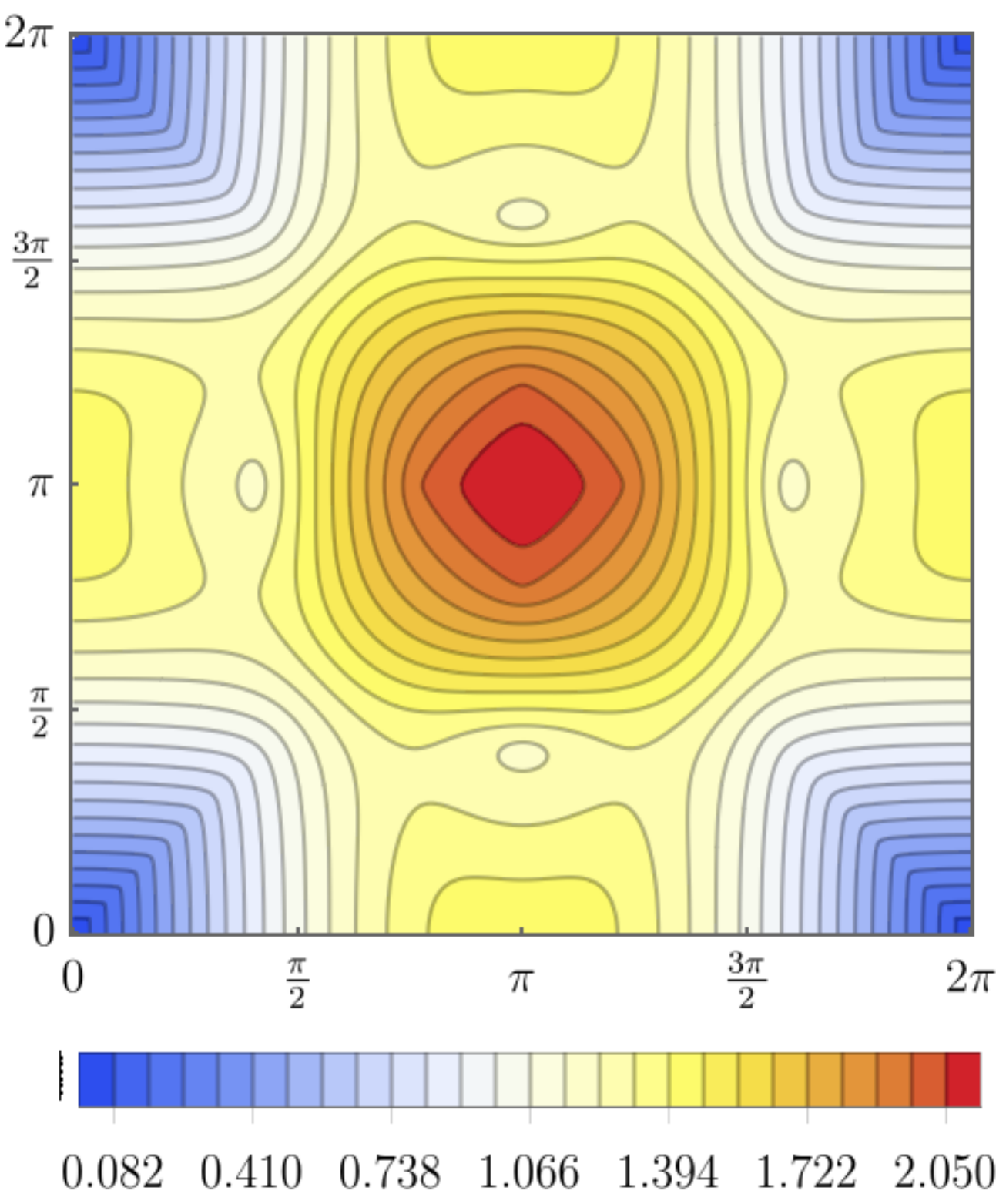}
    \end{subfigure}%
    \begin{subfigure}{0.2\textwidth}
        \centering
        \caption{\label{fig:surf_contour_3_10_Ray}}
        \includegraphics[width=1.00\linewidth]{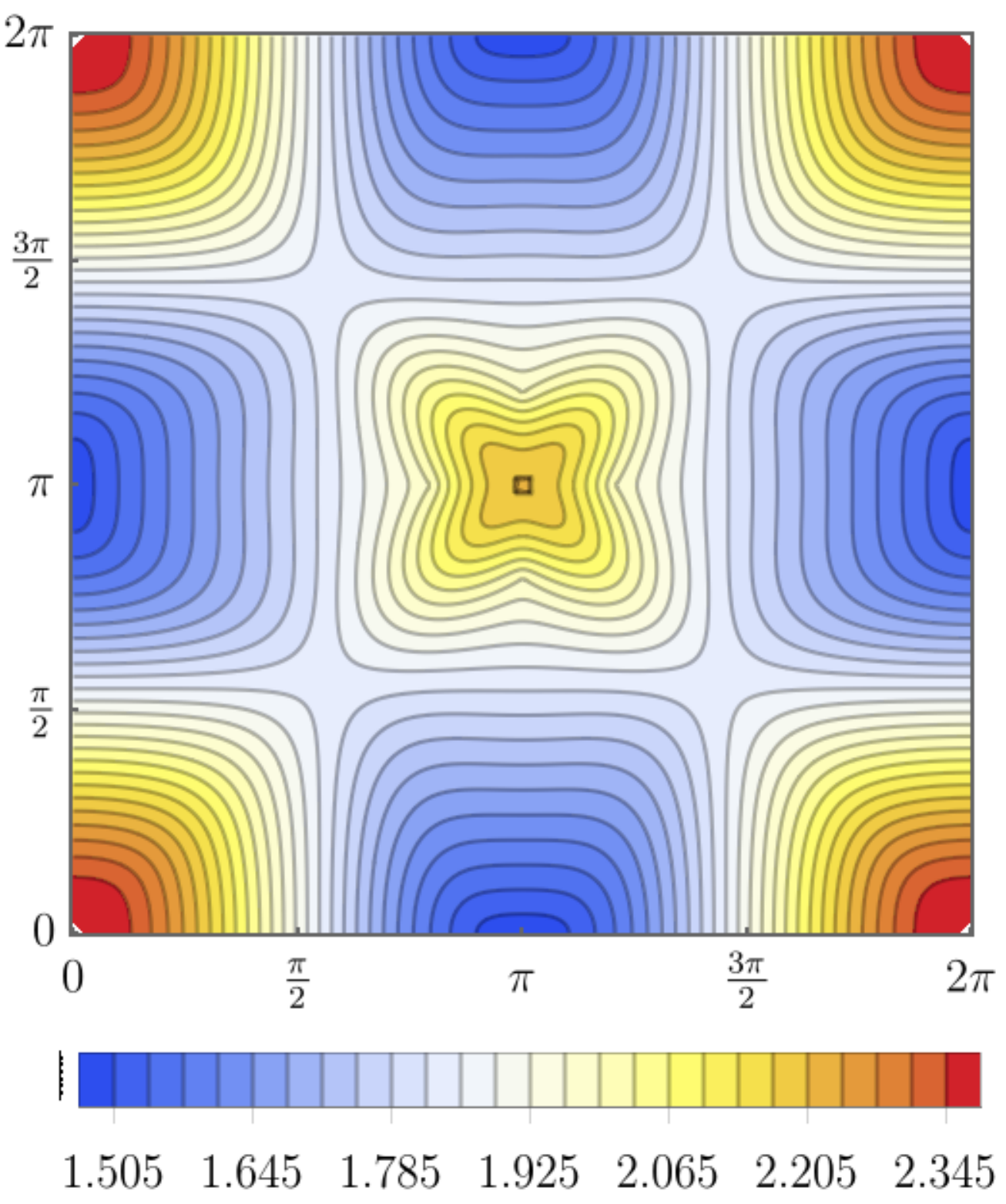}
    \end{subfigure}%
    \begin{subfigure}{0.2\textwidth}
        \centering
        \caption{\label{fig:surf_contour_4_10_Ray}}
        \includegraphics[width=1.00\linewidth]{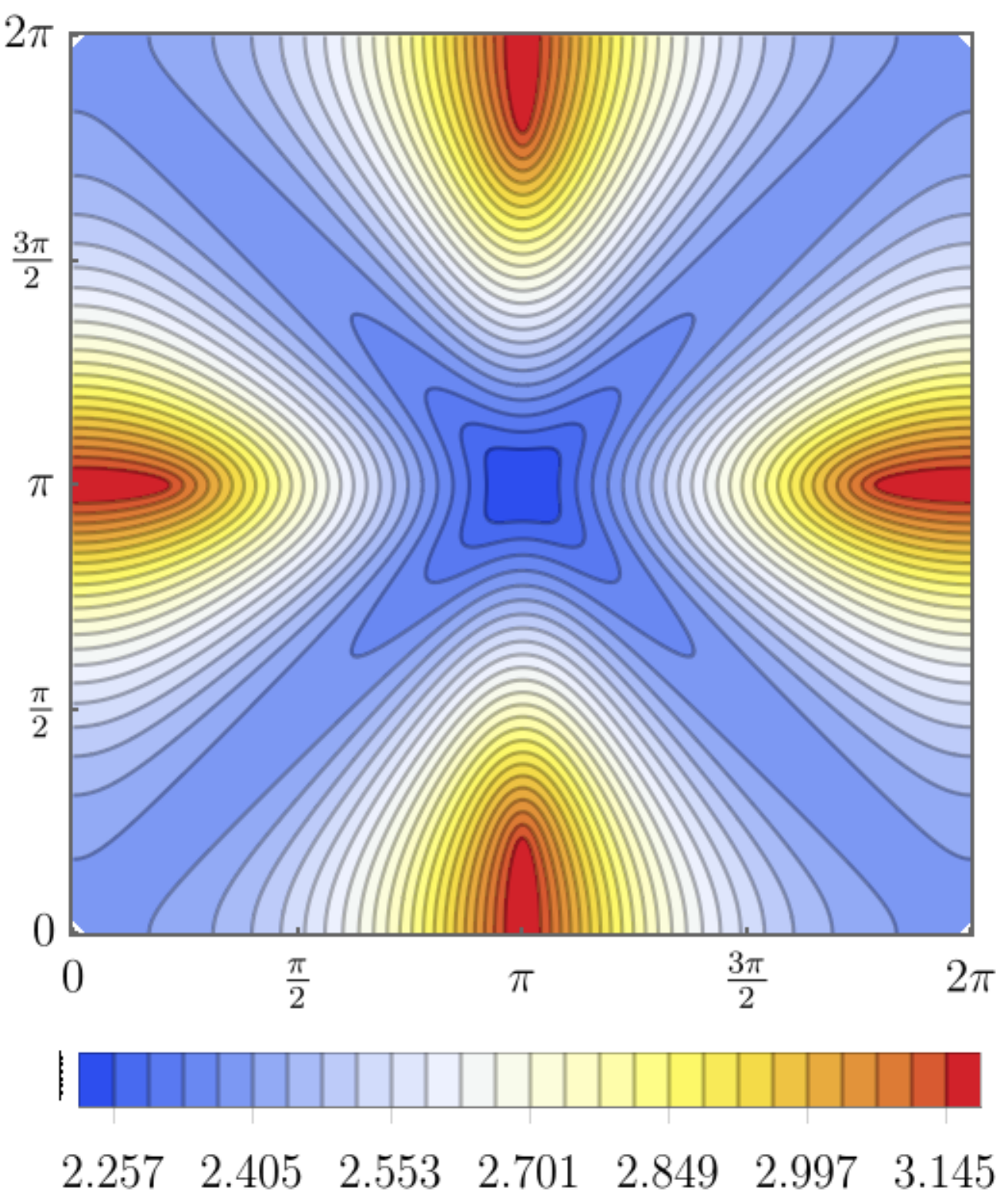}
    \end{subfigure}%
    \begin{subfigure}{0.2\textwidth}
        \centering
        \caption{\label{fig:surf_contour_5_10_Ray}}
        \includegraphics[width=1.00\linewidth]{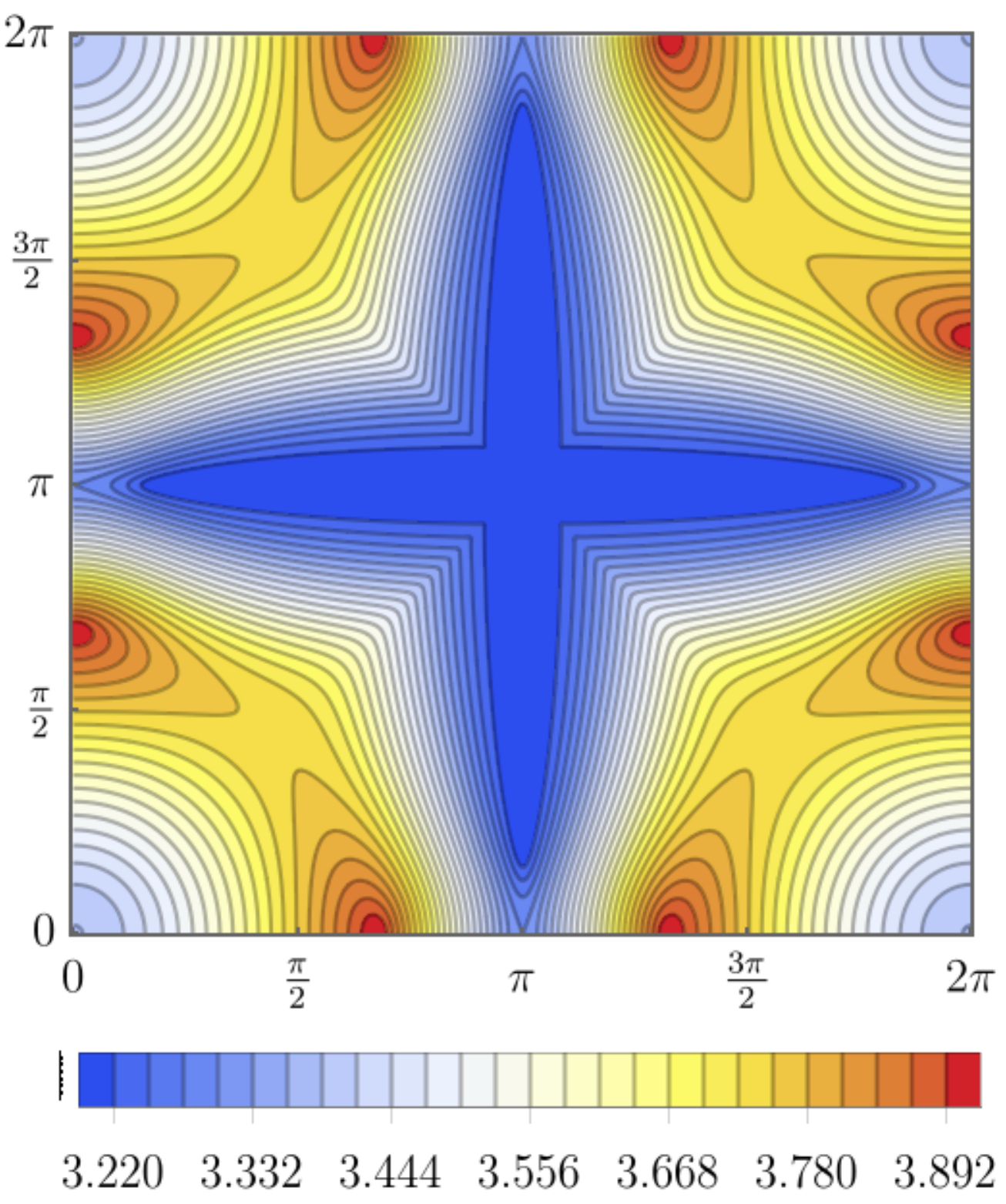}
    \end{subfigure}
\caption{\label{fig:surf_contour_10_Ray} 
The cubic anisotropy, occurring at all frequencies, is evident from the slowness contours (\subref{fig:surf_contour_1_10_Ray})--(\subref{fig:surf_contour_5_10_Ray}) and associated dispersion surfaces (\subref{fig:surf_1_10_Ray})--(\subref{fig:surf_5_10_Ray}), for a square grid of Rayleigh beams with slenderness $\lambda=10$. 
}
\end{figure}

\begin{figure}[htb!]
\centering
\begin{minipage}[c][]{0.5\textwidth}
    \begin{minipage}[c][]{0.5\textwidth}
	    \centering
	    \begin{subfigure}{\textwidth}
		    \centering
		    \includegraphics[width=\linewidth]{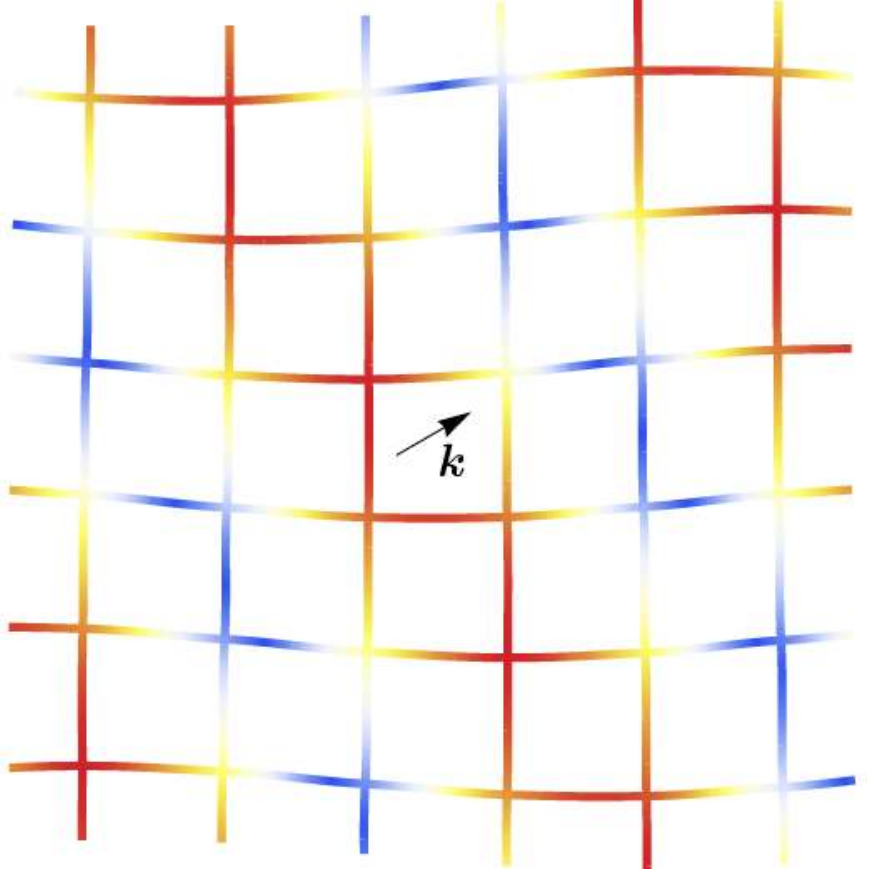}
	    \end{subfigure}
    \end{minipage}%
    \begin{minipage}[c][]{0.5\textwidth}
    \centering
	    \begin{subfigure}{0.5\textwidth}
	        \centering
	        \includegraphics[width=\linewidth]{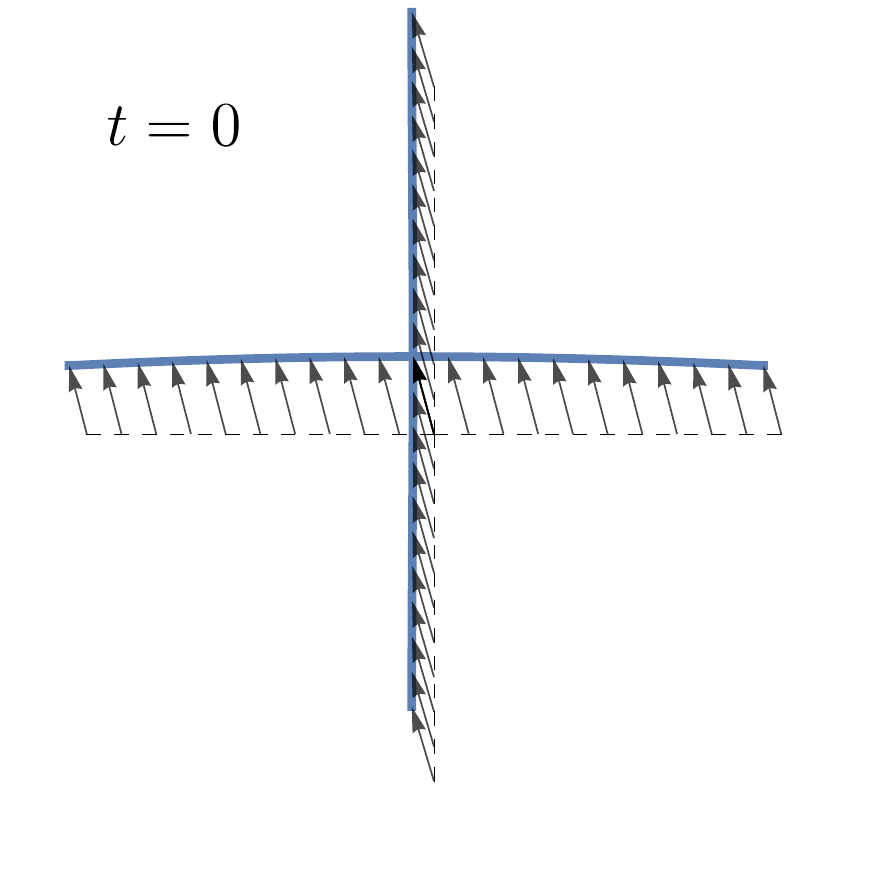}
	    \end{subfigure}%
	    \begin{subfigure}{0.5\textwidth}
	        \centering
	        \includegraphics[width=\linewidth]{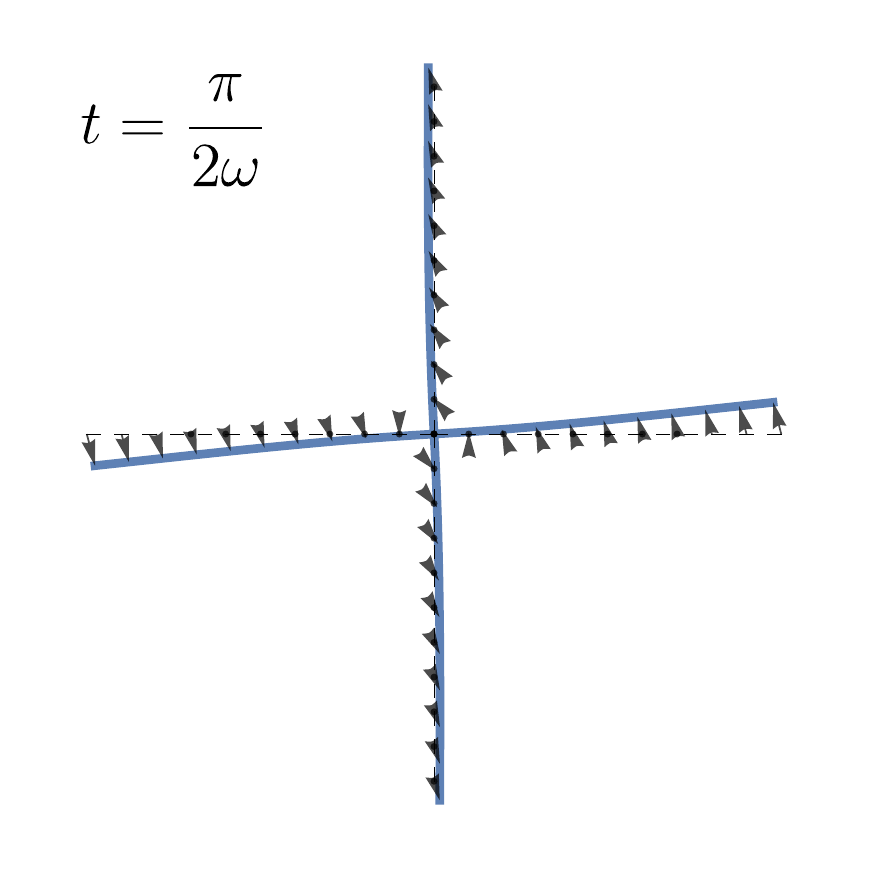}
	    \end{subfigure}
        \begin{subfigure}{0.5\textwidth}
	        \centering
	        \includegraphics[width=\linewidth]{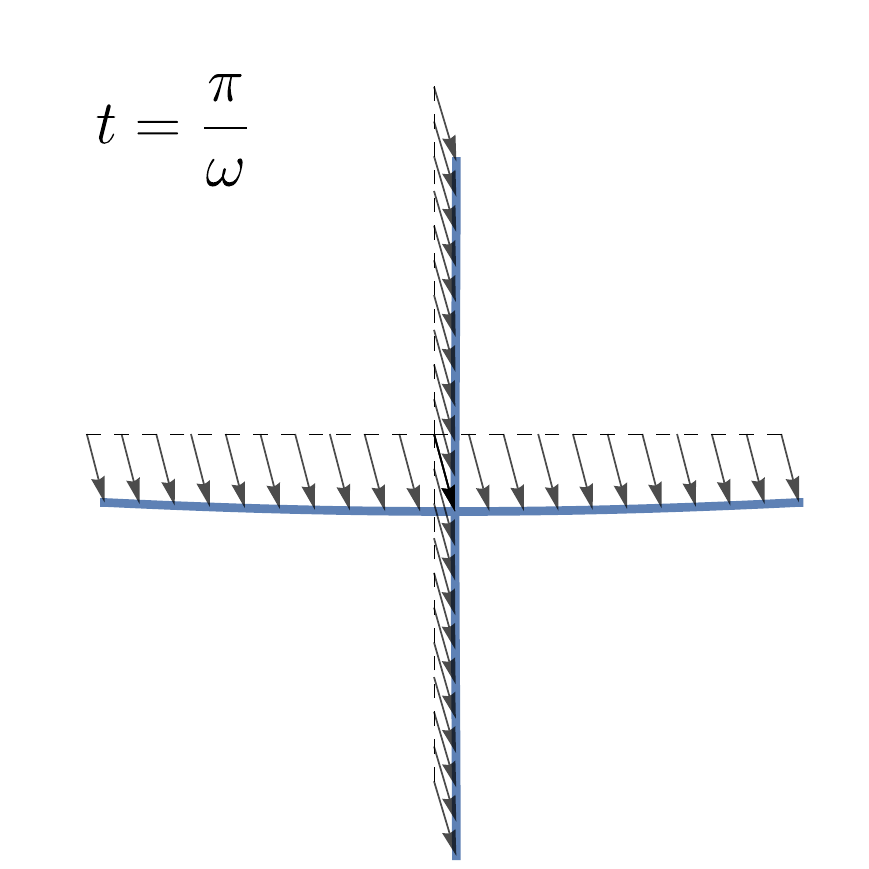}
	    \end{subfigure}%
	    \begin{subfigure}{0.5\textwidth}
	        \centering
	        \includegraphics[width=\linewidth]{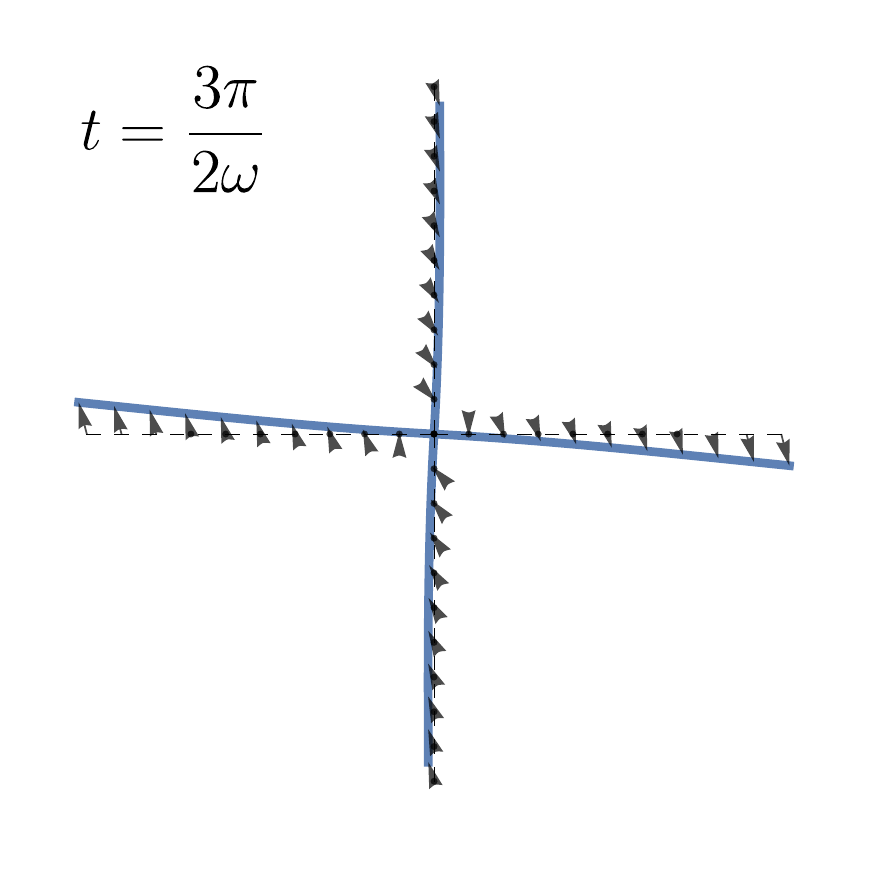}
	    \end{subfigure}
    \end{minipage}
	\subcaption{\label{fig:mode_1_5}Waveform at $P_1\!:\!\{K_1,K_2,\Omega\}\!=\!\{\frac{\sqrt{3}}{2},\frac{1}{2},0.230\}$.}
\end{minipage}%
\begin{minipage}[c][]{0.5\textwidth}
    \begin{minipage}[c][]{0.5\textwidth}
	    \centering
	    \begin{subfigure}{\textwidth}
		    \centering
		    \includegraphics[width=\linewidth]{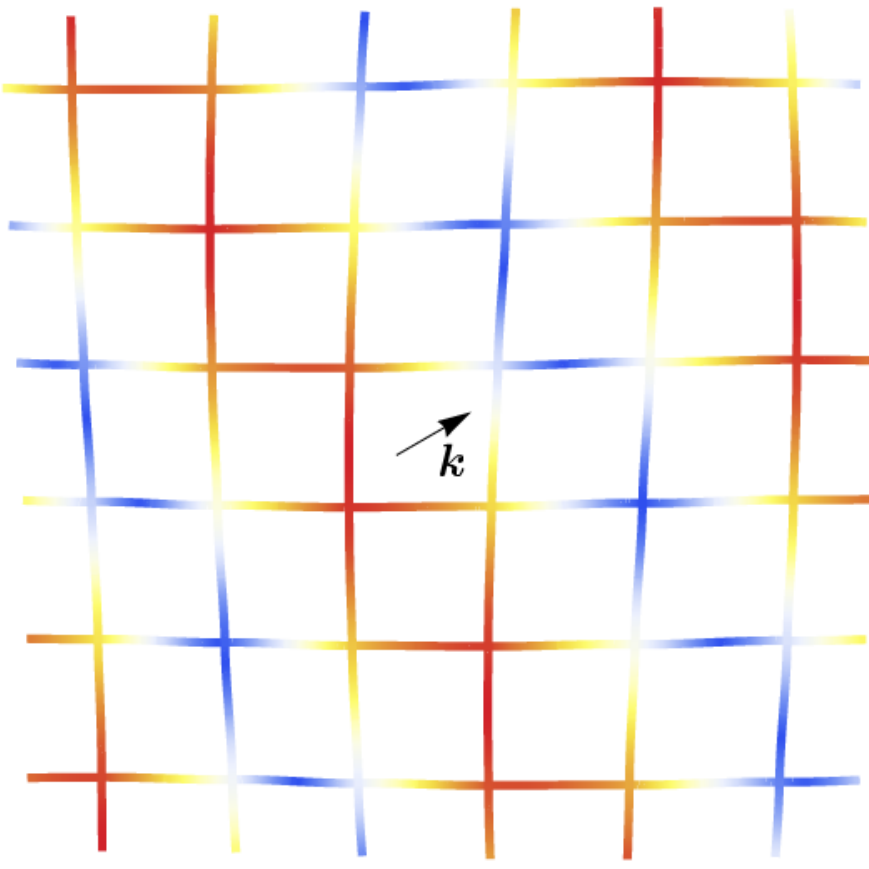}
	    \end{subfigure}
    \end{minipage}%
    \begin{minipage}[c][]{0.5\textwidth}
    \centering
	    \begin{subfigure}{0.5\textwidth}
	        \centering
	        \includegraphics[width=\linewidth]{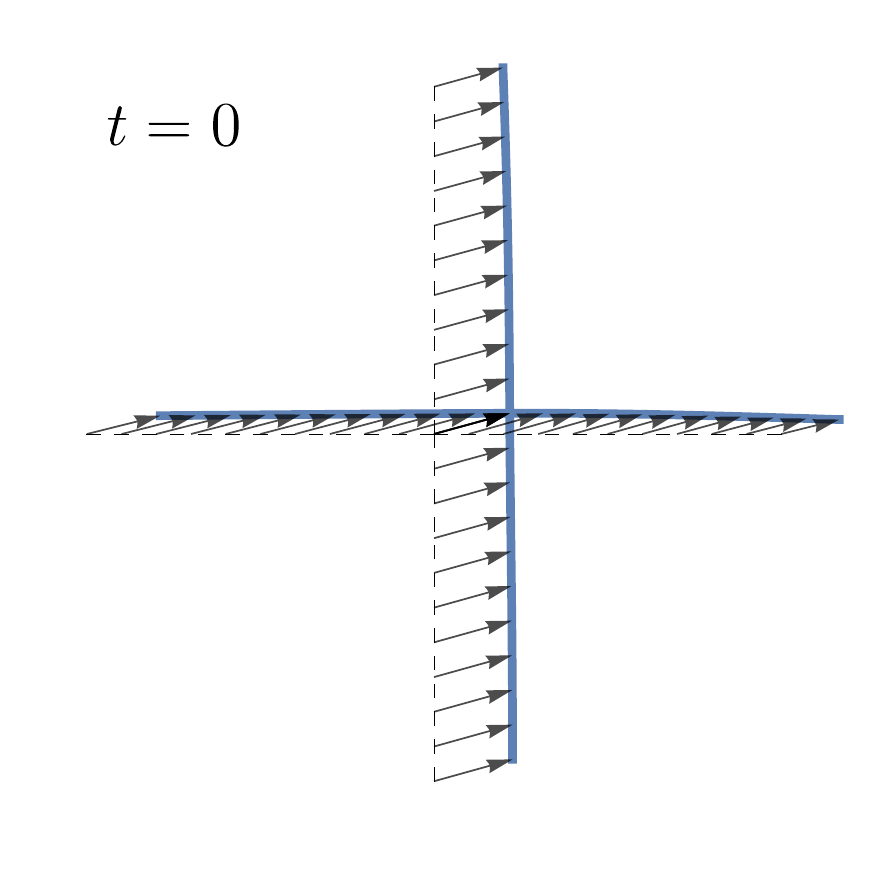}
	    \end{subfigure}%
	    \begin{subfigure}{0.5\textwidth}
	        \centering
	        \includegraphics[width=\linewidth]{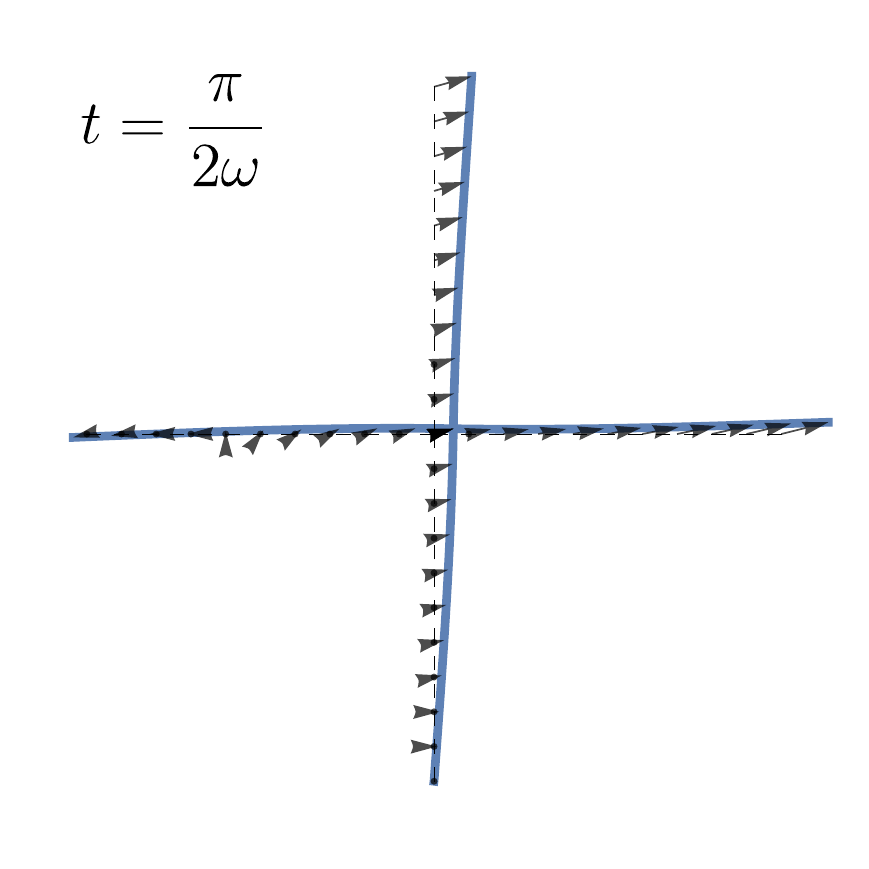}
	    \end{subfigure}
        \begin{subfigure}{0.5\textwidth}
	        \centering
	        \includegraphics[width=\linewidth]{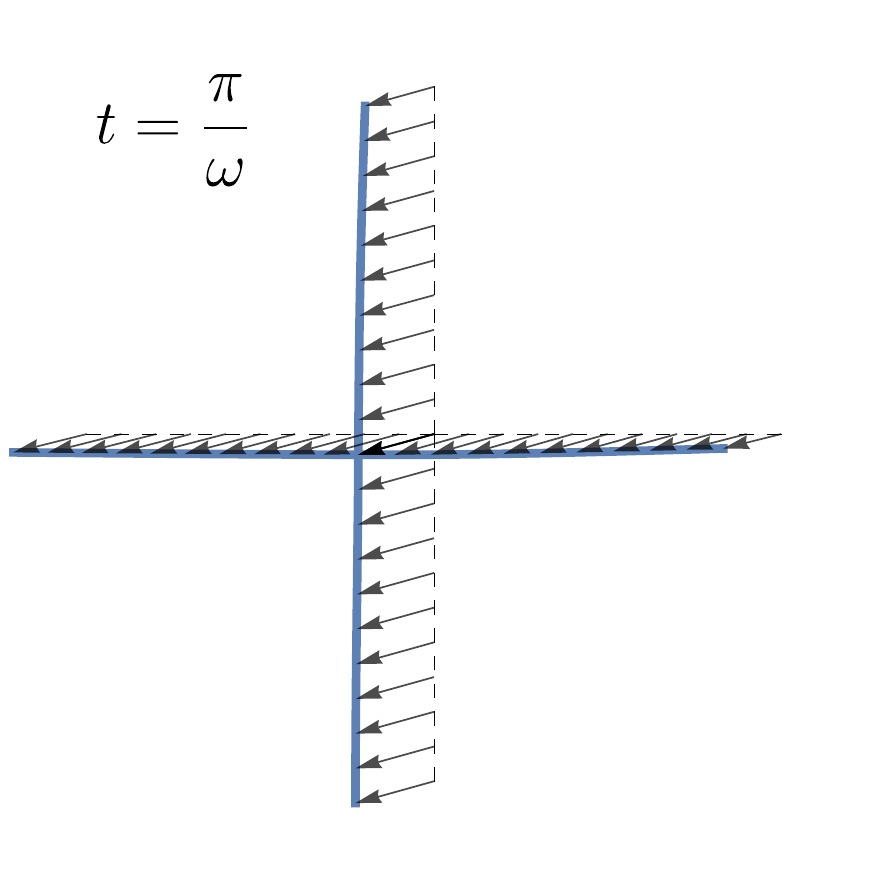}
	    \end{subfigure}%
	    \begin{subfigure}{0.5\textwidth}
	        \centering
	        \includegraphics[width=\linewidth]{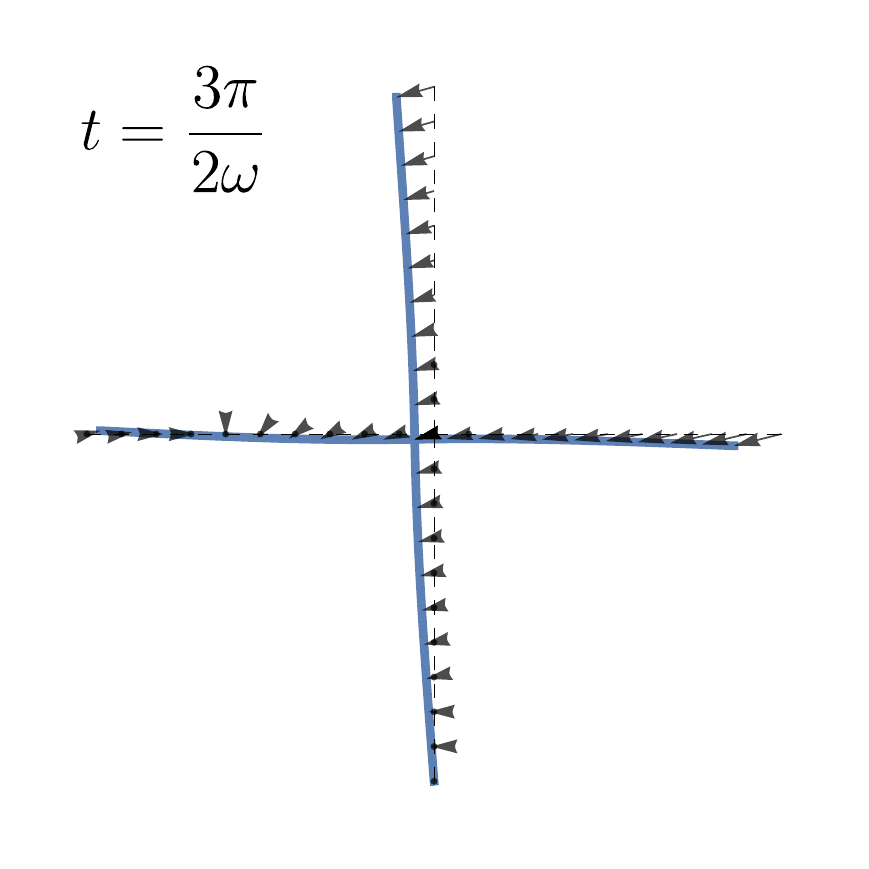}
	    \end{subfigure}
    \end{minipage}
	\subcaption{\label{fig:mode_2_5}Waveform at $P_2\!:\!\{K_1,K_2,\Omega\}\!=\!\{\frac{\sqrt{3}}{2},\frac{1}{2},0.325\}$.}
\end{minipage}
\caption{\label{fig:mode_1_2_5}
Low-frequency (long-wavelength) waveforms for a square Rayleigh beam grid, computed on the acoustic branches (the two lowest dispersion surfaces). 
The locations of the corresponding points $P_i$ on the dispersion surfaces are represented in Figs.~\ref{fig:surf_contour_1_5_Ray}~and~\ref{fig:surf_contour_2_5_Ray}.}
\end{figure}

\begin{figure}[htb!]
\centering
\begin{minipage}[c][]{0.5\textwidth}
    \begin{minipage}[c][]{0.5\textwidth}
	    \centering
	    \begin{subfigure}{\textwidth}
		    \centering
		    \includegraphics[width=\linewidth]{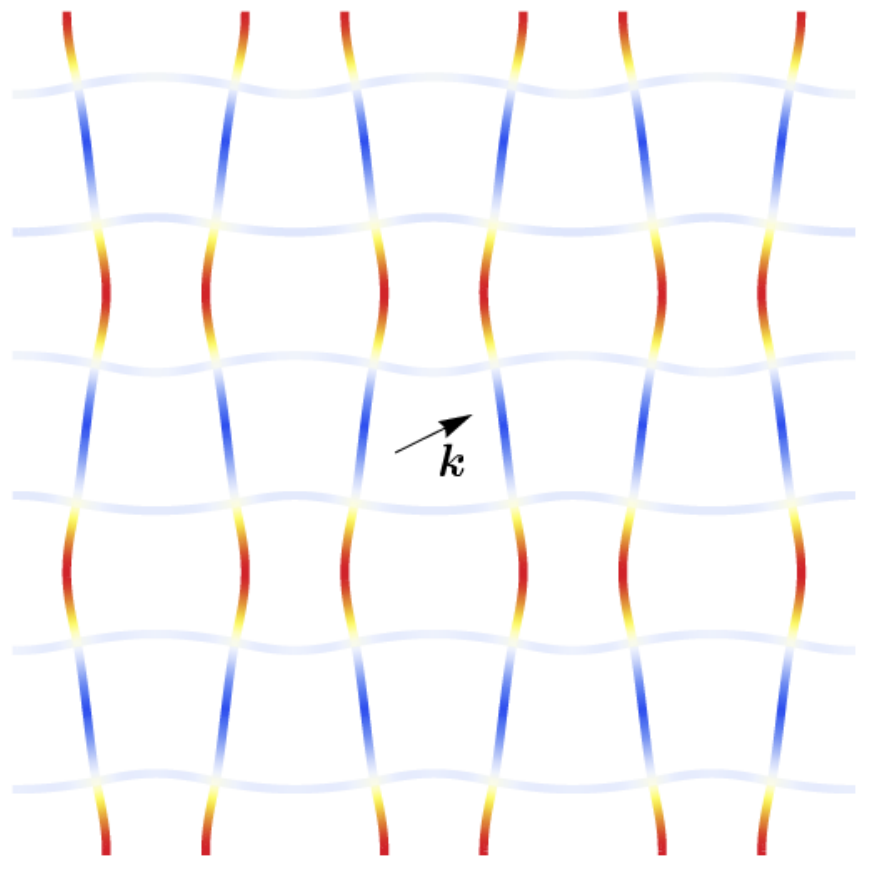}
	    \end{subfigure}
    \end{minipage}%
    \begin{minipage}[c][]{0.5\textwidth}
    \centering
	    \begin{subfigure}{0.5\textwidth}
	        \centering
	        \includegraphics[width=\linewidth]{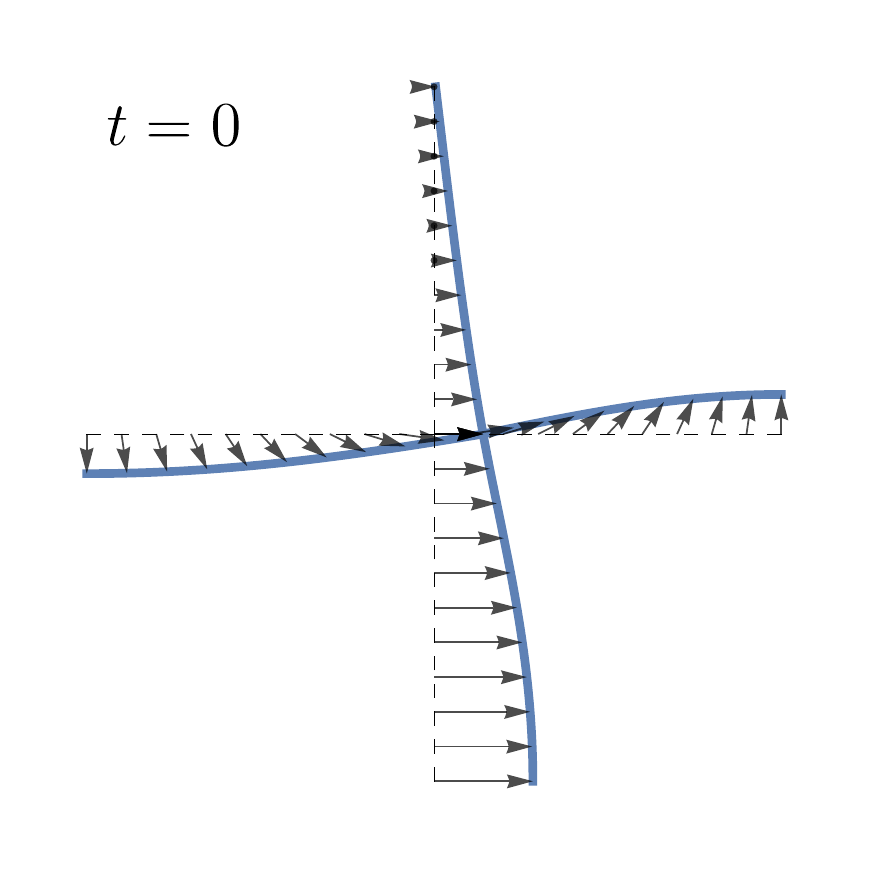}
	    \end{subfigure}%
	    \begin{subfigure}{0.5\textwidth}
	        \centering
	        \includegraphics[width=\linewidth]{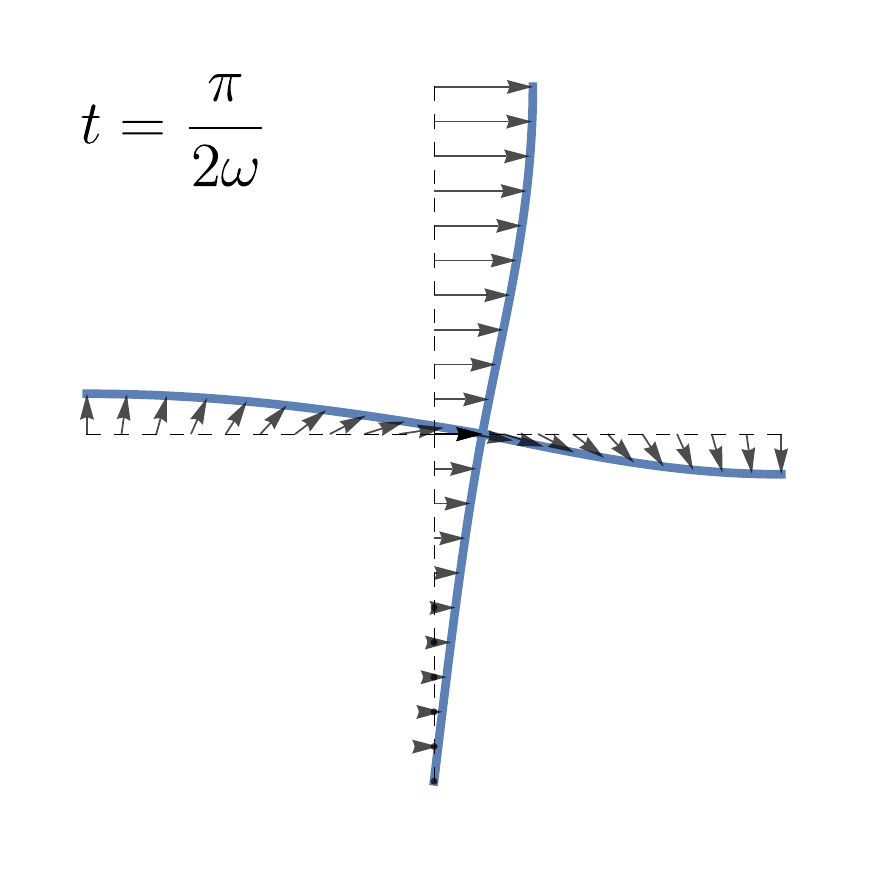}
	    \end{subfigure}
        \begin{subfigure}{0.5\textwidth}
	        \centering
	        \includegraphics[width=\linewidth]{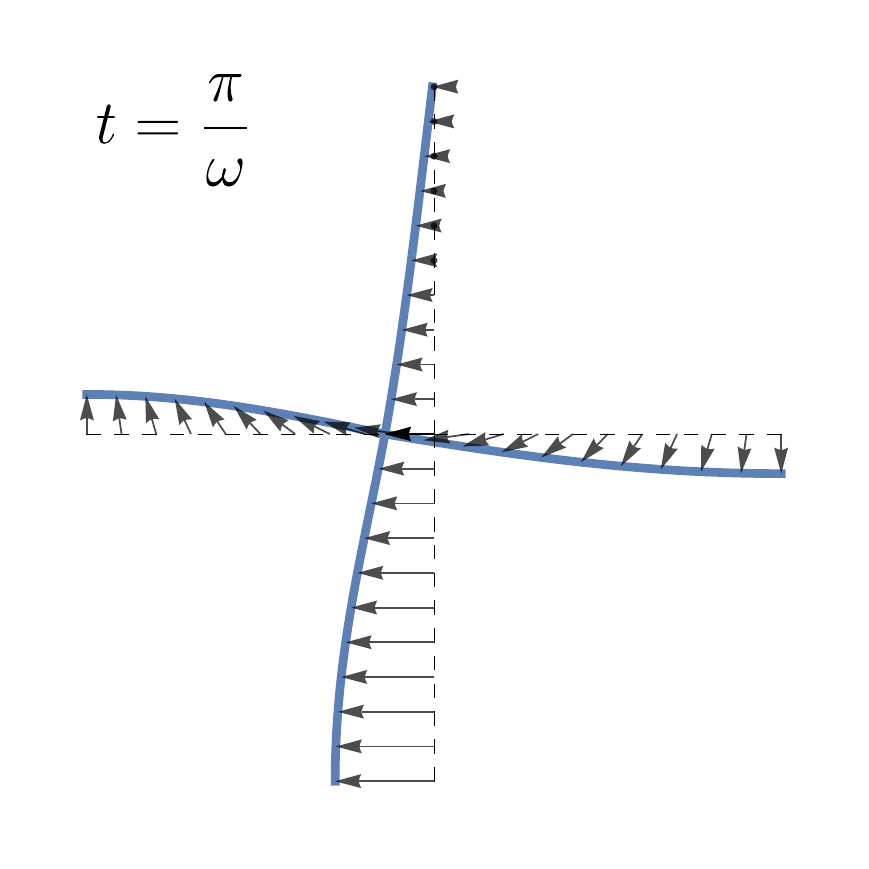}
	    \end{subfigure}%
	    \begin{subfigure}{0.5\textwidth}
	        \centering
	        \includegraphics[width=\linewidth]{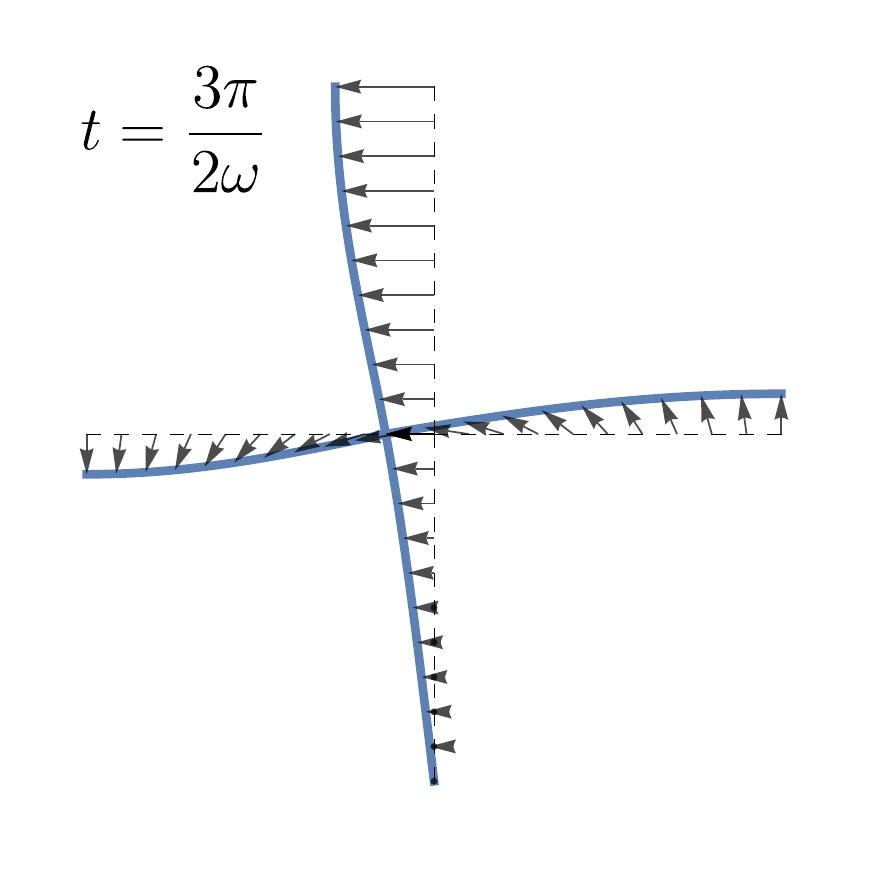}
	    \end{subfigure}
    \end{minipage}
	\subcaption{\label{fig:mode_10_5}$1^\textrm{st}$ waveform at $P_3\!:\!\{K_1,K_2,\Omega\}\!=\!\{\pi,1.555,0.812\}$.}
\end{minipage}%
\begin{minipage}[c][]{0.5\textwidth}
    \begin{minipage}[c][]{0.5\textwidth}
	    \centering
	    \begin{subfigure}{\textwidth}
		    \centering
		    \includegraphics[width=\linewidth]{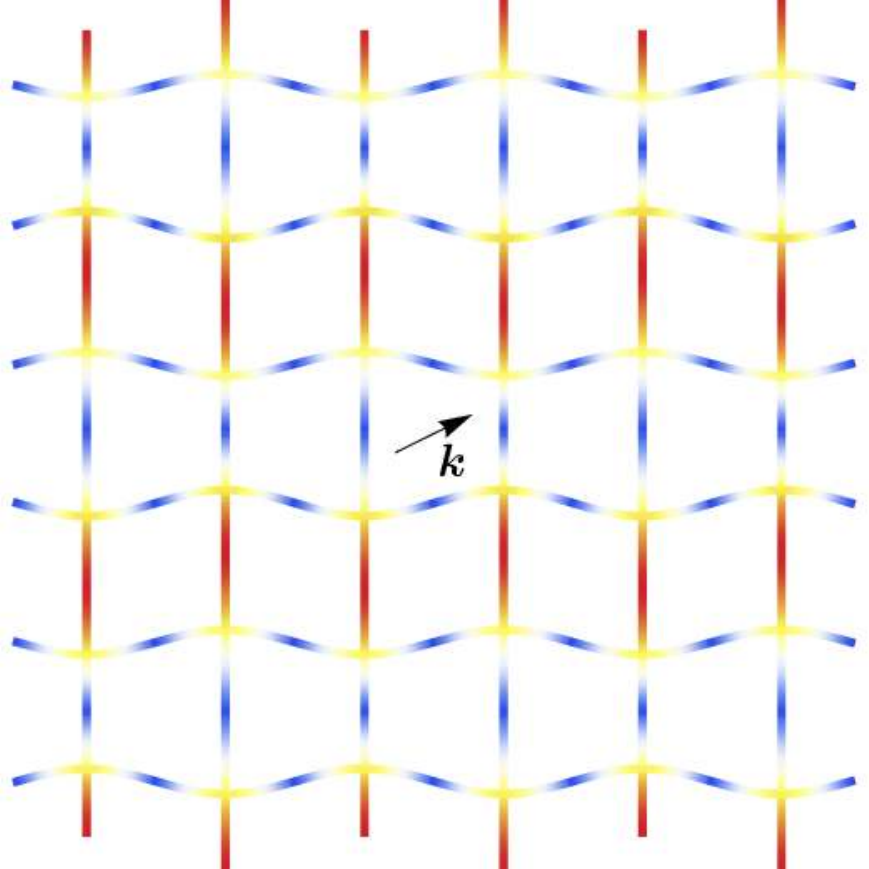}
	    \end{subfigure}
    \end{minipage}%
    \begin{minipage}[c][]{0.5\textwidth}
    \centering
	    \begin{subfigure}{0.5\textwidth}
	        \centering
	        \includegraphics[width=\linewidth]{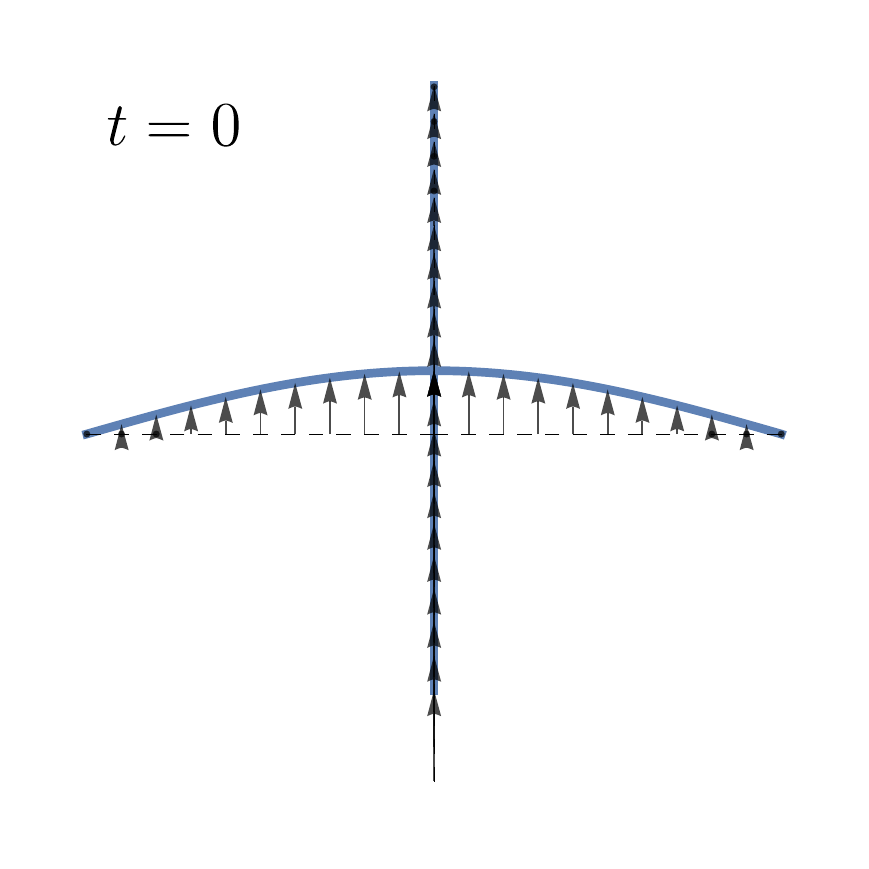}
	    \end{subfigure}%
	    \begin{subfigure}{0.5\textwidth}
	        \centering
	        \includegraphics[width=\linewidth]{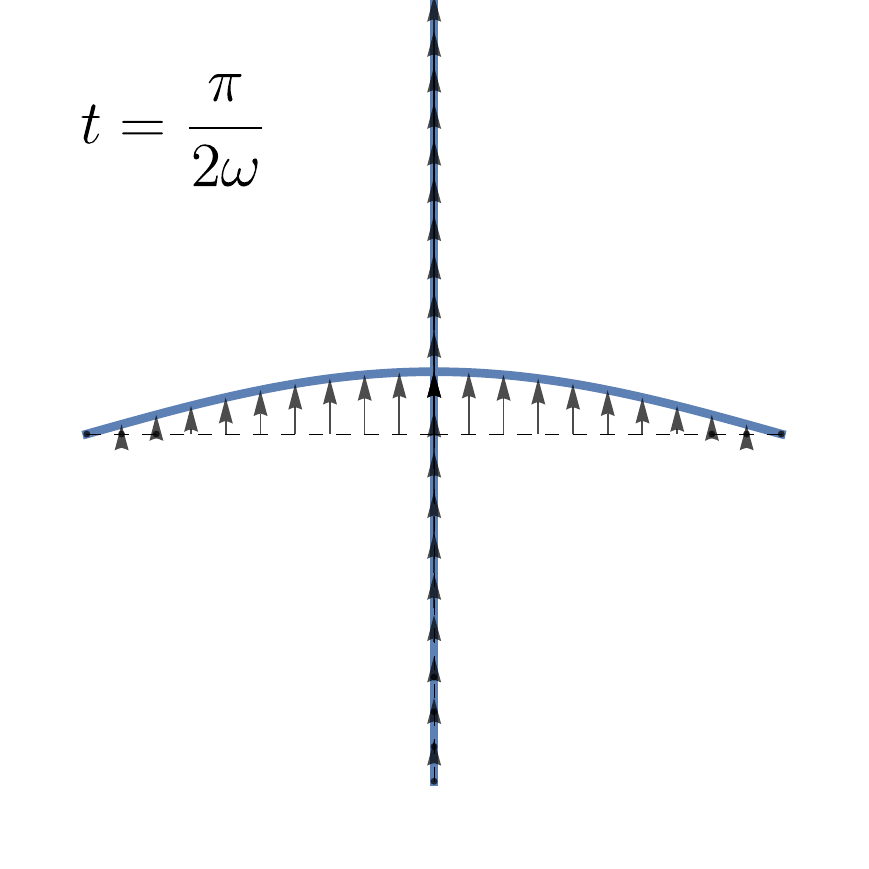}
	    \end{subfigure}
        \begin{subfigure}{0.5\textwidth}
	        \centering
	        \includegraphics[width=\linewidth]{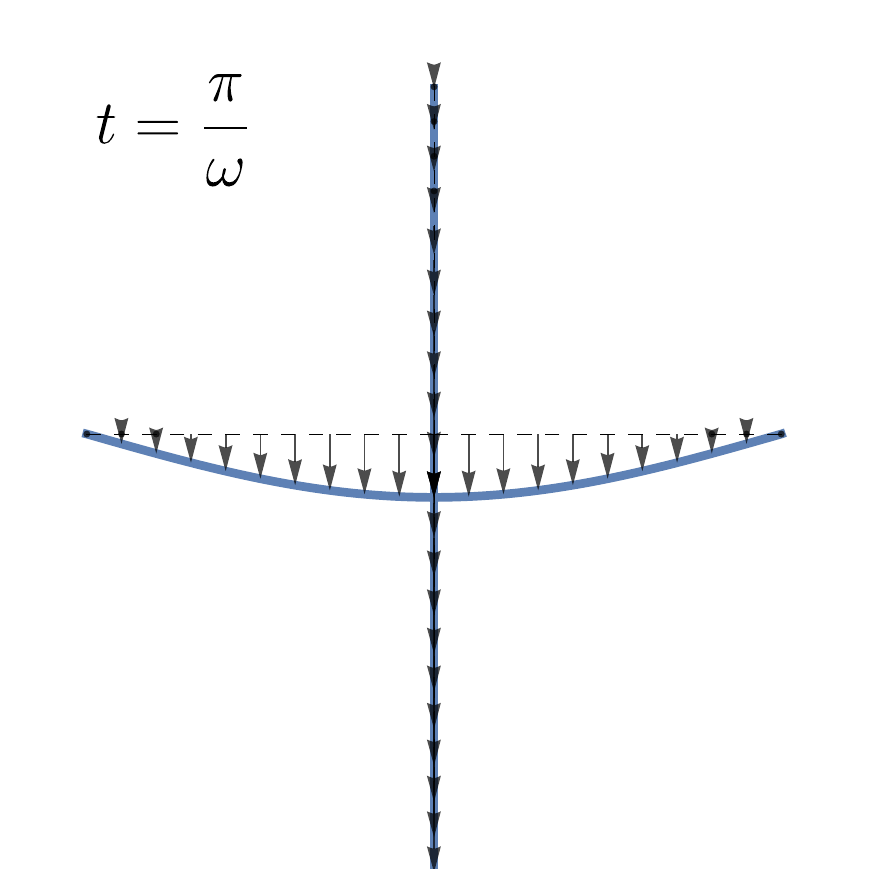}
	    \end{subfigure}%
	    \begin{subfigure}{0.5\textwidth}
	        \centering
	        \includegraphics[width=\linewidth]{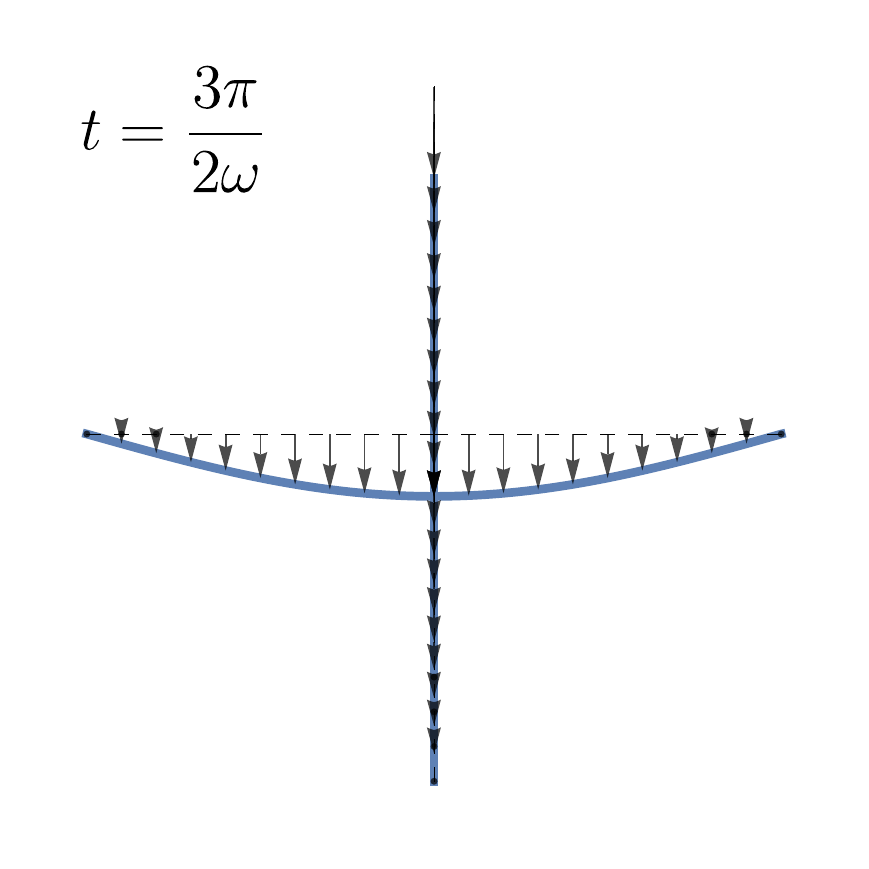}
	    \end{subfigure}
    \end{minipage}
	\subcaption{\label{fig:mode_11_5}$2^\textrm{nd}$ waveform at $P_3\!:\!\{K_1,K_2,\Omega\}\!=\!\{\pi,1.555,0.812\}$.}
\end{minipage}
\caption{\label{fig:mode_10_11_5}
Waveforms in a square grid of Rayleigh beams at the Dirac vertex, which connects the first and the second dispersion surface. Both waves propagate vertically, but are standing horizontally and, in particular, the vertical beams transmit pure flexural vibrations (\subref{fig:mode_10_5}) and pure axial waves (\subref{fig:mode_11_5}).}
\end{figure}

\begin{figure}[htb!]
\centering
\begin{minipage}[c][]{0.5\textwidth}
    \begin{minipage}[c][]{0.5\textwidth}
	    \centering
	    \begin{subfigure}{\textwidth}
		    \centering
		    \includegraphics[width=\linewidth]{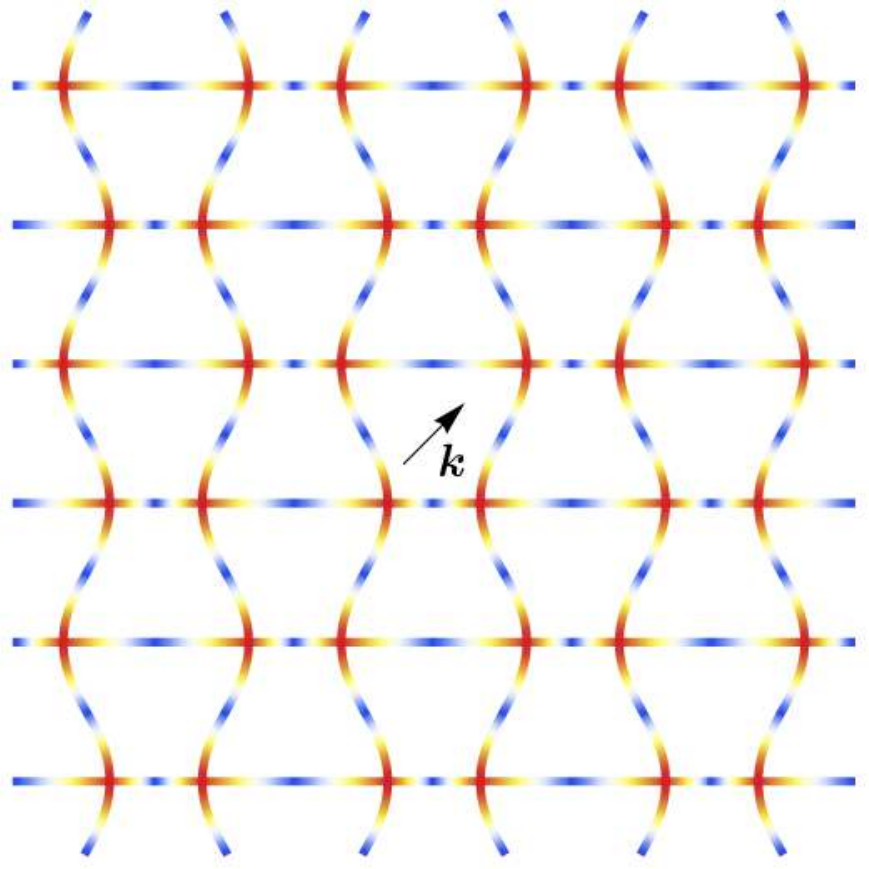}
	    \end{subfigure}
    \end{minipage}%
    \begin{minipage}[c][]{0.5\textwidth}
    \centering
	    \begin{subfigure}{0.5\textwidth}
	        \centering
	        \includegraphics[width=\linewidth]{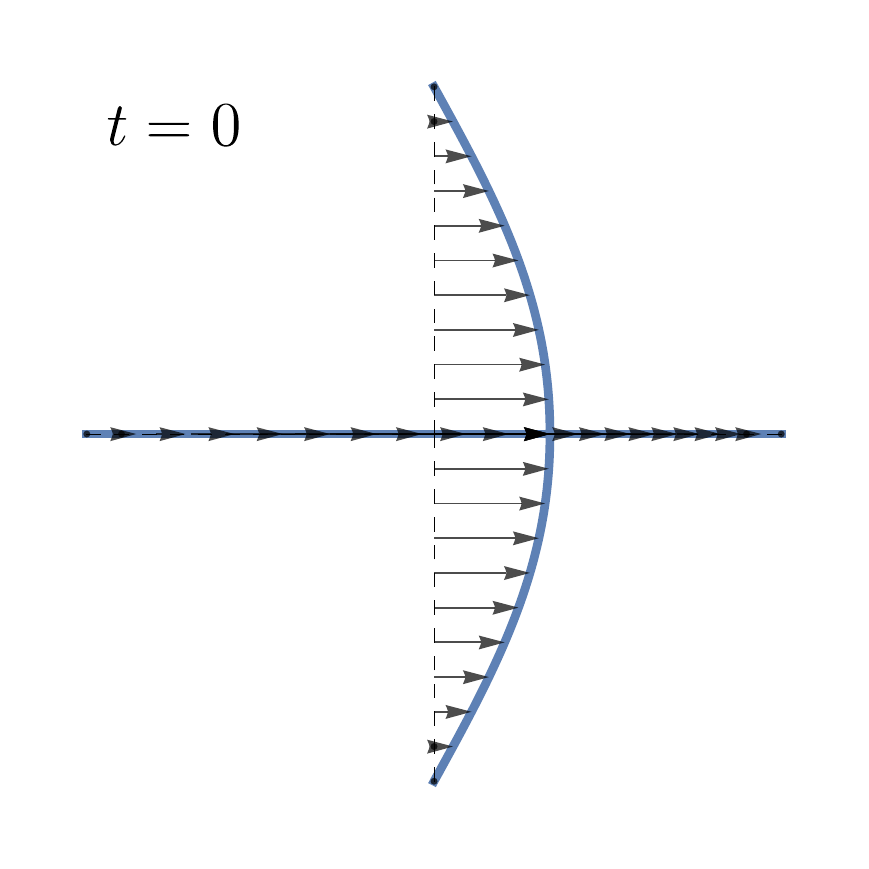}
	    \end{subfigure}%
	    \begin{subfigure}{0.5\textwidth}
	        \centering
	        \includegraphics[width=\linewidth]{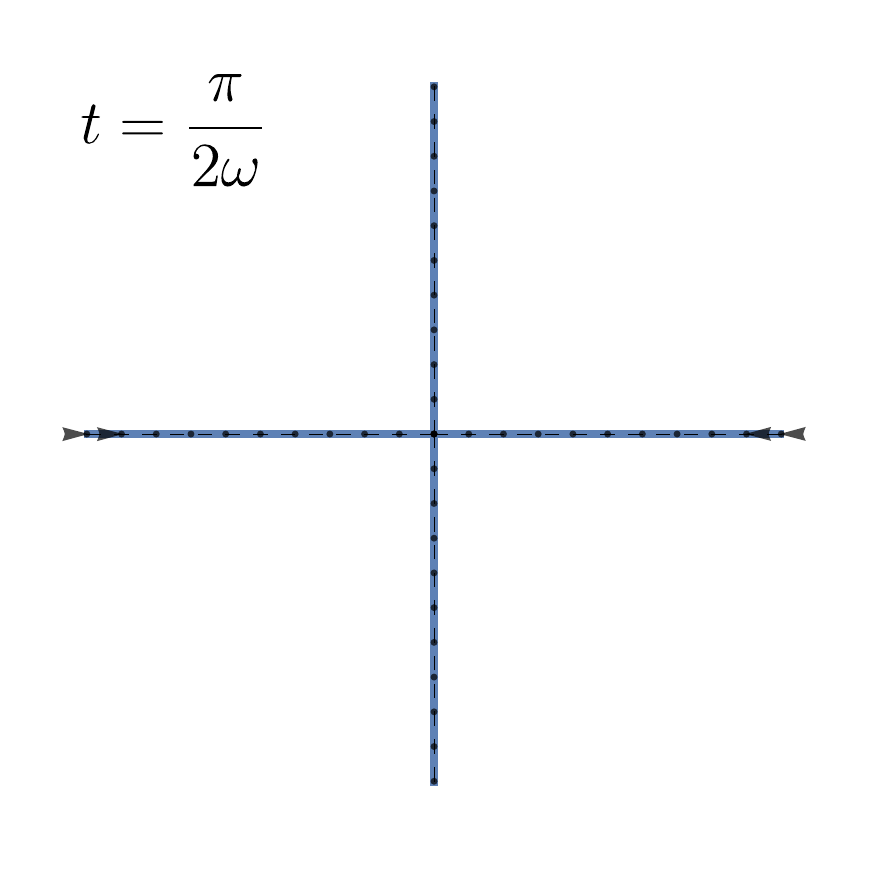}
	    \end{subfigure}
        \begin{subfigure}{0.5\textwidth}
	        \centering
	        \includegraphics[width=\linewidth]{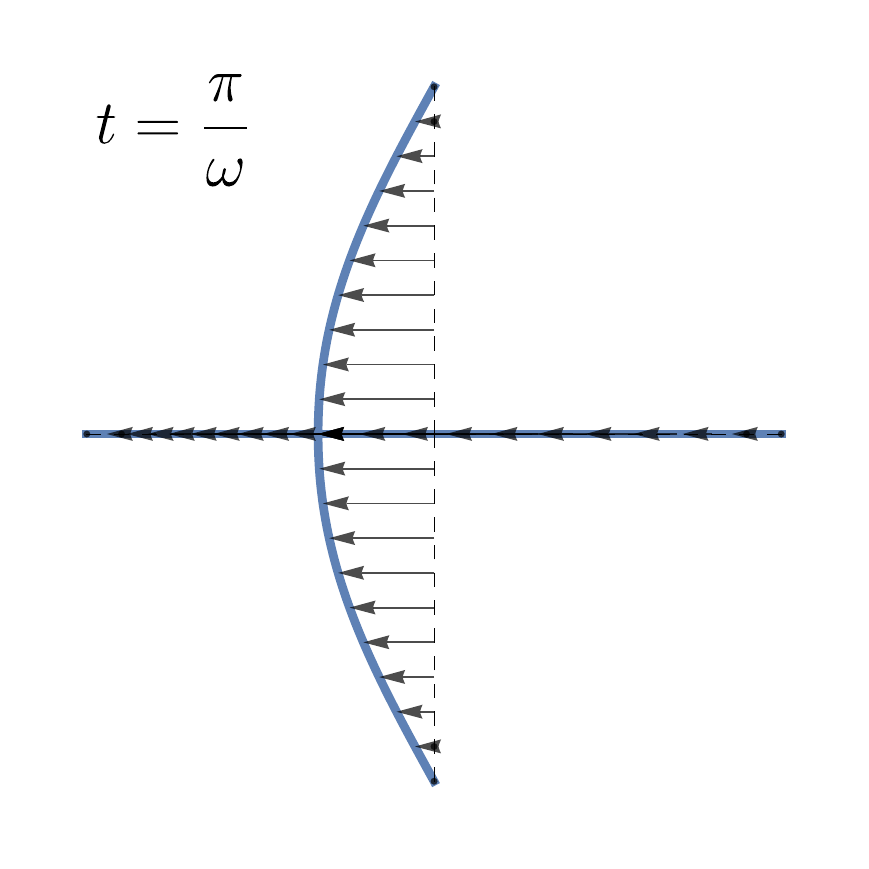}
	    \end{subfigure}%
	    \begin{subfigure}{0.5\textwidth}
	        \centering
	        \includegraphics[width=\linewidth]{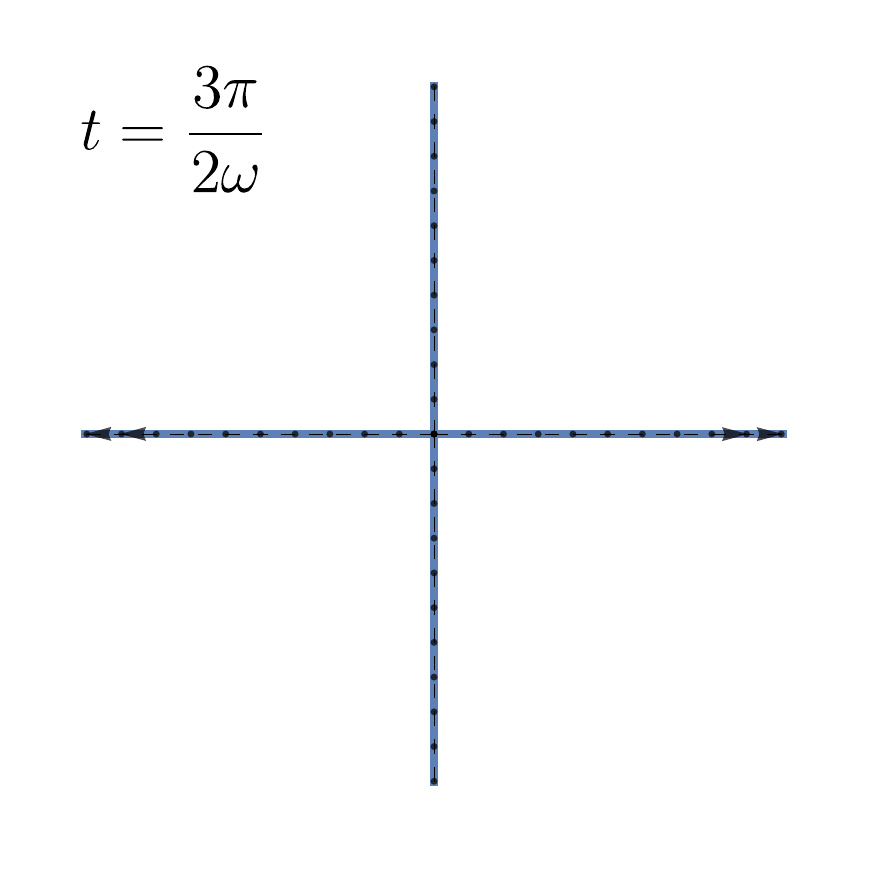}
	    \end{subfigure}
    \end{minipage}
	\subcaption{\label{fig:mode_6_5}Waveform at $P_5\!:\!\{K_1,K_2,\Omega\}\!=\!\{\pi,\pi,1.193\}$.}
\end{minipage}%
\begin{minipage}[c][]{0.5\textwidth}
    \begin{minipage}[c][]{0.5\textwidth}
	    \centering
	    \begin{subfigure}{\textwidth}
		    \centering
		    \includegraphics[width=\linewidth]{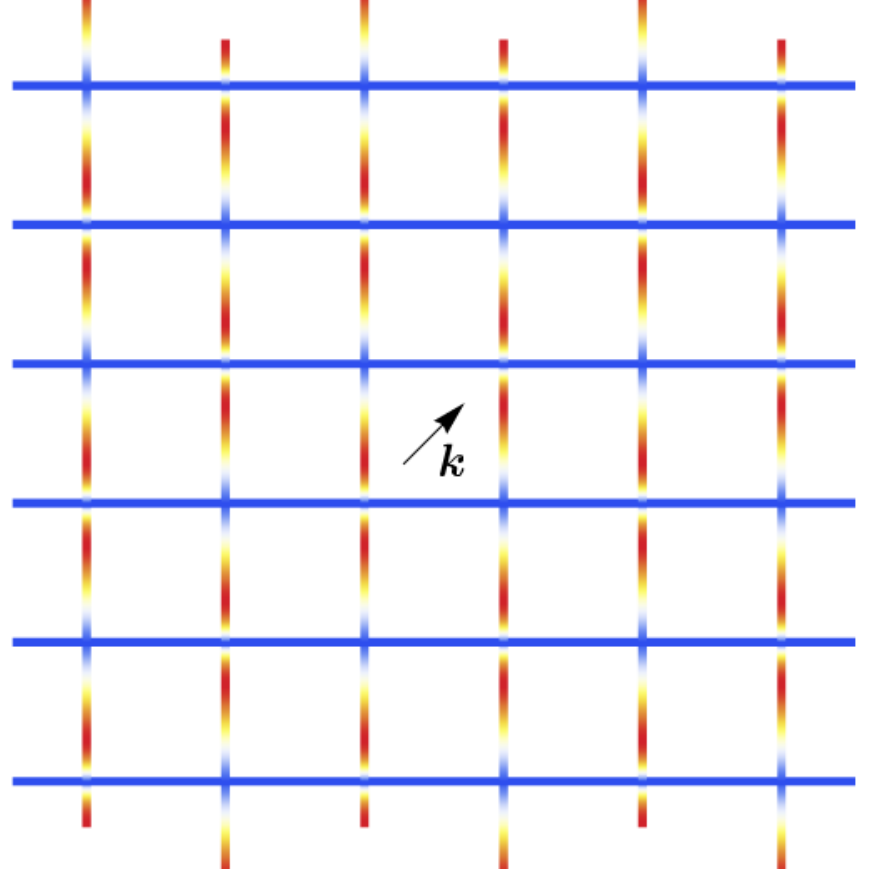}
	    \end{subfigure}
    \end{minipage}%
    \begin{minipage}[c][]{0.5\textwidth}
    \centering
	    \begin{subfigure}{0.5\textwidth}
	        \centering
	        \includegraphics[width=\linewidth]{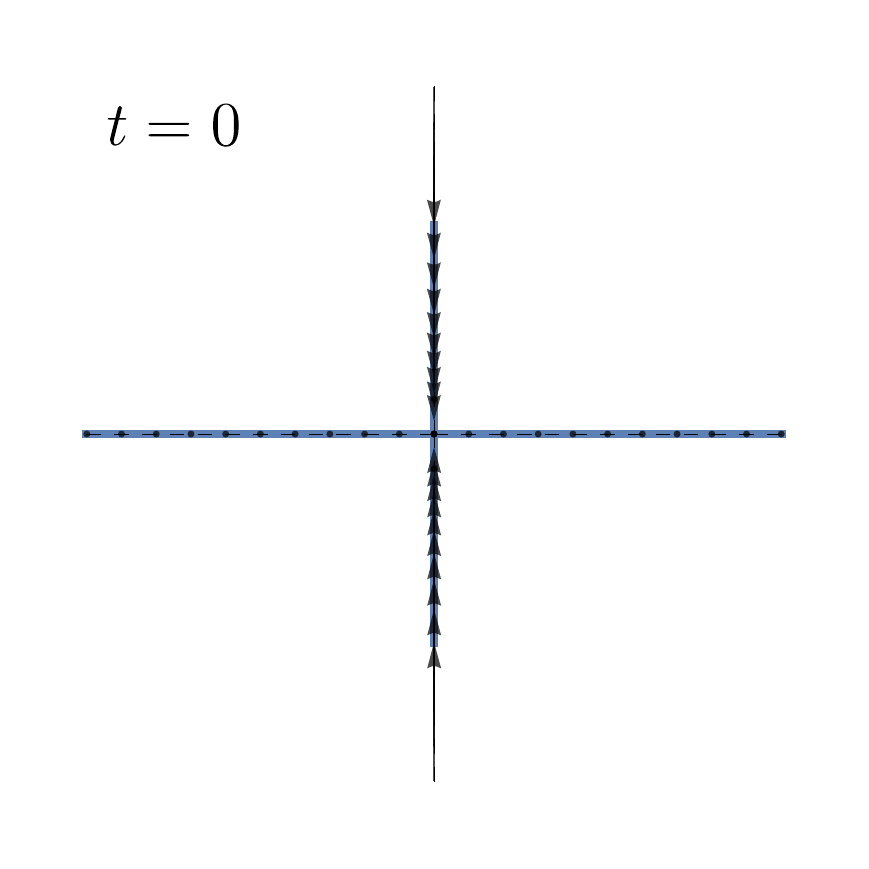}
	    \end{subfigure}%
	    \begin{subfigure}{0.5\textwidth}
	        \centering
	        \includegraphics[width=\linewidth]{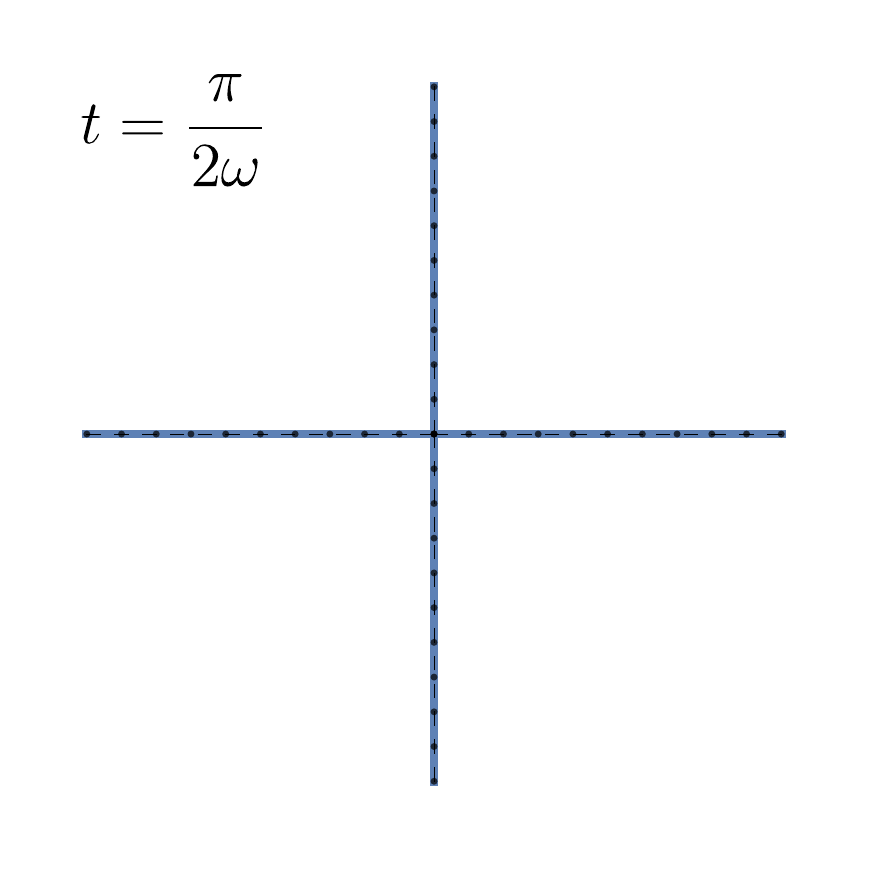}
	    \end{subfigure}
        \begin{subfigure}{0.5\textwidth}
	        \centering
	        \includegraphics[width=\linewidth]{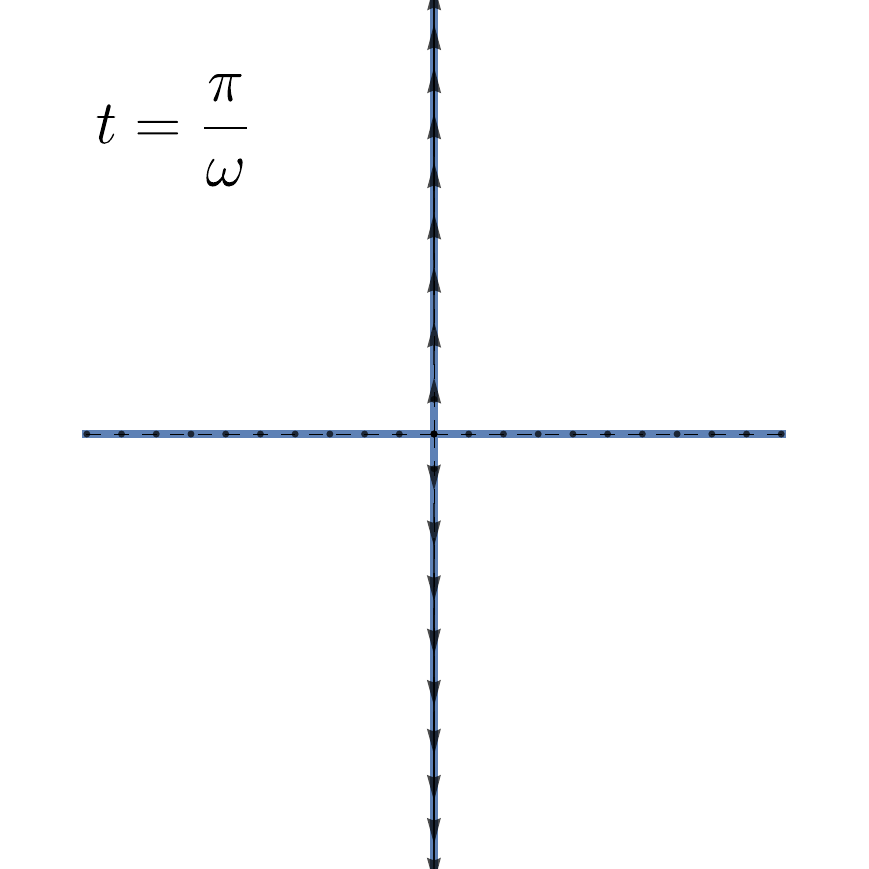}
	    \end{subfigure}%
	    \begin{subfigure}{0.5\textwidth}
	        \centering
	        \includegraphics[width=\linewidth]{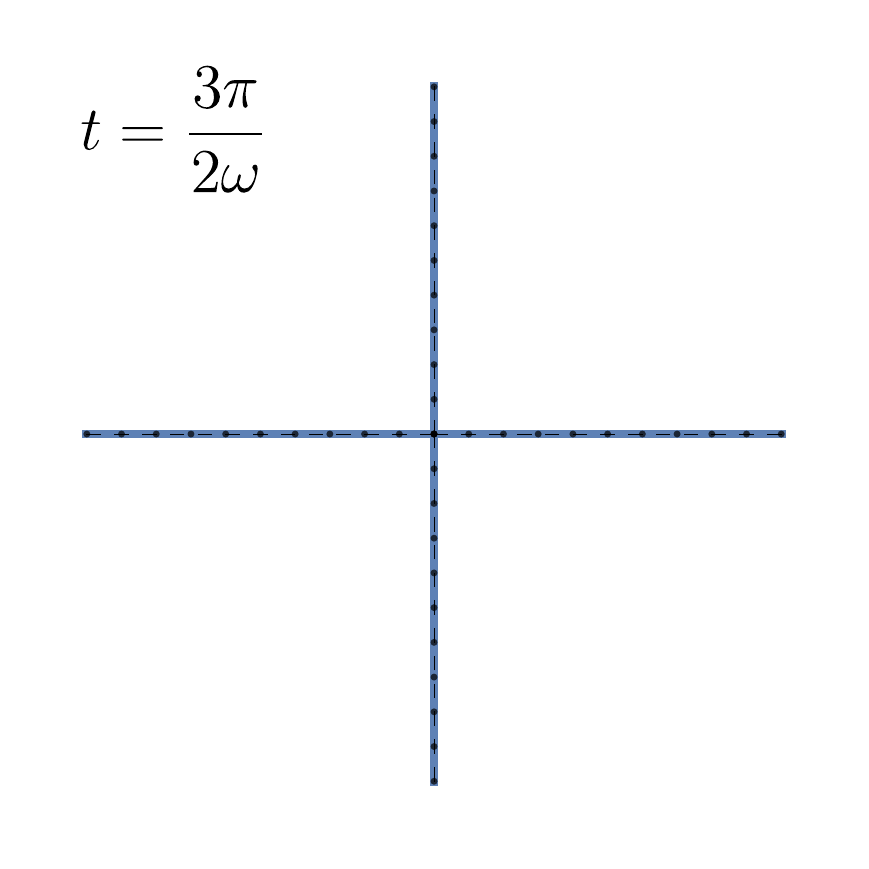}
	    \end{subfigure}
    \end{minipage}
	\subcaption{\label{fig:mode_8_5}Waveform at $P_6\!:\!\{K_1,K_2,\Omega\}\!=\!\{\pi,\pi,1.592\}$.}
\end{minipage}
\caption{\label{fig:mode_6_8_5}
Standing waves occurring in a square grid of Rayleigh beams, corresponding to the double root connecting the second to the third~(\subref{fig:mode_6_5}) and the fourth to the fifth~(\subref{fig:mode_8_5}) dispersion surface.
The vertical beams are subject to a purely flexural deformation while the horizontal beams exhibit pure axial vibrations, so that nodal points are located at the midpoints of beams~(\subref{fig:mode_6_5}).
The waveform~(\subref{fig:mode_8_5}) occurs at a frequency $\Omega=\Omega_a=\lambda/\pi$, corresponding to the first axial vibration mode of a double-clamped beam. 
The motion of horizontal and vertical beams are completely decoupled and nodal lines are clearly visible.
Due to symmetry, the companion waveforms analogous to~(\subref{fig:mode_6_5}) and~(\subref{fig:mode_8_5}) can also propagate at the same frequencies, with reversed roles of the horizontal and vertical beams.}
\end{figure}

For the Rayleigh beam model and the two values of slenderness, $\lambda=5$ and $10$, the slowness contours of the first five dispersion surfaces have been computed and reported in Figs.~\ref{fig:surf_contour_5_Ray} and~\ref{fig:surf_contour_10_Ray}, complemented by the corresponding 3D views.
Along the contours pertaining to $\lambda=5$, points are marked (labeled $P_1, P_2,...$), for which the corresponding waveforms are shown in Figs.~\ref{fig:mode_1_2_5}--\ref{fig:mode_6_8_5}. The numerical values of the coordinates of these points are provided in Table~\ref{tab:points_modes}; in addition, the same points have also been indicated in Fig. 
\ref{fig:Brillouin_5}.
%

\begin{table}[htb!]
    \centering
    \begin{tabular}{ccccc}
        \toprule
        Point & $K_1$           & $K_2$     & $\Omega$  & Disp. Surface             \\ \midrule
        $P_1$ & $\sqrt{3}/2$    & $1/2$     & 0.230     & $1^\textrm{st}$           \\
        $P_2$ & $\sqrt{3}/2$    & $1/2$     & 0.325     & $2^\textrm{nd}$           \\
        $P_3$ & $\pi$           & $1.555$   & 0.812     & $1^\textrm{st}$ -- $2^\textrm{nd}$\\
        $P_4$ & $\pi$           & $\pi$     & $\Omega_f\approx0.8467$     & $1^\textrm{st}$           \\
        $P_5$ & $\pi$           & $\pi$     & 1.193     & $2^\textrm{nd}$ -- $3^\textrm{rd}$ \\
        $P_6$ & $\pi$           & $\pi$     & $\Omega_a\approx1.592$     & $4^\textrm{th}$ -- $5^\textrm{th}$ \\
        $P_7$ & $\pi$           & $\pi$     & 1.853     & $6^\textrm{th}$ \\
        \bottomrule
    \end{tabular}
    \caption{\label{tab:points_modes}
    Location (in the $\{K_1,K_2,\Omega\}$-space) of the points on the dispersion surfaces (for a Rayleigh beam lattice with slenderness $\lambda=5$) at which the corresponding waveforms have been computed and reported in Figs.~\ref{fig:mode_1_2_5}--\ref{fig:mode_6_8_5}. 
    The points are also marked in Figs.~\ref{fig:Brillouin_5} and~\ref{fig:surf_contour_5_Ray}. 
    Note that points corresponding to double roots connect two surfaces.}
\end{table}

The first property that clearly emerges from the shape of the contours is the cubic symmetry in the quasi-static (low-frequency) response, inherited by the symmetry of the square grid itself.
In particular, considering the two lowest surfaces (Figs.~\ref{fig:surf_contour_1_5_Ray} and~\ref{fig:surf_contour_2_5_Ray} or~\ref{fig:surf_contour_1_10_Ray} and~\ref{fig:surf_contour_2_10_Ray}) in the neighbourhood of $\{K_1,K_2\}=\{0,0\}$, the contours perfectly match the linear dispersion of the acoustic branches of a classical Cauchy continuum endowed with a cubic material symmetry. It is in fact recalled that the effective elastic parameters (Young modulus $E^*$, Poisson's ratio $\nu^*$, shear modulus $G^*$, and mass density $\rho^*$) of a 2D continuum equivalent to a square beam grid are \citep{gibson_1988}
\begin{equation*}
    E^* = {E A}/(2 l_1), \qquad \nu^*=0, \qquad G^* = 6 {E I}/(2 l_1)^3, \qquad \rho^* = 2 {\rho A}/(2l_1), 
\end{equation*}
so that the velocities of the pressure and shear waves propagating in the effective continuum are
\begin{equation*}
    v_{p,0^\circ} = \sqrt{\frac{E^*}{\rho^*}}, \qquad v_{s,0^\circ} = \sqrt{\frac{G^*}{\rho^*}}, 
\end{equation*}
in the direction parallel to the principal axes and
\begin{equation*}
    v_{p,45^\circ} = \sqrt{\frac{(1+\nu^*)E^*+2 G^*}{2 \rho^*}}, \qquad v_{s,45^\circ} = \sqrt{\frac{(1-\nu^*)E^*}{2 \rho^*}}
\end{equation*}
in the direction inclined at $45^\circ$ with respect to the principal axes (which, for a cubic material, are the only directions corresponding to de-coupling of pressure and shear waves).
%

On the first two low-frequency branches (where the wavelength of the propagating waves is much larger than the size of the lattice unit cell), the structured medium is expected to exhibit a continuum-like dynamic response, displaying cubic anisotropy.
This behaviour is clearly demonstrated by the associated waveforms shown in Fig.~\ref{fig:mode_1_2_5}, where it can be noticed from the insets showing the unit cell that the modulation of the Floquet-Bloch wave is essentially uniform.
Moreover, the comparison between Fig.~\ref{fig:mode_1_5}~and~\ref{fig:mode_2_5} shows that the anisotropy induces a sort of `mixing' of the `shear' and `pressure' waves as the amplitudes are neither parallel nor orthogonal to the wave vector, in agreement with the cubic symmetry.

As we consider higher frequencies, the dispersion becomes nonlinear and the geometry of the slowness contours changes dramatically.
Non-convex slowness contours are evident in the proximity of the top of the first dispersion surface, displaying two orthogonal preferential directions inclined at $45^\circ$ with respect to the orientation of the beams (see Fig.~\ref{fig:surf_contour_1_5_Ray}).
This non-convex pattern occurs again on the third surface, but with different preferential directions, which are now aligned parallel to the beams of the lattice (see Fig.~\ref{fig:surf_contour_3_5_Ray}).
%

The peculiarities of the high-frequency dynamics can be associated not only to the features of the slowness contours, but also to the actual waveforms corresponding to double roots, Dirac cones, and standing waves.
In this regard, the Dirac vertex is considered, which is present at the point $P_3\!:\,\{K_1,K_2,\Omega\}=\{\pi,1.555,0.812\}$ connecting the two lowest dispersion surfaces.
The two waveforms related to this double root are depicted in Fig.~\ref{fig:mode_10_11_5}, where it can be seen that both waves \textit{propagate vertically, but are standing horizontally}, so that they highlight the difference between the phase velocity and the group velocity of Floquet-Bloch waves.
Specifically, a better understanding of these waves can be obtained by considering the motion of horizontal and vertical beams separately:
the latter beams (Fig.~\ref{fig:mode_10_5}) are subject to purely flexural vibrations, so that the junctions do not displace vertically, while in the waveform shown in Fig.~\ref{fig:mode_11_5} these beams undergo a purely axial motion. 
On the other hand, the dynamics of horizontal beams is characterized by nodal points (where displacement remains constantly null), which in the waveform reported in Fig.~\ref{fig:mode_11_5} are located at the midpoint of the beams for both the axial and flexural waves, while in Fig.~\ref{fig:mode_10_5} the nodes of the transverse and the axial displacement are located at the junctions and at the midpoints, respectively. 
Due to the cubic symmetry, three Dirac points analogous to that considered above are present for the same frequency at the Bloch vectors $\{K_1,K_2\}=\{\pi,\pi\pm(\pi-1.555)\}$ and $\{K_1,K_2\}=\{\pi\pm(\pi-1.555),\pi\}$ (see Figs.~\ref{fig:surf_contour_1_5_Ray}~and~\ref{fig:surf_contour_2_5_Ray}).

It is important to observe that the Dirac points are very different from those occurring for out-of-plane vibrations, where triple roots are found whose waveforms are purely standing waves~\citep{Piccolroaz_2017}.
In contrast, the in-plane vibrations associated to the Dirac vertex exhibit what can be called \lq \textit{unidirectional propagation}', as the waves propagate along one direction but are standing along the other.
%

%

\begin{figure}[htb!]
\centering
\begin{minipage}[c][]{0.5\textwidth}
    \begin{minipage}[c][]{0.5\textwidth}
	    \centering
	    \begin{subfigure}{\textwidth}
		    \centering
		    \includegraphics[width=\linewidth]{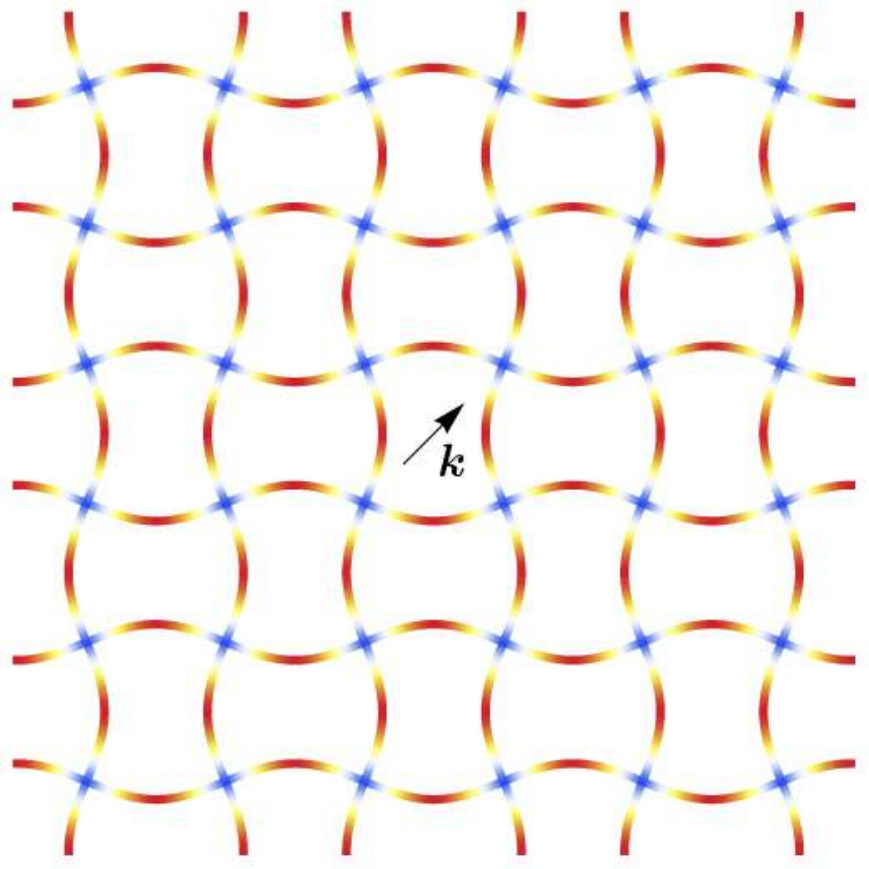}
	    \end{subfigure}
    \end{minipage}%
    \begin{minipage}[c][]{0.5\textwidth}
    \centering
	    \begin{subfigure}{0.5\textwidth}
	        \centering
	        \includegraphics[width=\linewidth]{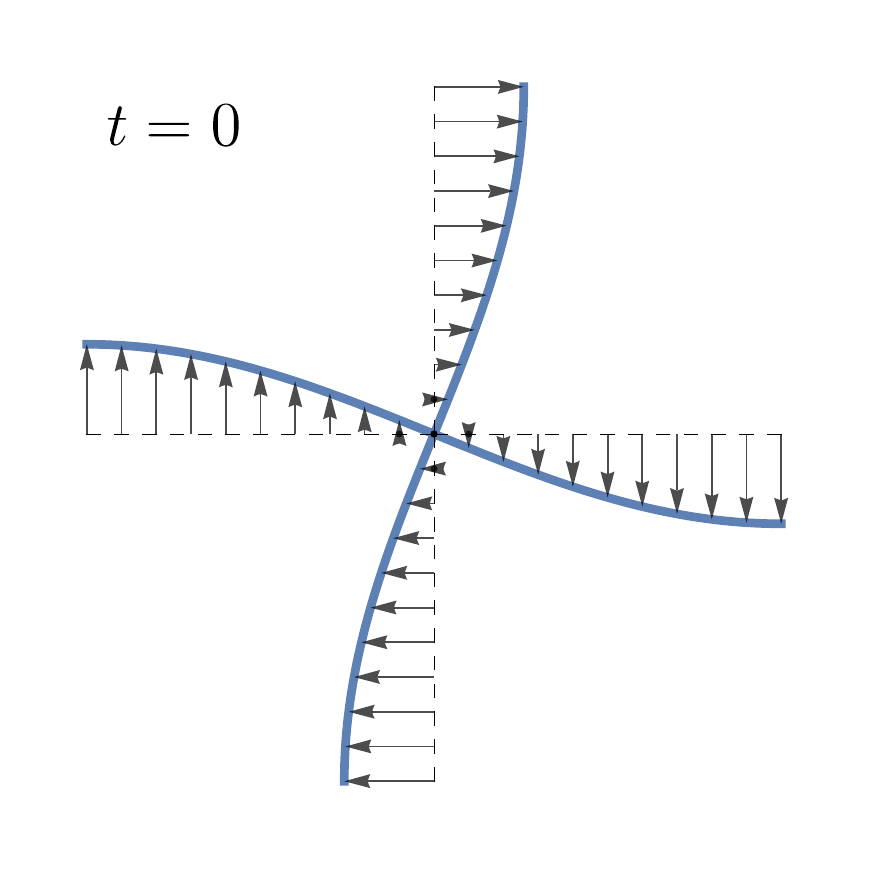}
	    \end{subfigure}%
	    \begin{subfigure}{0.5\textwidth}
	        \centering
	        \includegraphics[width=\linewidth]{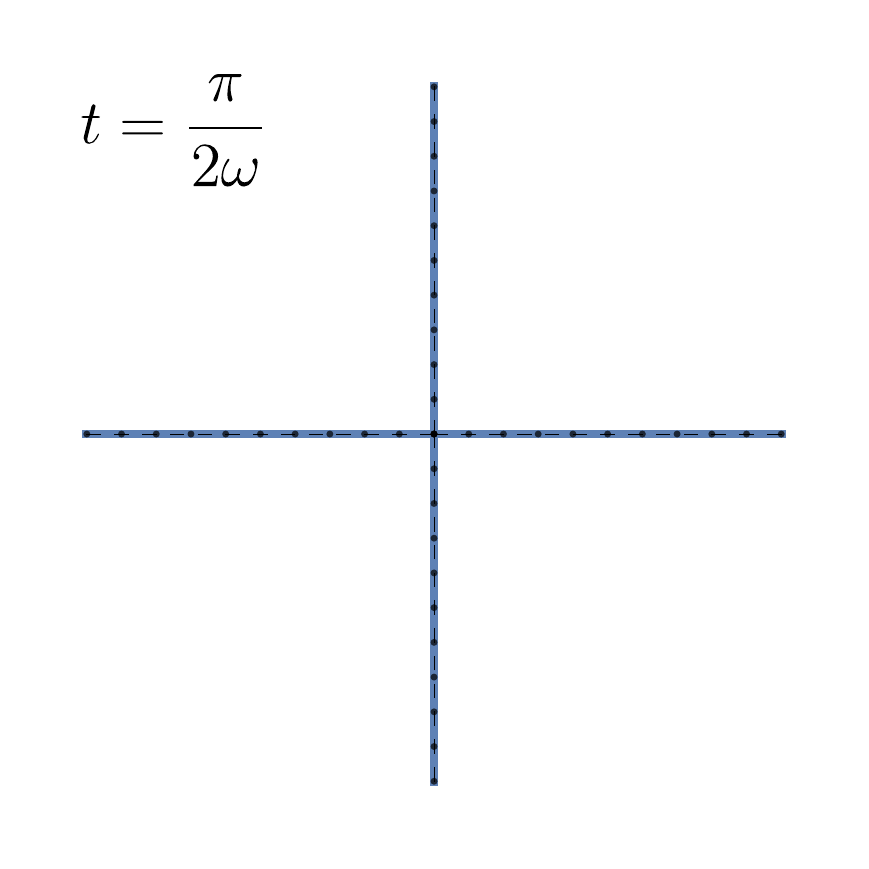}
	    \end{subfigure}
        \begin{subfigure}{0.5\textwidth}
	        \centering
	        \includegraphics[width=\linewidth]{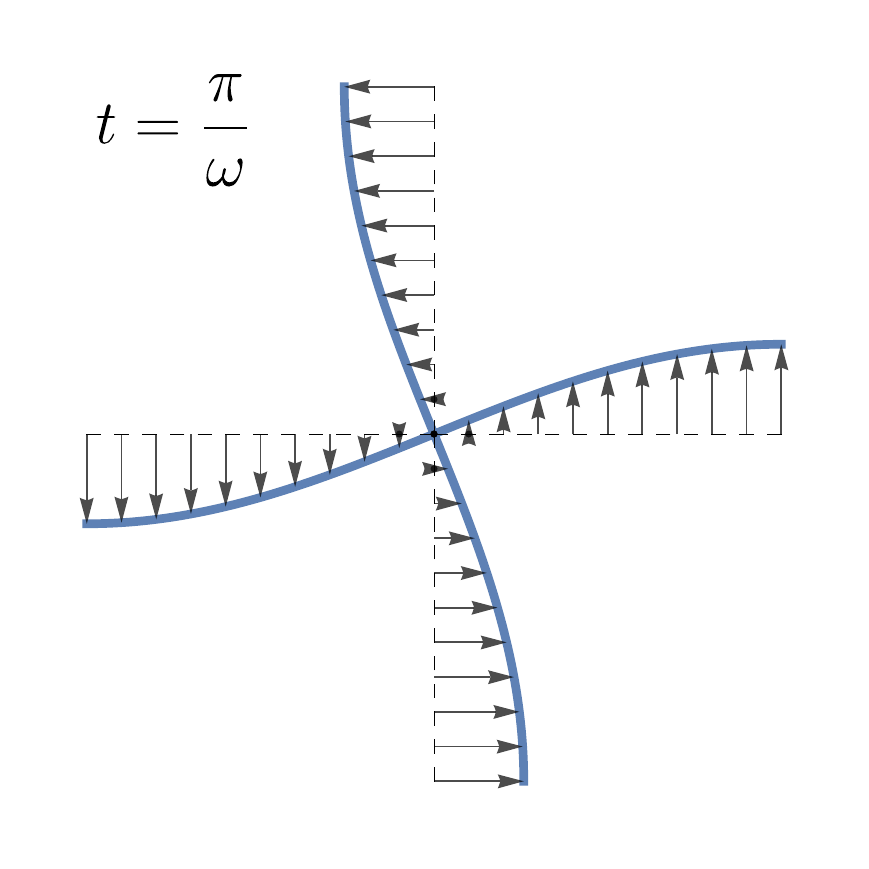}
	    \end{subfigure}%
	    \begin{subfigure}{0.5\textwidth}
	        \centering
	        \includegraphics[width=\linewidth]{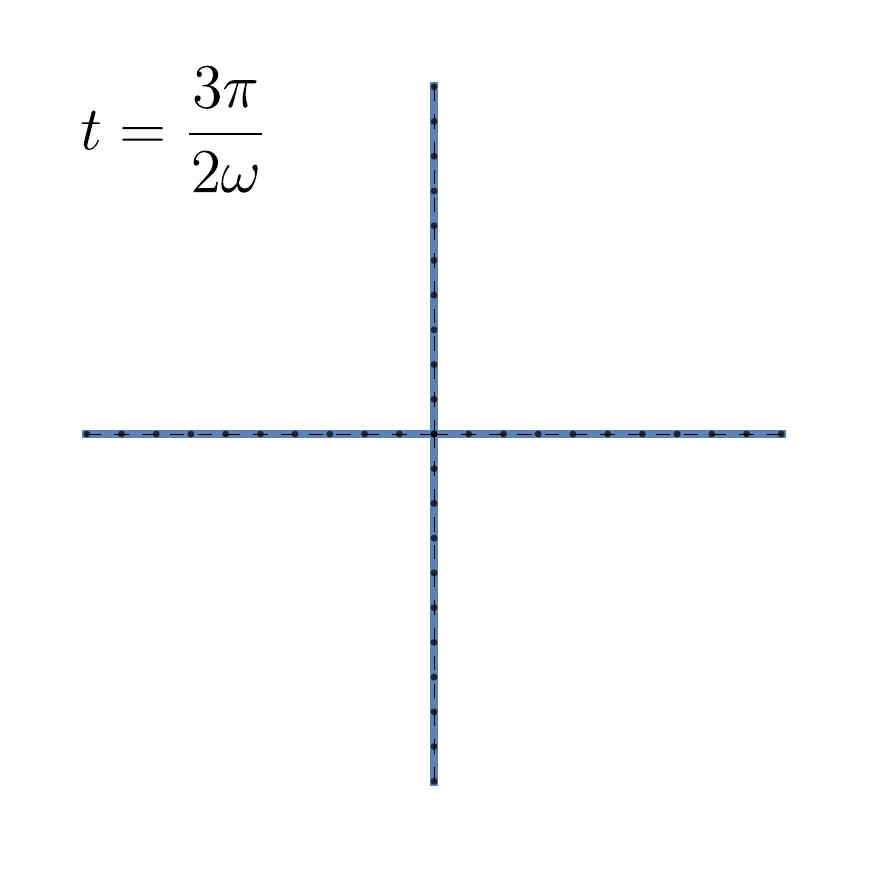}
	    \end{subfigure}
    \end{minipage}
	\subcaption{\label{fig:mode_5_5}Waveform at $P_4\!:\!\{K_1,K_2,\Omega\}\!=\!\{\pi,\pi,0.847\}$.}
\end{minipage}%
\begin{minipage}[c][]{0.5\textwidth}
    \begin{minipage}[c][]{0.5\textwidth}
	    \centering
	    \begin{subfigure}{\textwidth}
		    \centering
		    \includegraphics[width=\linewidth]{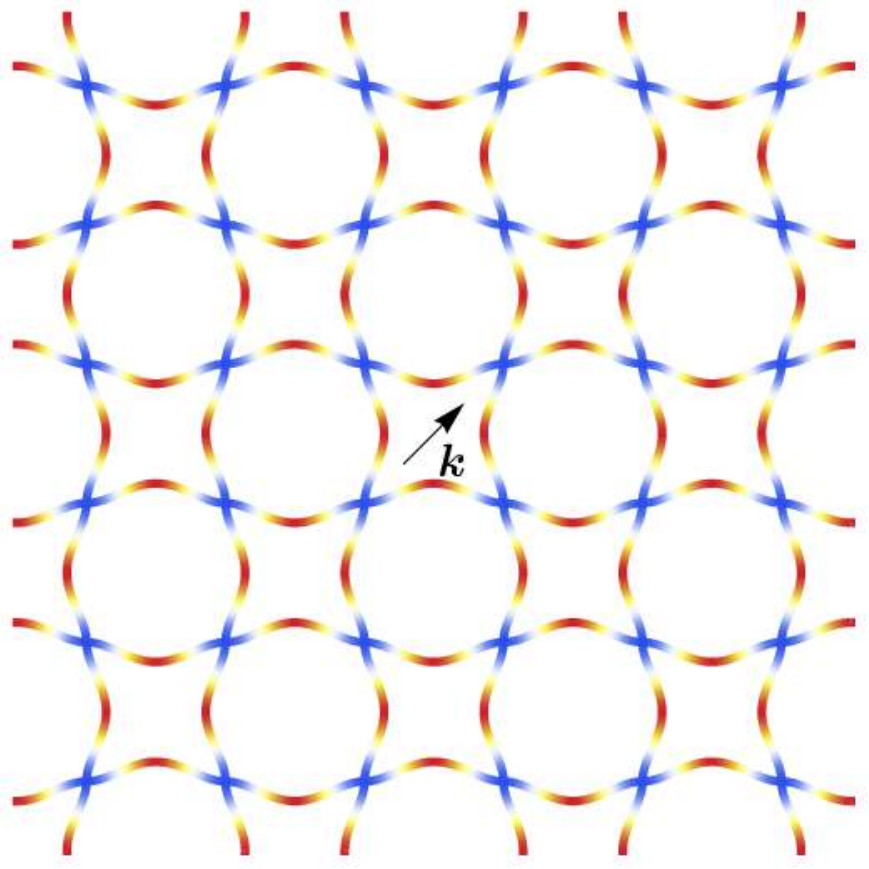}
	    \end{subfigure}
    \end{minipage}%
    \begin{minipage}[c][]{0.5\textwidth}
    \centering
	    \begin{subfigure}{0.5\textwidth}
	        \centering
	        \includegraphics[width=\linewidth]{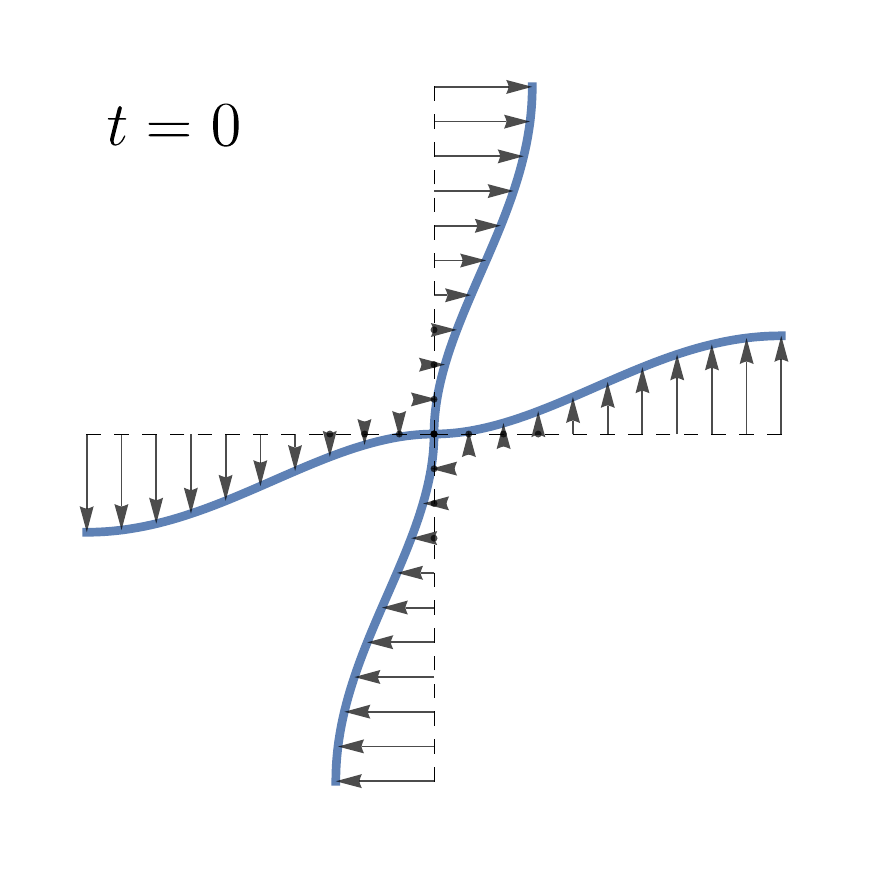}
	    \end{subfigure}%
	    \begin{subfigure}{0.5\textwidth}
	        \centering
	        \includegraphics[width=\linewidth]{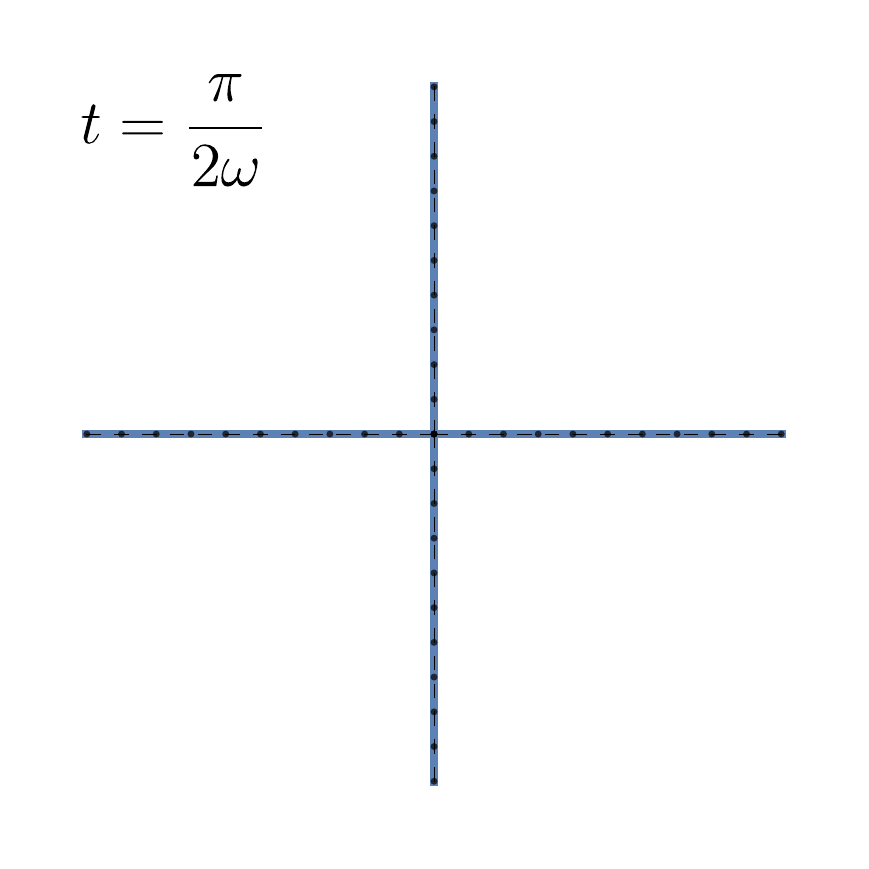}
	    \end{subfigure}
        \begin{subfigure}{0.5\textwidth}
	        \centering
	        \includegraphics[width=\linewidth]{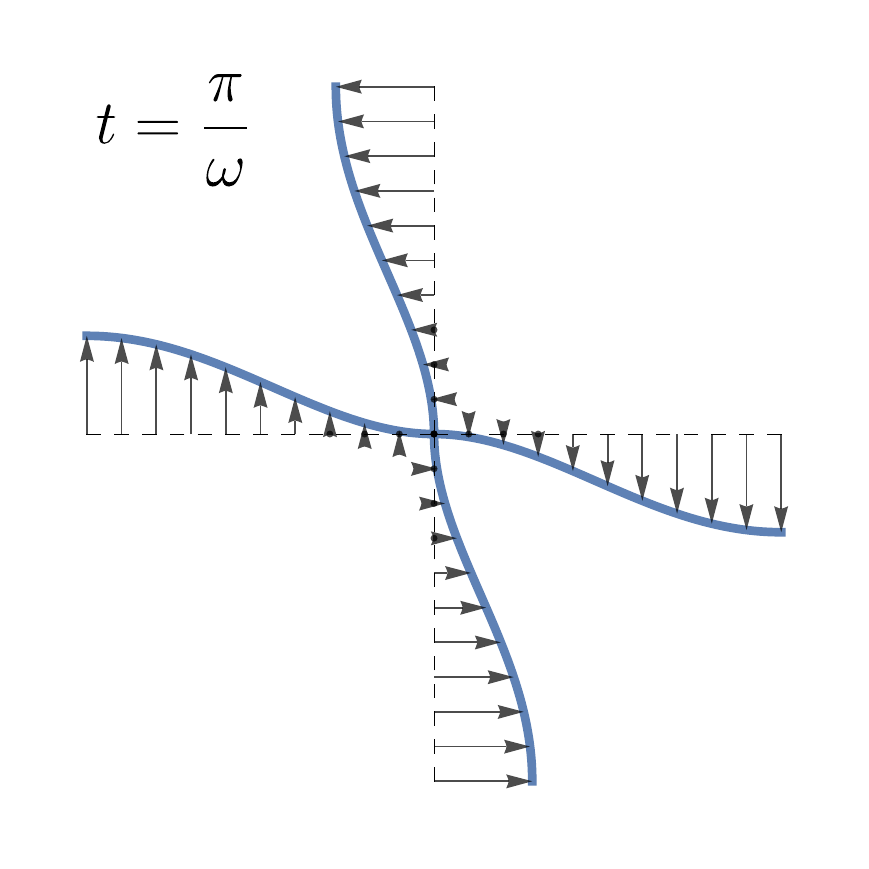}
	    \end{subfigure}%
	    \begin{subfigure}{0.5\textwidth}
	        \centering
	        \includegraphics[width=\linewidth]{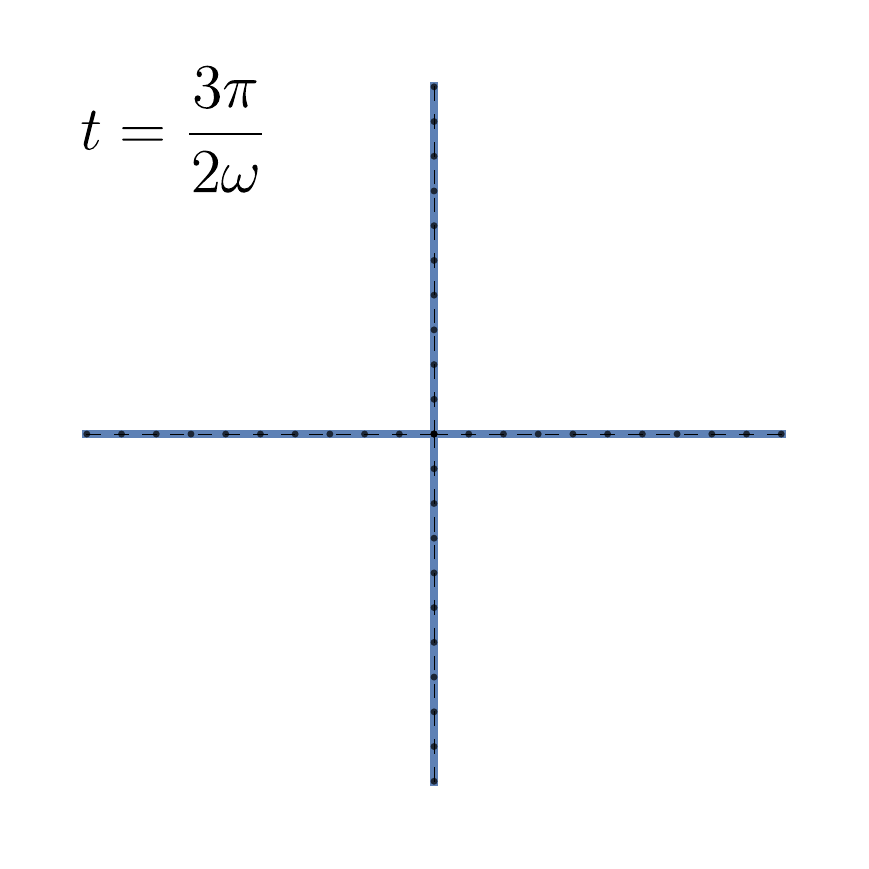}
	    \end{subfigure}
    \end{minipage}
	\subcaption{\label{fig:mode_12_5}Waveform at $P_7\!:\!\{K_1,K_2,\Omega\}\!=\!\{\pi,\pi,1.853\}$.}
\end{minipage}
\caption{\label{fig:mode_5_12_5}
Purely flexural standing waves occurring in a square grid of Rayleigh beams on a stationary point of the first~(\subref{fig:mode_5_5}) and the sixth~(\subref{fig:mode_12_5}) dispersion surface, at a frequency corresponding to the first flexural mode of a simply supported~(\subref{fig:mode_5_5}) and a double-clamped~(\subref{fig:mode_12_5}) Rayleigh beam.
Nodal points are located at the junctions for both cases, but these joints may have a rotational motion~(\subref{fig:mode_5_5}), as opposed to the situation where they are totally fixed~(\subref{fig:mode_12_5}).}
\end{figure}

At the points $P_5$ and $P_6$, two double roots are found, where the dispersion surfaces do not present a cone-like geometry, rather they seem to exhibit stationary points which would imply the presence of standing waves.
This is confirmed by the corresponding pairs of waveforms computed at these points and  in fact revealing sets of nodal points distributed along the two orthogonal directions (Fig.~\ref{fig:mode_6_8_5}).
In particular, the two eigenmodes corresponding to $P_5$ (Fig.~\ref{fig:mode_6_5}) are characterized by a peculiar combination of a purely flexural motion along one set of beams and a purely extensional deformation along the orthogonal set, with nodal points  located at the beams' midpoints.
On the other hand, the waveforms corresponding to $P_6$ (Fig.~\ref{fig:mode_8_5}) involve only purely axial standing waves along one direction and nodal lines along the other, so that the junctions remain fixed and the lattice vibrates with a frequency $\Omega=\Omega_a=\lambda/\pi$, corresponding to the first axial mode of a double-clamped beam.

Purely flexural standing waves are found at the points $P_4$ and $P_7$ where, respectively, the first and sixth branches of the dispersion relation become stationary.
The corresponding waveforms, represented in Fig.~\ref{fig:mode_5_12_5}, show nodal points at the junctions, so that each beam oscillates according to the first flexural vibration mode of a double-pinned beam in Fig.~\ref{fig:mode_5_5}, or of a double-clamped beam in Fig.~\ref{fig:mode_12_5}.

The above-reported investigation will be useful in the next section for the interpretation and prediction of the lattice dynamics induced by a time-harmonic point load.

\section{Forced vibration of a grid of Rayleigh beams}
\label{sec:forced_vibration}

The relation between the dynamic response of a grid of Rayleigh beams and the Floquet-Bloch analysis performed in the previous section can be investigated through the analysis of the vibrations induced by a time-harmonic source (a concentrated force or moment) in a lattice of infinite extent. 
To this purpose, a square grid of Rayleigh beams is numerically solved using the Comsol Multiphysics$^{\circledR}$ f.e.m. program in the frequency response mode. 
A square finite-size computational window with $(N-1) {\times} (N-1)$ unit cells is considered, where $N=161$ is the number of nodes in each direction, with a perfectly matched layer (PML) along the boundaries, to simulate an infinite lattice. By tuning the damping in the boundary layers, the outgoing waves can be completely absorbed, so that reflection is not generated in the interior domain.The physical parameters for the numerical computations  are chosen to be identical to those used in the previous Section~\ref{sec:dispersion_properties}. 

Since the in-plane problem is vectorial, different types of loading are considered, namely a concentrated in-plane (the vector defining the moment is orthogonal to the plane of the grid) moment and a concentrated in-plane force, applied to the central junction.
For a given loading and a given dimensionless angular frequency $\Omega$, the complex displacement field, with components $u=u_R+iu_I$ and $v=v_R+iv_I$, is computed. 
The results are plotted in terms of the total displacement associated to the real parts, $\delta_{R}(x,y,\Omega)=\sqrt{u_R^2+v_R^2}$.
For the sake of brevity, the total displacement associated to the imaginary parts $\delta_{I}(x,y,\Omega)=\sqrt{u_I^2+v_I^2}$ is omitted.

The numerical simulations are complemented with a Fourier analysis of the nodal displacements, with the purpose of providing a clear connection between the forced response of the Rayleigh beam lattice and the Floquet-Bloch analysis performed in the previous sections.

For a given dimensionless angular frequency $\Omega$, the two-dimensional fast Fourier transform is applied to the nodal displacement field, $u_{pq}=u(x_p,y_q)$ and $v_{pq}=v(x_p,y_q)$, where $(x_p,y_q)$ are the coordinates of the $(pq)$-node in the grid. This gives the transformed fields $U_{rs}=\mF[u_{pq}]$ and $V_{rs}=\mF[v_{pq}]$, where the transform is defined as follows
\begin{equation}
\begin{aligned}
X_{rs} = \mF[x_{pq}] &= \frac{1}{N^2} \sum_{p=1}^{N} \sum_{q=1}^{N} x_{pq} e^{-\frac{2\pi i}{N} (p-1) (r-1)} e^{-\frac{2\pi i}{N} (q-1) (s-1)} \\
&= \frac{1}{N^2} \sum_{p=1}^{N} \sum_{q=1}^{N} x_{pq} e^{- i (p-1) K_1} e^{- i (q-1) K_2} = X(K_1,K_2), \quad \forall \, r,s \in\{1,...,N\}
\end{aligned}
\end{equation}
in which $K_1 = \frac{2\pi}{N}(r-1)$ and $K_2 = \frac{2\pi}{N}(s-1)$ are the components of the dimensionless wave vector appearing in Eq.~\eqref{eq:bvp_unit_cell_bloch}.

The fast Fourier transform provides the spectrum of Bloch plane waves composing the forced dynamic response of the beam grid. Specifically, $|U(K_1,K_2)|$ and $|V(K_1,K_2)|$ are the amplitudes of a plane wave with wave vector $\{K_1,K_2\}$, such that the physical displacement field can be represented as the superposition of all the $N^2$ plane waves of the spectrum.
For each numerical simulation performed at a given frequency, the density plot of the quantity $\sqrt{|U|^2+|V|^2}$ is reported superimposed to the slowness contour computed at the same frequency with the Floquet-Bloch technique.

\subsection{Concentrated time-harmonic moment: wave localization and isotropization} 
\lb{avanti}

\begin{figure}[htb!]
\centering
\includegraphics[width=\textwidth]{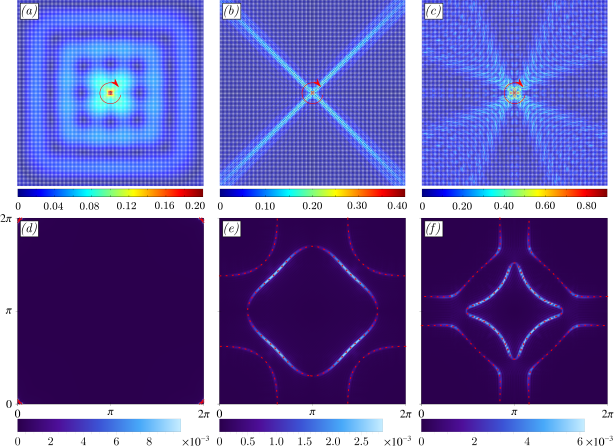}
\caption{\label{fig:LowRegime}
Total displacement field (upper part, a-b-c) and corresponding Fourier transform (lower part, d-e-f) during vibrations of a Rayleigh grid of beams excited by a time-harmonic concentrated moment (applied in the plane in a low-frequency interval, $0<\Omega<\Omega_f=0.8467$). 
The slowness contour evaluated from the Floquet-Bloch analysis is superimposed in red spots. 
\\
(a) and (d), $\Omega=0.025$, at low frequency the wave pattern is typical of a continuous material with cubic anisotropy (note that the slowness contour and the Bloch spectrum are confined at the corners of the figure(d)). 
\\
(b) and (e), $\Omega=0.65$, a strong vibration localization along directions inclined at $\pm 45^\circ$. 
\\
(c) and (f), $\Omega=0.8$ the inner cross-shaped slowness contour is the most excited by the applied load; however, its re-entrant curved edges lead to a fan of preferential directions, developing around the lines inclined at $\pm 45^\circ$.}
\end{figure}

\begin{figure}[htb!]
\centering
\includegraphics[width=\textwidth]{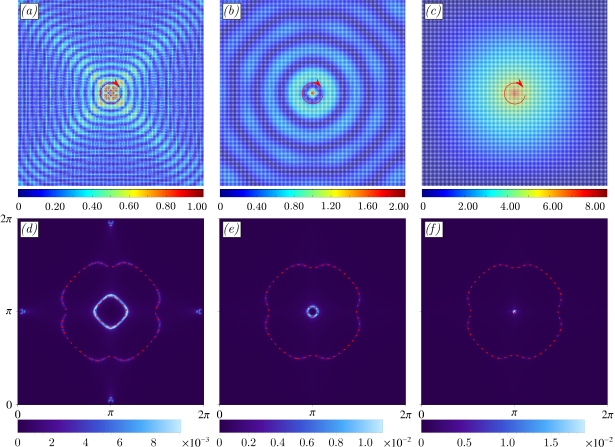}
\caption{\label{fig:Vertex}
Total displacement field (upper part, a-b-c) and corresponding Fourier transform (lower part, d-e-f) during vibrations of a Rayleigh grid of beams excited by a time-harmonic concentrated moment (applied in the plane at frequencies near 
the stationary point $P_4$ in Fig.~\ref{fig:surf_contour_5_Ray}, $\Omega_f = 0.8467$, of the first dispersion surface). 
The slowness contour evaluated  from the Floquet-Bloch analysis is superimposed in red spots.
\\
(a) and (d), $\Omega=0.83628$ , the inner diamond-shaped slowness contour is excited by the applied load, producing waves with squared wavefront; two preferential vibration directions inclined at $\pm 45^\circ$ are still visible. 
\\
(b) and (e), $\Omega=0.8455$, at a frequency very close to the resonant point $P_4$, the inner slowness contour shrinks to a little circle and the waves assume an almost circular wavefront when close to the source, while these assume an octagonal shape far away from the source. 
\\
(c) and (f), $\Omega_f=0.8467$, at the resonant frequency the inner slowness contour shrinks to a point, the corresponding evanescent waveform is typical of a resonant mode.
Note that, as the resonant frequency $\Omega_f$ is approached, the lattice response exhibits a remarkable `isotropization' with wavefronts becoming circular.
}
\end{figure}

\begin{figure}[htb!]
\centering
\includegraphics[width=\textwidth]{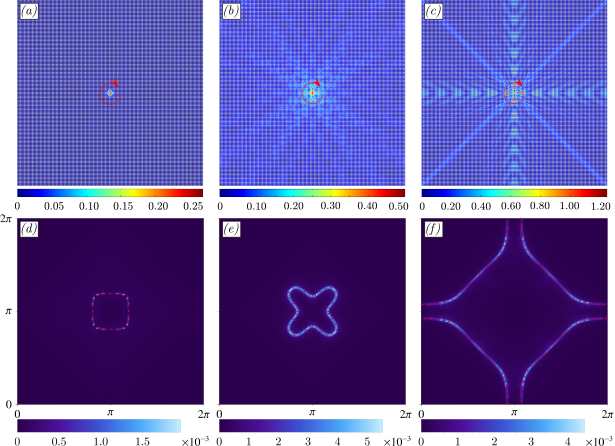}
\caption{\label{fig:MiddleRegime}
Total displacement field (upper part, a-b-c) and corresponding Fourier transform (lower part, d-e-f) during vibrations of a Rayleigh grid of beams excited by a time-harmonic concentrated moment (applied in the plane in 
an intermediate frequency regime $\Omega_f=0.8467<\Omega <\Omega_a=1.5915$, where only one dispersion surface is intersected). 
The slowness contour evaluated from the Floquet-Bloch analysis is superimposed in red spots.
\\
(a) and (d), $\Omega=1.1$, the applied concentrated moment does not produce any visible wave propagation. 
\\
(b) and (e), $\Omega=1.21$, the Bloch waves of an \lq X-shaped' slowness contour are almost uniformly excited, giving rise to several preferential directions inclined around the directions $\pm 45^\circ$. 
\\
(c) and (f), $\Omega=1.3$, the preferential vibration directions are vertical, horizontal and inclined $\pm 45^\circ$.
}
\end{figure}

\begin{figure}[htb!]
\centering
\includegraphics[width=\textwidth]{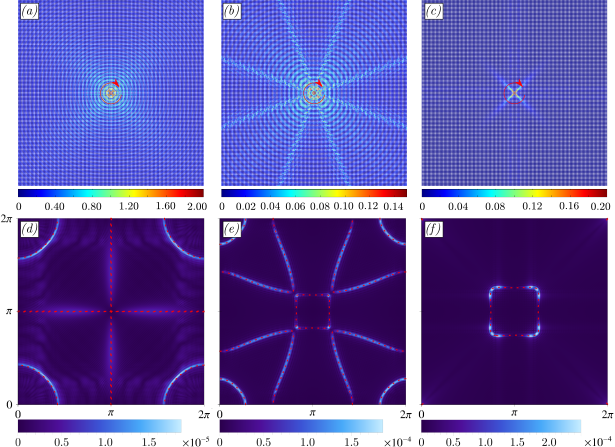}
\caption{\label{fig:HighRegime}
Total displacement field (upper part, a-b-c) and corresponding Fourier transform (lower part, d-e-f) during vibrations of a Rayleigh grid of beams excited by a time-harmonic concentrated moment (applied in the plane in a high-frequency interval, $\Omega \geq \Omega_a=1.5915$). 
The slowness contour evaluated from the Floquet-Bloch analysis is superimposed in red spots.
\\
(a) and (d), $\Omega_a=1.5915$, since the applied concentrated moment cannot excite axial waves, the associated cross-like slowness contour is not highlighted in the Fourier transform, so that almost isotropic waves are generated, which are associated to the rounded slowness contour. 
\\
(b) and (e), $\Omega=1.67$, the slowness contours have a complex geometry, including square and rounded segments, together with eight symmetrically distributed branches; the corresponding waveform shows eight preferential directions with rounded wavefronts.
\\
(c) and (f), $\Omega=1.7326$, a strong vibration localization along directions inclined at $\pm 45^\circ$ is clearly visible, also highlighted by the corresponding Fourier transform.}
\end{figure}

A grid of Rayleigh beams is investigated when forced by a time-harmonic concentrated moment, acting at a node and pulsating at a given dimensionless angular frequency $\Omega$, in the range of frequencies analyzed in Section~\ref{sec:dispersion_properties}.
The results are reported in Figs.~\ref{fig:LowRegime}--\ref{fig:HighRegime} in terms of the total displacement associated to the real parts, $\delta_{R}$.
Each numerical simulation is accompanied by the Fourier transform of the complex displacement field, shown in the lower part of the figure, where a red dotted line indicates the slowness contours obtained with the Floquet-Bloch analysis at the considered frequency $\Omega$ (see Fig.~\ref{fig:surf_contour_5_Ray}). 

Three different frequency intervals are investigated, namely: a low frequency regime, from $\Omega = 0$ up to the vertex $P_4$ of the first dispersion surface, at $\Omega_f = 0.8467$ (results are reported in Figs.~\ref{fig:LowRegime} and~\ref{fig:Vertex}); an intermediate frequency regime, between the points $P_4$ and $P_6$, at $\Omega_a=1.5915$, where the propagation of axial waves prevails (results are reported in Fig.~\ref{fig:MiddleRegime}); and finally a high frequency regime from the point $P_6$ up to higher frequencies (results are reported in Fig.~\ref{fig:HighRegime}). 
The transition between the low and the intermediate frequency regimes deserves a special attention, because here the first dispersion surface shows a stationary point (point $P_4$, being either a maximum or a minimum, depending on the slenderness ratio).
This frequency corresponds to the resonant mode occurring at $\Omega_f = 0.8467$  and represented by the pure flexural standing wave in  Fig.~\ref{fig:mode_5_5}, so that three different frequencies close to this point are investigated (results are reported in Fig.~\ref{fig:Vertex}).
%

The results of the numerical simulations for the low-frequency regime $\Omega \in (0,\Omega_f=0.8467)$ are reported in Fig.~\ref{fig:LowRegime}.
For a given frequency $\Omega$ in this range, two dispersion surfaces are always intersected. 
The Fourier transform of the nodal displacements of the forced lattice, shown in the lower part of the figure, displays the spectrum of Bloch plane waves composing the dynamic response, which nicely corresponds to the slowness contours (red dotted lines) obtained through the Floquet-Bloch analysis in Section~\ref{sec:dispersion_properties}. 
The long-wavelength regime for $\Omega=0.025$ is shown in Fig.~\ref{fig:LowRegime}a, where the wave pattern with square wavefronts is typical of a material with cubic symmetry. 
Increasing the frequency up to $\Omega=0.65$, the dynamic response exhibits a strong localization along two preferential directions at $\pm45^\circ$ with respect to the horizontal axis, Fig.~\ref{fig:LowRegime}b. 
The corresponding Fourier transform, reported in Fig.~\ref{fig:LowRegime}e, clearly highlights the excited Bloch waves, among the ones predicted by the slowness contours at the same frequency (red dotted lines). 
It is evident that the applied pulsating moment excites waves along the two inclined preferential directions, whereas waves with the `isotropic' shape corresponding to the rounded slowness contour are not generated.
Approaching the stationary point of the first dispersion surface (point $P_4$ in Fig.~\ref{fig:surf_contour_5_Ray}), at the frequency $\Omega=0.8$, a less marked but still visible diagonal localization is observed, together with a propagation along the principal axes of the lattice, see Fig.~\ref{fig:LowRegime}c. 
Note that, while the slowness contour is convex in Fig.~\ref{fig:LowRegime} (e), it becomes concave in part (f). 
The re-entrant curved edges lead to a fan of preferential propagation directions around lines inclined at $\pm 45^\circ$. 
The appearance of Bloch waves corresponding to the second slowness contour justifies the weak propagation along the principal axes.
%

In the proximity of the stationary point of the first dispersion surface (occurring at $\Omega_f = 0.8467$), a sudden change in the response of the lattice is observed, so that a narrow range of frequencies is analyzed and reported in Fig.~\ref{fig:Vertex}. 
Part (a) of this figure shows the displacement field for a pulsating moment with frequency $\Omega = 0.83628$, where the applied moment excites mostly Bloch waves corresponding to the inner diamond-shaped slowness contour visible in Fig.~\ref{fig:Vertex}d. Waves with squared wavefront are produced, while the two preferential directions inclined at $\pm 45^\circ$ still remain visible.
Immediately below the stationary point of the first dispersion surface, at $\Omega=0.8455$, the wave pattern becomes similar to the response of an isotropic material. Indeed, at this frequency, the inner slowness contour shrinks and becomes almost circular,  Fig.~\ref{fig:Vertex}e; correspondingly, the waves produced by the applied moment show an almost circular wavefront, when they are close to the source, while they assume an {\it octagonal shape}, when far away from the source and present a increased wavelength, when compared to the lower frequencies.
At the stationary point of the first dispersion surface, corresponding to the frequency $\Omega_f = 0.8467$, the total displacement field $\delta_R$, reported in Fig.~\ref{fig:Vertex}c, shows an evanescent wave pattern, typical of a resonant mode, so that the inner slowness contour reduces to a point,  Fig.~\ref{fig:Vertex}f. The Bloch eigenmode corresponding to this point is identified through the Floquet-Bloch analysis (Section~\ref{sec:dispersion_properties}) as a purely flexural standing wave, in which the junctions of the grid exhibit a pure rotational motion, Fig.~\ref{fig:mode_5_5}, which explains the observed resonant wave pattern.
%

Fig.~\ref{fig:MiddleRegime} shows the dynamic response of the lattice in the intermediate frequency regime, between the stationary point of the first dispersion curve, $\Omega_f = 0.8467$, and the fourth dispersion surface, $\Omega_a = 1.5915$. 
For $\Omega=1.10$ the slowness contour intersects the second dispersion surface. The total displacement field $\delta_R$, reported in Fig.~\ref{fig:MiddleRegime}a, shows an evanescent waveform prevailing at this frequency. The corresponding Fourier transform, Fig.~\ref{fig:MiddleRegime}d, confirms that the applied moment excites only weakly Bloch waves. 
At the lower part of the third dispersion surface, for $\Omega=1.21$, the displacement field shows a waveform with several preferential directions inclined at $\pm 45^\circ$, Fig.~\ref{fig:MiddleRegime}b. This pattern is in agreement with the corresponding \lq X-shaped' slowness contour shown in Fig.~\ref{fig:MiddleRegime}e.
At $\Omega=1.3$, Fig.~\ref{fig:MiddleRegime}c, localization is observed along preferential directions inclined at $\pm 45^\circ$, together with a characteristic `herringbone' pattern along the principal axes of the lattice ($0^\circ/90^\circ$), in agreement with the Bloch waves excited at this frequency,  Fig.~\ref{fig:MiddleRegime}f.
%

The dynamic response of the lattice in the high frequency regime, is reported in Fig.~\ref{fig:HighRegime}, 
starting from the troughs of the fourth dispersion surface at the frequency $\Omega_a = 1.5915$. 
In this regime the Floquet-Bloch analysis predicts the propagation of axial waves along the ligaments of the lattice , Fig.~\ref{fig:mode_8_5}. 
At the frequency $\Omega_a = 1.5915$, corresponding to the troughs of the fourth dispersion surface, the dynamic response of the lattice shows almost circular wavefronts with only a weak preferential direction of propagation inclined at $\pm45^\circ$, Fig.~\ref{fig:HighRegime}a. The corresponding Fourier transform,  Fig.~\ref{fig:HighRegime}d, highlights that the excited Bloch waves correspond to points of the third dispersion surface, having almost circular slowness contours. This wave pattern can be deduced from the Floquet-Bloch analysis because the vibration eigenmodes pertaining to the troughs (associated with the cross-like slowness contour) consist of purely extensional standing waves, Fig.~\ref{fig:mode_8_5}, which cannot be excited by a time-harmonic moment, so that an almost isotropic wave propagation prevails, associated with the rounded slowness contours.
At $\Omega=1.67$ the wave pattern reported in Fig.~\ref{fig:HighRegime}b shows four fans (spanning an angle of 45$^\circ$) of preferential directions with rounded wavefronts. This waveform is the result of the complex geometry of the slowness contours, as illustrated in Fig.~\ref{fig:HighRegime}e, which includes square and rounded contours together with eight symmetrically distributed branches.
Finally, at the frequency $\Omega = 1.7326$, an unexpected strong localization is observed, along directions inclined at $\pm 45^\circ$, Fig.~\ref{fig:HighRegime}c. Here the slowness contours would predict preferential directions along the principal axes of the lattice ($0^\circ/90^\circ$), but the Fourier transform reported in Fig.~\ref{fig:HighRegime}f shows that the excited Bloch waves correspond to the corners of the squared slowness contour, which explains the observed preferential vibration directions.

\subsection{Concentrated time-harmonic force: vibration channelling and localization}
\lb{indietro}

\begin{figure}[htb!]
\centering
\includegraphics[width=\textwidth]{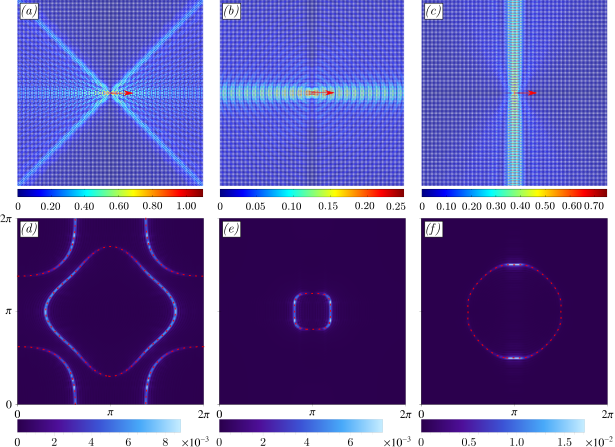}
\caption{\label{fig:forza_LF}
Total displacement field (upper part, a-b-c) and corresponding Fourier transform (lower part, d-e-f) during vibrations of a Rayleigh grid of beams excited by a horizontal time-harmonic concentrated force (applied in the plane in the frequency interval $0<\Omega <\Omega_a=1.5915$). 
The slowness contour evaluated from the Floquet-Bloch analysis is superimposed in red spots. 
\\
(a) and (d), $\Omega = 0.65$, an \lq X-shaped' localization of vibrations around two preferential directions inclined at $\pm 45^\circ$; waves propagate within the left and right sectors, while waves are not visible in the upper and lower sectors. 
\\
(b) and (e), $\Omega = 1.10$, waves possess a rhombus-shaped wavefront and an amplitude localized along the horizontal axis. 
\\
(c) and (f), $\Omega=1.273$, waves are strongly localized along the vertical direction (orthogonal to the force).
}
\end{figure}

\begin{figure}[htb!]
\centering
\includegraphics[width=\textwidth]{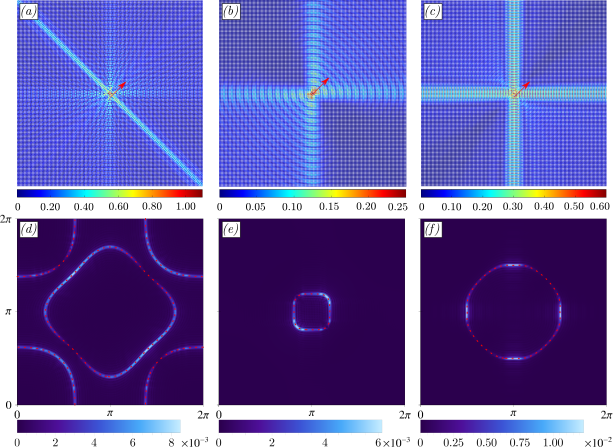}
\caption{\label{fig:forza_LF_45}
Total displacement field (upper part, a-b-c) and corresponding Fourier transform (lower part, d-e-f) during vibrations of a Rayleigh grid of beams excited by a time-harmonic concentrated force (inclined at 45$^\circ$ and applied in the plane in the frequency interval $0<\Omega<\Omega_a=1.5915$).
The slowness contour evaluated from the Floquet-Bloch analysis is superimposed in red spots. 
\\
(a) and (d), $\Omega = 0.65$, waves are strongly localized along preferential directions inclined at $-45^\circ $ (perpendicular to the applied force); rapidly decaying waves are also visible along vertical and horizontal directions
\\
(b) and (e), $\Omega = 1.10$, waves propagate within the first and third quadrant with vertical and horizontal preferential directions. 
\\
(c) and (f), $\Omega=1.273$, a symmetrical cross-shaped wave localization is visible, where waves propagate within the second and fourth quadrant.
}
\end{figure}

\begin{figure}[htb!]
\centering
\includegraphics[width=\textwidth]{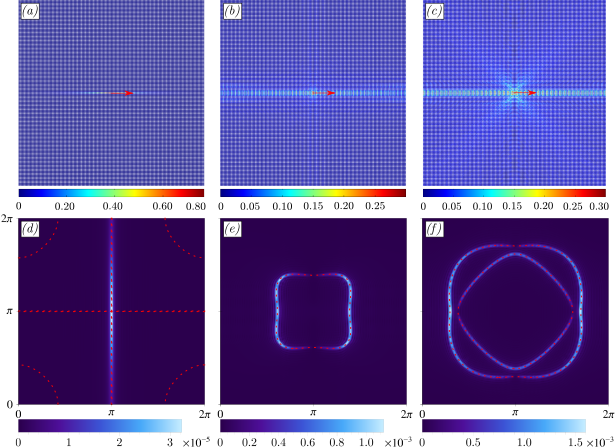}
\caption{\label{fig:forza_HF}
Total displacement field (upper part, a-b-c) and corresponding Fourier transform (lower part, d-e-f) during vibrations of a Rayleigh grid of beams excited by a horizontal time-harmonic concentrated force (applied in the plane in a high-frequency regime, $\Omega \geq \Omega_a=1.5915$). 
The slowness contour evaluated from the Floquet-Bloch analysis is superimposed in red spots. 
\\
(a) and (d), $\Omega_a = 1.5915$, an extremely localized wave pattern is visible, which involves only horizontal beams. 
\\
(b) and (e), $\Omega = 1.85$, waves localized along the horizontal axis are visible. 
\\
(c) and (f), $\Omega = 2.10$, combination of a prevalent horizontal localization associated to the activation of the outer contour and an `X-shaped' wave pattern produced by the Bloch waves belonging to the inner slowness contour.}
\end{figure}

The dynamic response is analyzed of a square grid of Rayleigh beams (with $\lambda =5$) subject to a a time-harmonic in-plane force (with different inclinations: horizontal or at 45$^\circ$) applied to a node. 
Total displacement fields (upper parts, a-b-c) and corresponding Fourier transform (lower parts, d-e-f) 
are reported in Figs.~\ref{fig:forza_LF}--\ref{fig:forza_HF}, together with the slowness contours evaluated from the Floquet-Bloch analysis, superimposed with red spots to facilitate comparisons. 
%

%

For the frequency $\Omega=0.65$, the total displacement field $\delta_R$, reported in Fig.~\ref{fig:forza_LF}a, looks different when compared to the displacement produced by a nodal moment (Fig.~\ref{fig:LowRegime}b). 
In particular, in addition to preferential propagation directions inclined at $\pm 45^\circ$, which produce an \lq X-shaped' vibration localization, other directions of propagation emerge, exhibiting a distinctive `herringbone' wave pattern along the horizontal axis.
A comparison between Figs.~\ref{fig:forza_LF} (a) and (d) and Fig.~\ref{fig:LowRegime} (b) and (e) (all pertaining to the same frequency $\Omega = 0.65$) shows the presence in the case of the concentrated force of Bloch waves corresponding to the second dispersion surface with rounded slowness contour, a circumstance which explains the propagation in directions other than $\pm 45^\circ$.

Increasing the dimensionless angular frequency to $\Omega=1.10$, waves are localized along the horizontal axis, as illustrated in Fig.~\ref{fig:forza_LF}b.
For an higher frequency, $\Omega=1.273$, a behaviour peculiar of the Rayleigh beam lattice is observed, namely, the propagation becomes strongly localized in the direction perpendicular to the direction of the applied force, as clearly shown in Fig.~\ref{fig:forza_LF}c. 
Although the slowness contour at this frequency has an almost circular shape, the Fourier transform, Fig.~\ref{fig:forza_LF}f, highlights that the activated Bloch waves are localized at the ends of the vertical diameter, which explains the observed strong localization.

Effects related to the directionality of the pulsating force can be appreciated through a comparison between Fig.~\ref{fig:forza_LF} and Fig.~\ref{fig:forza_LF_45}, where the frequency-dependent interaction is visible between the vibration patterns produced by the two in-plane components of the pulsating force.
For instance, for $\Omega=0.65$ the wave pattern produced by a point force inclined at $45^\circ$ with respect to the horizontal axis (Fig.~\ref{fig:forza_LF_45}a) is characterized by a strong localization along the preferential direction at $-45^\circ$, whereas the preferential direction at $+45^\circ$, present when the force is horizontal (Fig.~\ref{fig:forza_LF}a), disappears. Rapidly decaying waves are also visible along vertical and horizontal directions.
At the frequency $\Omega=1.10$, the rhombus-shaped wavefronts visible in Fig.~\ref{fig:forza_LF}b are not affected by the inclination of the load, but the combination of the two force components generates a wave pattern characterized by an absence of propagation in the second and fourth quadrant and, at the same time, by an amplification of the response in first and third quadrant (see Fig.~\ref{fig:forza_LF_45}b).
A comparison between Figs.~\ref{fig:forza_LF}c and~\ref{fig:forza_LF_45}c, at $\Omega=1.273$, shows that the total displacement field $\delta_R$ produced by the inclined load displays a prevalent propagation in the second and fourth quadrant, while a negligible response is observed in the first and third quadrant.
%

At the frequency $\Omega = \Omega_a = 1.5915$, corresponding to the troughs of the fourth dispersion surface, the dynamic response of the lattice to a nodal force is drastically different from that generated by a nodal moment (compare Fig.~\ref{fig:HighRegime}a to Fig.~\ref{fig:forza_HF}a). 
While an activation of Bloch waves in the third dispersion surface with rounded slowness contour (Fig.~\ref{fig:HighRegime}a and d) are observed for an applied moment, a nodal force generates axial waves involving only the horizontal and/or (depending on the direction of the force) the vertical beams connected to the junction where the force is applied. This extremely localized wave pattern is linked to the \lq cross-shaped' slowness contour, as shown by the Fourier transform in Fig.~\ref{fig:forza_HF}d.

Finally, the dynamic behaviour of the beam grid, when a time-harmonic nodal force is applied, pulsating at high frequency, $\Omega=1.85$, is reported in Fig.~\ref{fig:forza_HF}b. At this frequency, the slowness contour has a squared shape similar to that of Fig.~\ref{fig:forza_LF}e at $\Omega = 1.10$.
Correspondingly, also the wave pattern is similar, showing an horizontal preferential vibration direction.
When the frequency increases to the value $\Omega=2.10$, Fig.~\ref{fig:forza_HF}c, the displacement becomes strongly localized in the horizontal direction, while rapidly-decaying vibrations emerge with inclination $\pm 45^\circ$.

\subsection{Energy flow}
\label{sec:energy_flow}

The data obtained from the numerical simulations presented in the previous section are now analyzed to investigate the dynamic anisotropy of the beam grid in terms of the \textit{energy flow} through the lattice produced by the pulsating load. This aspect can be of significant interest for the control of wave propagation and energy channelling in metamaterials.

With the purpose of constructing a 2D vector field representation of the energy flow propagating through the beams of the lattice, 
the flow along a single beam is derived. Denoting with $s$ the local coordinate measured along the beam and increasing in the direction of the unit vector $\bt$, the conservation of energy for an arbitrary part of a beam in an integral form writes 
\begin{equation}
\label{eq:energy_conservation_integral}
\frac{d}{dt}\int_{s_1}^{s_2} (\mT(s,t) + \mE(s,t)) \,ds = (\Re{\ba_{\boldsymbol{t}}(s,t)} \scalp \Re{\dot{\bu}(s,t)})\Big\rvert_{s=s_1}^{s=s_2} + h(t), \qquad \forall s_1, s_2,
\end{equation}
where $\mT$ and $\mE$ are, respectively, the kinetic and elastic energy densities (functions of the coordinate $s$ and of the time $t$), while $\ba_{\boldsymbol{t}}$ is the vector collecting the internal forces acting on the cross-section with unit normal $\bt$, $\dot{\bu}$ collects the corresponding velocities and $h$ accounts for energy sources (for instance the power of external loads) and dissipation (for instance viscous damping) present along the interval $(s_1,s_2)$ of the beam.
The complex representation of the displacement field is used, so that the $\Re{}$ operator is needed.

In the absence of energy sources and dissipations, Eq.~\eqref{eq:energy_conservation_integral} expresses the balance between the rate of variation of the energy stored and the power done by the internal forces acting at the ends of any beam interval.
This power is expressed through the scalar product $\Re{\ba_{\boldsymbol{t}}}\scalp\Re{\dot{\bu}}$, regardless of the structural model employed for the beam  and it can be represented in an orthonormal basis $\{\bt,\bn,\be_3\}$ as follows
\begin{equation}
\lb{zorro}
\begin{aligned}
\Re{\ba_{\boldsymbol{t}}}\scalp\Re{\dot{\bu}} &= \Re{(N\,\bt+V\,\bn+M\,\be_3)} \scalp \Re{(\dot{u}_t\,\bt+\dot{u}_n\,\bn+\dot{\varphi}\,\be_3)}, \\
&= \Re{N}\,\Re{\dot{u}_t} + \Re{V}\,\Re{\dot{u}_n} + \Re{M}\,\Re{\dot{\varphi}},
\end{aligned}
\end{equation}
where $N$, $V$ and $M$ are, respectively, the axial force, the shear force and the bending moment, while the axial, transverse and rotational velocities are denoted as $\dot{u}_t$, $\dot{u}_n$ and $\dot{\varphi}$.
As expression (\ref{zorro}) defines the instantaneous energy flux flowing in the $-\bt$ direction, the \textit{instantaneous energy flow} on a single beam is defined as 
\begin{equation}
\label{eq:energy_flow_instantaneous}
\bq(s,t) = - (\Re{N}\,\Re{\dot{u}_t} + \Re{V}\,\Re{\dot{u}_n} + \Re{M}\,\Re{\dot{\varphi}})\, \bt,
\end{equation}
where the dependence on coordinate $s$ and the time $t$ is now highlighted.

For time-harmonic response of the beam lattice, it is convenient to evaluate the time average of the energy flow~\eqref{eq:energy_flow_instantaneous} over one period of oscillation, so that the `effective' energy transmitted is obtained.
A well-known result of complex variable calculus \citep{brillouin_1946} yields
\begin{equation}
\label{eq:energy_flow}
\langle\bq(s,t)\rangle = -\bt\, \frac{\omega}{2\pi} \int_{0}^{2\pi/\omega} \Re{\ba_{\boldsymbol{t}}} \scalp \Re{\dot{\bu}} \,dt = -\frac{1}{2} \Re{(\ba_{\boldsymbol{t}} \scalp \dot{\bu}^*)}\, \bt,
\end{equation}
where the symbol $^*$ denotes the complex conjugate and $\langle\,\rangle$ the time average operator.
Furthermore, it is worth noting that, for time-harmonic motion, the time average of the energy flow is also constant in $s$ when applied loads and dissipation are absent, $h=0$, a property which can be easily obtained by localizing Eq.~\eqref{eq:energy_conservation_integral}
\begin{equation}
\label{eq:energy_conservation_local}
\deriv{}{t}(\mT(s,t) + \mE(s,t)) = -\deriv{}{s}(\bq(s,t) \scalp \bt),
\end{equation}
and then averaging on time both sides to obtain 
\begin{equation*}
\left\langle\deriv{}{t}(\mT + \mE)\right\rangle = \frac{\omega}{2\pi} (\mT + \mE)\Big\rvert_{t=0}^{t=2\pi/\omega} = 0 ,
\end{equation*}
(where the left-hand side vanishes due to the time-harmonic assumption), so that
\beq
\deriv{}{s}(\langle\bq(s,t)\rangle \scalp \bt) = 0,
\eeq
which proves the average energy flow to be independent of $s$ and therefore to coincide with its mean value $\langle\bq\rangle$.
%

\begin{figure}[htb!]
\centering
\includegraphics[width=\textwidth]{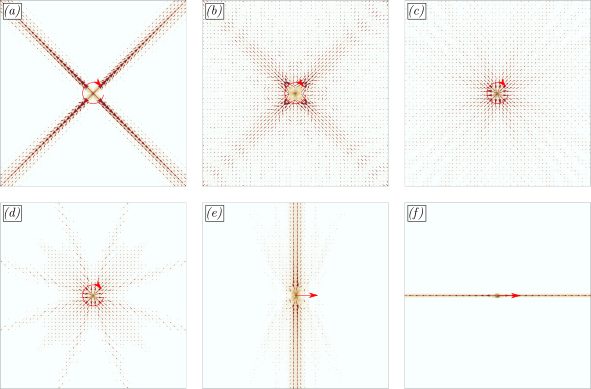}
\caption{\label{fig:energy_flow}
Different forms of vibration localization visible on the vectorial representation (the figure with the associated displacement field is reported in parenthesis) of the energy flow produced by a pulsating concentrated moment at the frequencies: 
(a) $\Omega=0.65$ (Fig.~\ref{fig:LowRegime}b);
(b) $\Omega=0.8363$ (Fig.~\ref{fig:Vertex}a);
(c) $\Omega=1.5915$, (Fig.~\ref{fig:HighRegime}a);
(d) $\Omega=1.67$, (Fig.~\ref{fig:HighRegime}b); by an horizontal force at 
(e) $\Omega=1.273$  (Fig.~\ref{fig:forza_LF}c); and by a horizontal force at 
(f) $\Omega=1.5915$ (Fig.~\ref{fig:forza_HF}a).
} 
\end{figure}
%

\begin{figure}[htb!]
\centering
\includegraphics[width=0.5\textwidth]{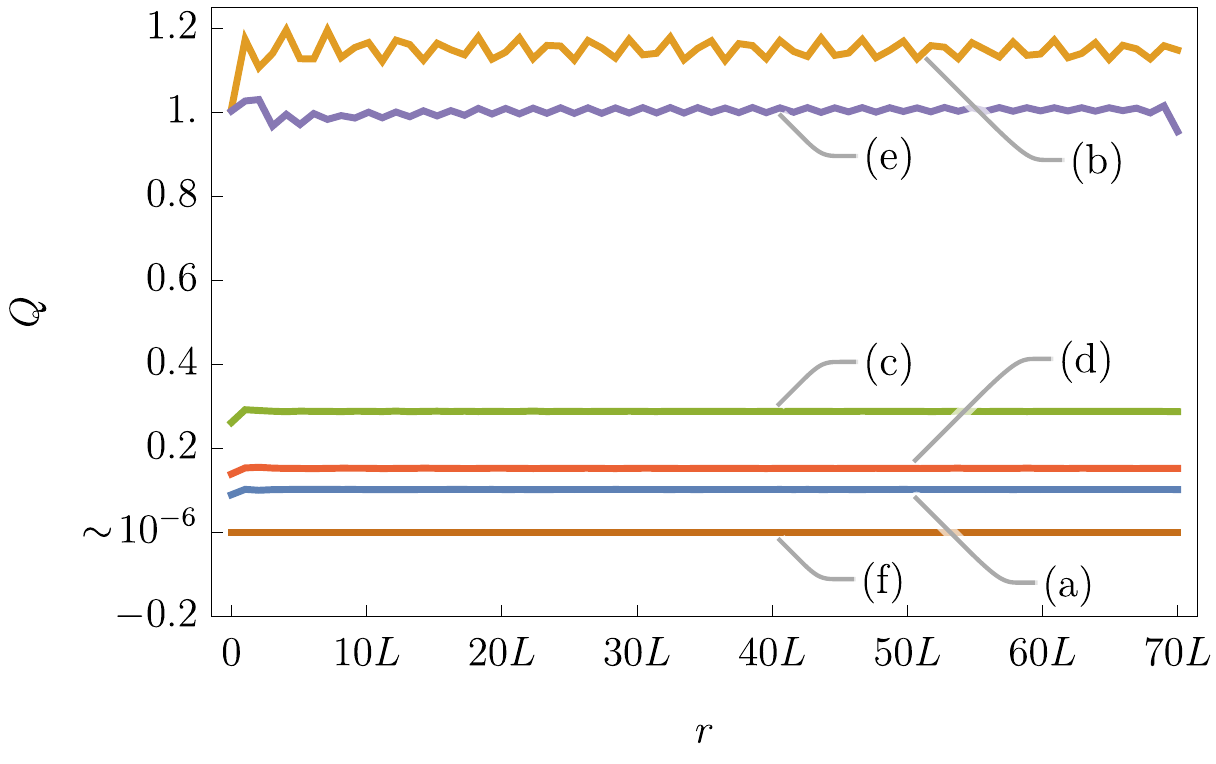}
\caption{\label{fig:FlussiQ}
Outgoing energy flux $Q$ across a circular path centred at the loading point. This flux should be independent of the radius $r$, so that the weak oscillations shown in the graph represent a verification of the accuracy of the numerical solutions obtained in Section.~\ref{avanti} and~\ref{indietro}. 
Labels (a)--(f) refer to the corresponding energy flow vectorial plots reported in Fig~\ref{fig:energy_flow}.}
\end{figure}

Eq.~\eqref{eq:energy_flow} can be computed on each beam of the grid, thus providing the vector field of the energy flow, given in Fig.~\ref{fig:energy_flow} for the cases of an applied concentrated moment or force, which is considered in the previous two sections.
Moreover, the outgoing flux $Q$ across a circular path (of radius $r$ and outward unit normal $\bn$), centered at the loading point, 
\begin{equation}
Q(r) = \int_{0}^{2\pi} \langle\bq\rangle \scalp \bn\, r\, d\theta,
\end{equation}
is reported in Fig.~\ref{fig:FlussiQ}. 
As the energy conservation requires the flux $Q$ to be independent of the radius, this independence is used to verify the accuracy of the simulations as well as to compare the amount of mechanical power absorbed by the lattice for different frequencies and loads.

The comparison between the vectorial representations reported in Fig.~\ref{fig:energy_flow} and the corresponding displacement fields (referenced in the captions) clearly shows that the directions of the energy flow are in nice agreement with the wave patterns computed in the previous sections.
Considering the case of concentrated moment, the symmetry of the load produces a peculiar rotational symmetry in the directions of propagation of the energy, exhibiting different degrees of localization, which depend on the frequency.
Comparing, for instance, Fig.~\ref{fig:energy_flow}a and~\ref{fig:energy_flow}d, the energy flows along four and eight radial preferential directions, respectively, and in both cases the intensity of the flow decreases with the distance from the load due to the corresponding increase of the length of the wavefront.

Figs.~\ref{fig:energy_flow}b and~\ref{fig:energy_flow}c show that the anisotropy of the energy flow is less significant at the frequencies $\Omega=0.8363$ and $\Omega=1.5915$, where, in fact, the Fourier transforms indicate the prevalence of wave vectors corresponding to almost circular slowness contours (Figs.~\ref{fig:Vertex}d and~\ref{fig:HighRegime}d).

The case of applied force (Figs.~\ref{fig:energy_flow}e and~\ref{fig:energy_flow}f) differs strongly from the case of applied moment, as the in-plane load breaks the rotational symmetry.
This is clearly evident in Figs.~\ref{fig:energy_flow}e, where the pulsating horizontal force induces an energy flow propagating in the vertical direction, forming two symmetric `triangular' streams of decaying intensity.
Another interesting effect emerges at the frequency of the axial waves $\Omega = \Omega_a = \lambda/\pi \approx 1.5915$, for which the energy transmitted by the force exhibits an extremely localized unidirectional propagation, as shown in Fig.~\ref{fig:energy_flow}f, where the force is applied horizontally and the energy flows along a strongly localized `channel' without attenuation.

\section{Concluding remarks}
\label{sec:conclusions}

Localization of vibration in various complex forms (\lq channels' or \lq X-', \lq cross-', \lq star-' shaped narrow modes), anisotropic --but also isotropic-- wave propagation, and Dirac cones and flat bands in the dispersion surfaces have been shown to be possible at various frequencies, through Floquet-Bloch exact treatment and numerical analysis of a rectangular grid of Rayleigh elastic beams with diffused mass. 
The presented results demonstrate that these effects can be designed by tuning the aspect ratio of the grid, the slenderness and the rotational inertia of the beams. 
Therefore, additive manufacturing technologies can in principle be used to produce microstructured materials with engineered vibrational properties.

\vspace{10mm}
{\small
\noindent
\textbf{Data accessibility.} This article has no additional data.

\noindent
\textbf{Authors' contributions.} All authors contributed equally to this work and gave their final approval for publication.

\noindent
\textbf{Competing interests.} We declare we have no competing interests.

\noindent
\textbf{Funding.} G.B., L.C., A.P., gratefully acknowledge financial support from the ERC Advanced Grant `Instabilities and nonlocal multiscale modelling of materials' ERC-2013-ADG-340561-INSTABILITIES.
D.B. thanks financial support from the PRIN 2015 `Multi-scale mechanical models for the design and optimization of micro-structured smart materials and metamaterials' 2015LYYXA8-006.
}

\printbibliography

\end{document}

%% file: definitions.tex
\newcommand{\beq}{\begin{equation}}
\newcommand{\eeq}{\end{equation}}
\newcommand{\lb}{\label}
\newcommand{\beqar}{\begin{eqnarray}}
\newcommand{\eeqar}{\end{eqnarray}}
\newcommand{\bit}{\begin{itemize}}
\newcommand{\eit}{\end{itemize}}
\newcommand{\benum}{\begin{enumerate}}
\newcommand{\eenum}{\end{enumerate}}
\newcommand{\barr}{\begin{array}}
\newcommand{\earr}{\end{array}}

\def\scalp{\mbox{\boldmath $\, \cdot \,$}}

\newcommand{\deriv}[2]{\frac{\partial #1}{\partial #2}}
\newcommand{\nderiv}[3]{\frac{\partial^{\,#3} #1}{\partial #2^{\,#3}}}

\def\XXint#1#2#3{{\setbox0=\hbox{$#1{#2#3}{\int}$}
   \vcenter{\hbox{$#2#3$}}\kern-.5\wd0}}

\def\b0{\mbox{\boldmath $0$}}
\def\ba{\mbox{\boldmath $a$}}

\def\bc{\mbox{\boldmath $c$}}

\def\be{\mbox{\boldmath $e$}}

\def\bk{\mbox{\boldmath $k$}}

\def\bn{\mbox{\boldmath $n$}}

\def\bq{\mbox{\boldmath $q$}}

\def\bt{\mbox{\boldmath $t$}}
\def\bu{\mbox{\boldmath $u$}}

\def\bA{\mbox{\boldmath $A$}}

\def\bK{\mbox{\boldmath $K$}}

\def\f0{\ensuremath{\mathbb{O}}}

\newcommand{\mE}{\ensuremath{\mathcal{E}}}
\newcommand{\mF}{\ensuremath{\mathcal{F}}}

\newcommand{\mT}{\ensuremath{\mathcal{T}}}

\def\Re{\mathop{\mathrm{Re}}}

